\documentclass[a4paper,11pt]{article}
\usepackage{jheppub} 

\usepackage{lineno}

\usepackage{tikz}
\usepackage{placeins}
\usepackage{tikz, subfigure}
\usetikzlibrary{arrows.meta, positioning, quotes,decorations.markings}
\usetikzlibrary{decorations.markings}
\usetikzlibrary{calc}

\usepackage[numbers,sort&compress]{natbib}
\usepackage{graphicx}
\graphicspath{ {./images/} }
\usepackage{amsmath}
\usepackage{amsthm}
\usepackage{amssymb}
\usepackage{mathtools}

\theoremstyle{definition}

\newtheorem{theorem}{Theorem}[section]

\usepackage{xspace}
\usepackage{mathtools}
\usepackage[all]{xy}
\usepackage{tikz-cd}
\usepackage[dvipsnames]{xcolor}

\usepackage{graphicx}
\graphicspath{ {./images/} }
\usepackage{hyperref}

\usepackage{mathtools,amssymb,scalerel}
\usepackage{physics}

\usepackage{parskip}
\usepackage{xspace}
\def\bbZ{\mathbb{Z}}

\newcommand{\Poincare}{Poincar\'e\xspace}
\newcommand{\Cech}{\v{C}ech}
\newcommand{\calZ}{\mathcal{Z}}
\newcommand{\calD}{\mathcal{D}}

\usepackage[most]{tcolorbox}
\usepackage{relsize}

\usepackage{slashed}
\usepackage{pdfpages}

\title{\boldmath On Gauging Finite Symmetries by Higher Gauging Condensation Defects}

\abstract{Based on the work by C{\'o}rdova-Costa-Hsin \cite{Cordova:2024mqg}, we propose an EFT-style, Lagrangian procedure to gauge finite 0-form symmetries in untwisted Dijkgraaf–Witten gauge theories on closed oriented manifolds using higher gauging condensation defects and point out its limitations.  Using this proposal, we construct effective actions of untwisted Dijkgraaf-Witten theories with Heisenberg gauge group over $\mathbb{Z}_p$ and show that the braiding data from Hopf link and the fusion rules match with the expected discrete gauge theories. We also study the symTFT implications of these effective Lagrangians and clarify their relations with higher group global symmetries. }

\author[a]{Yuan Xue,}
\affiliation[a]{Department of Physics, The University of Texas at Austin, Austin, TX 78712, USA}
\emailAdd{yuan\_xue@utexas.edu}

\author[b]{and Eric Y. Yang}
\affiliation[b]{Department of Physics, University of California, San Diego, 9500 Gilman Drive \# 0319
, La Jolla, CA 92093, USA }
\emailAdd{yuy073@ucsd.edu}

\author[c]{and Zipei Zhang}
\affiliation[b]{Department of Physics, University of California, San Diego, 9500 Gilman Drive \# 0319
, La Jolla, CA 92093, USA }
\emailAdd{ziz011@ucsd.edu}

\begin{document}
\maketitle
\flushbottom

\section{Introduction} \label{section - introduction}
     
     Symmetries are powerful tools in quantum field theory. From the modern generalized symmetry perspective \cite{Gaiotto:2014kfa}, global symmetries are implemented by a collection of topological defects.
     The appropriate mathematical structure to describe these symmetry defects is a higher fusion category \cite{Bah:2025oxi,Bhardwaj:2022yxj}. The generalized definition of global symmetry has led to novel constraints on renormalization group (RG) flows, IR phases of gauge theories \cite{Chang:2018iay, Cordova:2019jqi,Gaiotto:2017yup,Komargodski:2017keh, Gaiotto:2017tne,Gomis:2017ixy, Komargodski:2020mxz, Cordova:2018cvg}, and new implications on phenomenological models \cite{Cordova:2022qtz, Brennan:2020ehu,Aloni:2024jpb,Brennan:2024iau}. See \cite{Brennan:2023mmt,Schafer-Nameki:2023jdn, Shao:2023gho,Costa:2024wks} for introductions.

     An especially interesting class of generalized symmetries is called non-invertible symmetries, whose fusion rules are characterized by fusion coefficients valued in topological quantum field theories (TQFTs). In the earlier literature, non-invertible symmetries first appeared in the study of 2D conformal field theories (CFTs) \cite{Fuchs:2002cm,Fuchs:2003id,Fuchs:2004dz,Fuchs:2004xi, Fjelstad:2005ua, Moore:1989vd}. Recently, they have been realized in higher spacetime dimensions by field theoretic methods \cite{Choi:2021kmx, Choi:2022zal, Choi:2023pdp, Choi:2022fgx, Choi:2022rfe, Choi:2022jqy,Roumpedakis:2022aik}, categorical constructions \cite{Bhardwaj:2017xup,Bhardwaj:2022kot,Bhardwaj:2022maz,Lu:2024lzf,Lu:2022ver}, D-brane constructions \cite{Apruzzi:2021nmk,Apruzzi:2022rei, Apruzzi:2023uma}, and stabilizer code models \cite{Lu:2024ytl, Seifnashri:2025fgd,Gorantla:2024ocs,Seiberg:2024gek, Choi:2024rjm}. Among these constructions, many of the non-invertible symmetry defects were realized as higher gauging condensation defects \cite{Roumpedakis:2022aik}. Especially in $(2+1)$D TQFT, it was proposed that all 0-form symmetries arise from higher gauging condensation defects \cite{Roumpedakis:2022aik}. 

     Given topological nature of symmetry defects, it is helpful to develop a formalism that separates the symmetry defect actions from the dynamical degrees of freedom of a QFT. For example, consider a pure Yang-Mills theory in $D$ spacetime dimension with a simply connected $\mathrm{SU}(N)$ gauge group. One can define Wilson lines of arbitrary $\mathrm{SU}(N)$ representations, which can end on local operator transforming in the adjoint representation of $\mathrm{SU}(N)$. The Wilson lines that are endable on local operators are said to be screened. The set of all Wilson lines in this theory can be labeled by their restrictions to the $\mathbb{Z}_N$ center subgroup. The quotients of all $\mathbb{Z}_N$ charges by the central charges of the screened Wilson lines define the 1-form symmetry group, which is $\mathbb{Z}_N$ for the pure $\mathrm{SU}(N)$. The symmetry defects are implemented by codimension-2 Gukov-Witten operators supported on closed oriented submanifolds. The Gukov-Witten operators are in one-to-one correspondence with the conjugacy classes of the gauge group and not all of them are topological. In fact, consider a pure $G$ gauge theory, where $G$ is a compact gauge group. The topological Gukov-Witten operators are in one-to-one correspondence with the conjugacy classes contained in the centralizer $Z_G(G_0)$ of the identity component $G_0$ of the group \cite{Heidenreich:2021xpr}\footnote{It was pointed out in \cite{Heidenreich:2021xpr} that all these conjugacy classes have finite sizes for continuous compact Lie groups. For discrete groups, $Z_G(G_0)$ is reduced to the centralizer of the identity element of $G$, so $T_{[g]}^{\mathrm{GW}}$ are labeled by $G$ conjugacy classes and all of them are topological in the absence of matter.}. The action of a topological Gukov-Witten operator on a Wilson line is \cite{Rudelius:2020orz,Heidenreich:2021xpr}:
     \begin{equation}\label{eq - gauge theory EM linking}
         T_{[g]}^{\mathrm{GW}}(S^{D-2})W_{\rho}(\gamma) = \frac{\chi_\rho(g)}{\chi_{\rho}(1)}\text{size(g)}W_{\rho}(\gamma),
     \end{equation}
     where ``size(g)" denotes the size of the conjugacy class $[g]$, which is the quantum dimension of the topological Gukov-Witten operator. Going from the LHS and RHS, we simply shrink the topological Gukov-Witten operators to a point. The linking between $S^{D-2}$ and $\gamma$ is an obstruction to this shrinking and it is also responsible to the ratio $\chi_{\rho}(g)/\chi_{\rho}(1)$. The factor \text{size(g)} is the remnant of shrinking $T_{[g]}^{\mathrm{GW}}(S^{D-2})$ in the absence of $W_\rho(\gamma)$. Since $\chi_{\rho}(g)/\chi_{\rho}(1)$ is necessarily a U$(1)$ phase, the symmetry action can be equivalently expressed as the linking between the Wilson lines and 't Hooft operators in a $\mathbb{Z}_N$ discrete gauge theory in a TQFT defined on a cylinder $M_D\times I$. This TQFT is known as the symTFT \cite{Freed:2022qnc, Brennan:2024fgj,Apruzzi:2024htg, Jia:2025jmn,Antinucci:2024zjp,Bonetti:2024cjk,Pace:2025hpb,Apruzzi:2025hvs} of the pure $\mathrm{SU}(N)$ Yang-Mills theory. In this sense, we have achieved an embedding of the generalized symmetry action into a untwisted $\mathbb{Z}_N$ Dijkgraaf-Witten (DW) theory in $M_D\times I$. 

    The TQFTs modeling the symmetry defects are generally described by higher category theory and they are known as \textbf{\textit{fully extended local TQFTs}}\cite{Lurie2009OnTC, Lurie2006HigherTT,Baez:1995xq}\footnote{Typically, we do not need the full machinery for physics purposes, but all the manipulations of topological symmetry defects for finite symmetries should fall into this category.}.  A simple class of examples are \textbf{\textit{Dijkgraaf-Witten theories}}\cite{Dijkgraaf:1989pz}. It is well-known that an untwisted Dijkgraaf-Witten theory with finite abelian gauge group $\mathbb{Z}_N$ has a BF type action \cite{Banks:2010zn,Kapustin:2014gua}:
    \begin{equation} \label{eq - p-form BF Action}
        I = \frac{iN}{2\pi}\int_{M_D} \tilde{a}_{D-p-1}\wedge d a_p,
    \end{equation}
    in terms of U$(1)$ connections, or equivalently:
    \begin{equation}\label{eq - p-form BF Action Cocycle Description}
        I = \frac{2\pi i}{N}\int_{M_D} a_{D-p-1}\cup d a_p,
    \end{equation}
    in terms of $\mathbb{Z}_N$-valued cocycles. Recently, a Lagrangian description for untwisted Dijkgraaf-Witten theory with a $\mathbb{D}_4$ gauge group was proposed in \cite{Cordova:2024mqg,Cordova:2024jlk}:
    \begin{equation} \label{eq - D4 Action}
        I = i\pi\int_{M_D}\left(\tilde{a}_{D-1}\cup da_1+  \tilde{b}_{D-1}\cup db_1 + \tilde{c}_{D-1}\cup dc_1 + a_1 \cup \tilde{b}_{D-1}\cup c_1\right),
    \end{equation}
    where $\mathbb{D}_4$ is the dihedral group of order 8. Since $\mathbb{D}_4 \simeq (\mathbb{Z}_2\times \mathbb{Z}_2)\rtimes \mathbb{Z}_2$, where the twist sends $(1,0)$ to $(1,1)$, the untwisted $\mathbb{D}_4$ DW theory can be thought of the untwisted $\mathbb{Z}_2\times\mathbb{Z}_2$ DW theory with the said $\mathbb{Z}_2^{(0)}$ symmetry gauged. In the $\mathbb{Z}_2\times\mathbb{Z}_2$ gauge theory, the $\mathbb{Z}_2$ symmetry is generated by a higher gauging condensation defect \cite{Cordova:2024mqg}:
    \begin{equation}
        S(\Sigma) = \frac{1}{\abs{H_1(\Sigma,\mathbb{Z}_2)}}\sum_{\substack{\gamma\in H_1(\Sigma, \mathbb{Z}_2)\\
            \Gamma \in H_{D-2}(\Sigma, \mathbb{Z}_2)}}(-1)^{\langle \gamma , \Gamma\rangle  }W_1(\gamma) M_2(\Gamma).
    \end{equation}
    Since it generates a $\mathbb{Z}_2$ group-like symmetry, it necessarily follows the fusion rule $S(\Sigma)^2 = 1$. This implies that it can be expressed as a U$(1)$ phase by explicitly carrying out the sum in $S(\Sigma)$:
    \begin{equation}
        S(\Sigma) = e^{i\pi \int_{\Sigma}a_1\cup\tilde{b}_{D-2}}.
    \end{equation}
    To gauge this symmetry, one simply inserts $S(\Sigma)$ over all codimension-1 cycles of the spacetime. The \Poincare dual statement of this is the coupling of a discrete torsion term:
    \begin{equation}
        I_{\text{torsion}} = i\pi\int_{M_D} a_1\cup\tilde{b}_{D-2}\cup c_1, 
    \end{equation}
    where $c_1$ is a $\mathbb{Z}_2$-valued cochain. Finally, we need to ensure the flatness of $\mathbb{Z}_2$ background gauge field, which can be achieved by introducing a Lagrange multiplier:
    \begin{equation}
        I_c = i\pi \int \tilde{c}_{D-2}\cup dc_1
    \end{equation}
    to the action. Collecting all the terms, one obtains the action in Eq. \eqref{eq - D4 Action}.  

    This procedure is analogous to the gauging of continuous symmetries by the Noether procedure. For example, consider the Lagrangian for massive fermions in (3+1)D:
    \begin{equation}
        \mathcal{L}_{\text{Dirac}} = i\Bar{\Psi}i\slashed{\partial}\Psi - m\Bar{\Psi}\Psi,
    \end{equation}
    which has a $\mathrm{U}(1)$ global symmetry transforming fermions as $\Psi \mapsto e^{-i\alpha}\Psi$ and $\Bar{\Psi}\mapsto e^{i\alpha}\Psi$ with a Noether current $j^{\mu} = \Bar{\Psi} \gamma^\mu \Psi$. To gauge the $\mathrm{U}(1)$, we have a two-step procedure: (1) inserting the conserved current into the theory by coupling it to a $\mathrm{U}(1)$ background gauge field; (2) promoting the background gauge field to a dynamical gauge field. This produces the standard QED Lagrangian:
    \begin{equation}
        \mathcal{L}_{\text{Dirac}} = -\frac{1}{4}F_{\mu\nu}F^{\mu\nu} + i\Bar{\Psi}i\slashed{\partial}\Psi - m\Bar{\Psi}\Psi + e\Bar{\Psi}\gamma^\mu \Psi A_\mu.
    \end{equation}
    Going back to the discrete gauge theory case, by comparison it is then natural to identify the $U(1)$ phase representing the $\mathbb{Z}_2$ symmetry generator as an analog of the conserved current. Similarly, the discrete torsion term is just the insertion of the conserved current into the action. Finally, turning on the BF kinetic term $i\pi\int \tilde{c}_{M_D}\cup dc_1$ completes the gauging. 
    
    Despite of this interesting analogy, we must stress here that this similarity is purely formal. The derivation of a conserved current by the Noether procedure requires a continuous parameter parameterizing the symmetry group, which is absent for discrete symmetries. In principle, one simply cannot define a Noether current for a discrete symmetry.  Therefore, it should be unsurprising that this procedure might fail to describe the actual gauging procedure for a 0-form symmetry of a Dijkgraaf-Witten theory. One focus of this work is to describe the details of this Lagrangian description, outline an analysis of the discrete gauge theory after the gauging, and point out a rough range of validity. We stress that this manipulation should be understood as an \textbf{\textit{effective field theory description of symmetry gauging}} and it should not be taken as a canonical definition.

    Since higher gauging condensation defects constructed in \cite{Roumpedakis:2022aik, Cordova:2024jlk, Cordova:2024mqg} are only valid for abelian gauge theories, we will only focus on the gauging of finite abelian symmetries in terms finite abelian gauge theories. In this work, unless stated otherwise, we will explicitly use $\mathrm{U}(1)$ gauge fields. Let $H$ be the gauge group of the original theory $\mathcal{T}$ and $c_1$ be the background gauge field of some $\mathbb{Z}_K^{(0)}$ 0-form symmetry to be gauged. Since all finite abelian gauge groups admit unique prime decompositions up to isomorphisms, it suffices to consider $H = \mathbb{Z}_N$ or $H = \mathbb{Z}_N\times \mathbb{Z}_M$, where $N$ and $M$ two prime powers. When $H = \mathbb{Z}_N$, the $\mathbb{Z}_K^{(0)}$ gauged theory typically has the following action:
   \begin{equation}\label{eq - type-II action}
    I_{\mathcal{T}/\mathbb{Z}_K}=\frac{i}{2\pi}\int_{M_D}\left(N \tilde{a}_{D-2}\wedge d a_1 + M\tilde{c}_{D-2}\wedge d c_1 + \frac{p}{2\pi} a_1\wedge\tilde{a}_{D-2}\wedge c_1 \right),
    \end{equation}
    which we refer to as a \textbf{\textit{type-II action}}. When $H = \mathbb{Z}_N\times \mathbb{Z}_M$, the $\mathbb{Z}_K^{(0)}$ gauged theory typically has the following action:
    \begin{equation}\label{eq - type-I action}
    I_{\mathcal{T}/\mathbb{Z}_K}=\frac{i}{2\pi}\int_{M_D}\left( N\tilde{a}_{D-2}\wedge d a  + M\tilde{b}_{D-2}\wedge d b_1 + K\tilde{c}_{D-2}\wedge d c_1 + \frac{p}{2\pi} a_1\wedge\tilde{b}_{D-2}\wedge c_1 \right),
    \end{equation}
    which we refer to as a \textbf{\textit{type-I action}}. In both cases, $p$ is some integer subject to appropriate quantization conditions. We will show that type-II actions in terms of $U(1)$-valued gauge fields have intrinsic inconsistencies.

    Another focus of this work is to study the topological boundary conditions of type-I actions on $M_D\times I$. Here the type-I actions can be interpreted as the symTFT of higher form symmetries with a particular mixed anomaly. Due to the effective field theory nature of our analysis, not all conclusions drawn from the Lagrangian analysis can be directly mapped onto the symTFT concepts as canonically defined in \cite{Freed:2022qnc}. Nonetheless, the reduction of bulk gauge transformation to the topological boundary can be trusted. Moreover, a type-I action admits a \textbf{\textit{topological sigma model}} interpretation. Combining these two perspective, we can draw useful qualitative conclusions about the global symmetries realized by the type-I actions at the topological boundary.

    This paper is organized as follows. In Sec. \ref{section - discrete gauge theory}, we review relevant facts about TQFTs and higher gauging condensation defects. Especially, we will give a rather detailed review of topological sigma models and demonstrate how the operator manipulations in the physics literature echo with the categorical definition of TQFT in the math literature.  In Sec. \ref{section - D4 gauge theory}, we review the operator analysis of the $\mathbb{D}_4$ gauge theory action following \cite{Cordova:2024jlk, Cordova:2024mqg,He:2016xpi}. The main supporting evidences of the Lagrangian formulation are the operator fusion rules and evaluations of linking invariants that contain the character table of the $\mathbb{D}_4$ gauge group. We will carefully work out the computational details left out in these references. In Sec. \ref{section - various notions of gauging}, we generalize the Lagrangian description of finite symmetry gauging originally proposed in \cite{Cordova:2024jlk, Cordova:2024mqg}. We also provide further examples by constructing effective actions for untwisted Dijkgraaf-Witten theories with $H_3(\mathbb{Z}_p)$ gauge group in arbitrary spacetime dimensions, where $H_3(\mathbb{Z}_p)$ is the Heisenberg group over $\mathbb{Z}_p$ with prime $p$.  In Sec. \ref{section - type-I action analysis}, we perform a general analysis of type-I actions and their $q$-form generalizations on closed oriented manifolds. In Sec. \ref{section - type-I_symTFT}, we define Type-I actions on manifolds with boundaries and treat them as symTFTs. We study their physical interpretations by examining truncations of the bulk gauge transformation at the topological boundary. We point out the relations between type-I actions and higher group global symmetries by establishing a few simple no-go theorems. In Sec. \ref{section - type-II action analysis}, we study a concrete example of a type-II action in (3+1)D and show that it gives the correct on-shell constraints but a set of off-shell gauge transformations incompatible with the $U(1)$-variables. We will conclude in Sec. \ref{section - conclusion and discussion} and discuss possible future directions. In Appendix \ref{appendix - large gauge transformation}, we explicitly work out the quantization condition for the discrete torsion coefficient for both type-I and type-II actions. In Appendix \ref{appendix - small gauge transformations}, we outline a derivation of the off-shell gauge transformations for type-I actions. In Appendix \ref{appendix - gauging finite symmetry in DW}, we review a result from the math literature on finite symmetry gaugings in $(2+1)$D untwisted Dijkgraaf-Witten theories, which were used in the proposal for constructing effective actions of untwisted Dijkgraaf-Witten theories in arbitrary spacetime dimensions. Finally in Appendix \ref{appendix - Mackey's Theory}, we review the little group method from Clifford's theory that allows us to construct the character table for semi-direct products of finite abelian groups.

    \section{Discrete Gauge Theory and Higher Gauging Condensation Defects}\label{section - discrete gauge theory}

    In this section, we quickly review some relevant facts about discrete gauge theories and topological sigma models in Sec. \ref{subsection - discrete gauge theory review}. In Sec. \ref{subsection - condensation defects in untwisted DW theories}, we review the identification and fusion of higher gauging condensation defect relevant to the finite symmetry gauging in spacetime. In Sec. \ref{subsection - higher gauging condensation defects as 0-form symmetry defect}, we provide a heuristic definition of higher gauging condensation defects as 0-form symmetry defects and give a simple recipe for their construction.
    \subsection{Discrete Gauge Theories}\label{subsection - discrete gauge theory review}

    The class of TQFTs used to describe finite symmetry defects admits a categorical definition \cite{Atiyah:1989vu, Lurie2006HigherTT}. Specifically, a \textbf{\textit{$D$-dimensional fully extended local TQFT}} on an oriented manifold $M_D$ is a symmetric monoidal functor from the $D$-category of bordisms $\textbf{Bord}_D^{\xi}$ to a symmetric monoidal $D$-category $\mathcal{C}$\footnote{Technically speaking, both $\textbf{Bord}_D^{\xi}$ and $\mathcal{C}$ are $(\infty,D)$-categories. 
    }, with a choice of tangential structure $\xi$. This definition describes the interaction between topological defects on all possible submanifolds of $M_D$, where the submanifolds can have boundaries or even corners\footnote{A manifold with corner means the boundary of manifold has its own boundary.}. If we are only interested in defects supported on closed oriented submanifolds, then the TQFT we are examining is a \textbf{\textit{truncation}} of the original fully extended theory. In this work, we will always work with such a truncation. 

    Fully extended local TQFTs can be effectively studied by applying the cobordism hypothesis \cite{Baez:1995xq, Lurie2009OnTC}. To each closed oriented submanifold $X$ of $M_D$, we can associate a number called \textbf{\textit{the partition function}} $\calZ(X)$. \textbf{\textit{Locality}} means that we can break any $X$ into a collection of neighborhoods of points of $X$, and the evaluation of $\calZ(X)$ is done by evaluating the partition functions on these neighborhoods followed by gluing. Another powerful notion of the cobordism hypothesis is \textbf{\textit{duality}}, which, loosely speaking, is related to the existence of the \textit\textbf{{charge conjugation}} of a defect up to isomorphisms. See \cite{Bah:2025oxi} for further subtleties of this interpretation.  

    If a topological action is available, then we can find direct analogs of locality and duality in the usual operator manipulations in physics. For example, consider a topological action:
    \begin{equation}
        I = 2\pi i \int_{M_D}\mathcal{L}(\mathcal{F}),
    \end{equation}
    where $M_D$ is a closed oriented manifold,  $\mathcal{L}$ is a closed $D$-form, and $\mathcal{F}$ is a collection of differential-form-valued fields in the de Rham cohomology of $M_D$. Given $I$, we can derive a set of gauge transformations $\mathcal{F} \mapsto \mathcal{F} + \delta\mathcal{F}$ so that $\delta I=0$ on $M_D$. Since $\mathcal{F} \mapsto \mathcal{F} + \delta\mathcal{F}$ holds only on a local patch of $M_D$, we call it a \textbf{\textit{local (spacetime) gauge transformation}}. We can also pullback the spacetime transformation rules to a submanifold $X$ of $M_D$. The spectrum of admissable operators supported on $X$ is determined by the  worldvolume gauge invariance under this pullback. This echoes with the locality requirement. On the other hand, if a collection of algebraic data can be pulled back to $X$, we should also be able to pullback the same data to the orientation reversal $\overline{X}$. This echoes with the duality requirement\cite{Bah:2025oxi}. Finally, the fusion of two parallel $n$-dimensional defects should only produce a collection of $n$-dimensional defects \cite{Freed:2022qnc}. In operator manipulations, this requirement leads to the appearance of condensation defects. 

    Of course, we should not expect the heuristic physics manipulations to be capable of reproducing all features of a fully extended local TQFT. Here are a few possible subtleties:
    \begin{itemize}
        \item In the physics manipulations, we implicitly assume the existence of a path integral measure, which is not always guaranteed for the TQFT that we aim to model.
        \item In the physics manipulations, the gauge transformations $\mathcal{F}\mapsto \mathcal{F} + \delta \mathcal{F}$ are \textbf{\textit{off-shell gauge transformations}}. However, we can also apply the variational principle and derive a collection of transformations that deforms the equations of motion up to some consistency conditions, which we call \textbf{\textit{on-shell deformations}}. In practice, the equations of motions of the theory typically implement consistency conditions on the variables $\mathcal{F}$. If the off-shell gauge transformations do not agree with on-shell deformations, then we have ambiguities which must be eliminated by manually imposing further constraints. 
        \item For a specific TQFT that we aim to model, we can choose to define the local data $\mathcal{F}$ with different cohomology theories. However, changing from one cohomology description to another (for example, going from de Rham cohomology to simplicial cohomology) in general leads to a loss or introduction of extra information. Again, these differences must be manually tuned by introducing extra consistency conditions. 
    \end{itemize}

    Despite of all these disadvantages, the explicit Lagrangian descriptions are straightforward and offer more useful physical insights. 
    There are a few models where explicit Lagrangian descriptions are possible. They are the Dijkgraaf-Witten theories \cite{Dijkgraaf:1989pz} and their higher form generalizations. These theories provide an useful illustration of bosonic topological orders in $D=3,4$ \cite{Lan2017ClassificationO,Lan2018ClassificationO, Johnson-Freyd:2020usu}. On the other hand, for $G$ a finite discrete group, the symTFT for a 0-form $G$-symmetry can be realized as a DW theory with gauge group $G$ \cite{Freed:2022qnc, Brennan:2024fgj,Apruzzi:2024htg, Jia:2025jmn,Antinucci:2024zjp,Bonetti:2024cjk,Pace:2025hpb,Apruzzi:2025hvs}. Dijkgraaf-Witten theories are a simple example of the so-called \textbf{\textit{topological sigma models}}, where useful information of the TQFT can often be evaluated in terms of homotopy theory calculations. See \cite{Costa:2024wks} for an introduction. If a Lagrangian description of a topological sigma model is available, then we can use homotopy theory calculations as a cross check against the Lagrangian descriptions and introduce \textbf{\textit{regularizations}} when necessary. In this sense, the Lagrangian description should be understood as \textbf{\textit{an effective field theory (EFT)}} of the underlying topological sigma model. 
    
    Let us first review the definition of Dijkgraaf-Witten theories as topological sigma models. Recall that a gauge theory is defined by a principal $G$ bundle $P \xrightarrow{\pi} M_D$. For any $G$, there exists a universal covering space $EG$ that is contractible and admits a free $G$-action. Define the classifying space $BG = EG/G$. This naturally defines another principal $G$-bundle $EG \rightarrow BG$. The classifying space $BG$ satisfies the property that any $G$-bundle $P \rightarrow M_D$ can be realized as the pullback bundle by a map $f:X \rightarrow BG$:
    \begin{equation}
        \begin{tikzcd}
G \arrow{r} & P=f^*(EG)\arrow{d}  & EG\arrow{d} & \arrow{l} G\\%
{} & M_D \arrow{r}{\gamma}& BG & {}
\end{tikzcd}.
    \end{equation}
    The map $\gamma$ induces a pullback of the cohomological data $[\omega]\in H^{D}(BG,\mathrm{U}(1))$ to the physical spacetime $M_D$. The partition function is a sum over the homotopy classes of maps from $M_D$ to $BG$ weighted by some topological action:
    \begin{equation}
        \calZ_{\omega}^{BG}[M_D] = \frac{1}{\abs{G}^{b_0}}\sum_{[\gamma]: M_D \rightarrow BG} e^{2\pi i\langle \gamma^*\omega, [M_D]\rangle},
    \end{equation}
    where $\gamma^*\omega$ is the pullback action in $H^{D}(BG,\mathrm{U}(1))$, $b_0$ is the zero-th betti number of $M_D$, and $[M_D]$ is the fundamental class in $H^{D}(M_D,\mathbb{Z})$. The pairing is given by the integral $ \langle \gamma^*\omega, [M_D]\rangle = \int_{[M_D]}\gamma^*\omega$.

    The classifying space $BG$ is an example of Eilenberg-MacLane spaces. An Eilenberg-MacLane space $K(G,n)$ is specified by a discrete group $G$ and an integer $n$ so that:
    \begin{equation}
        \pi_k(K(G,n)) = \begin{cases}
            G, & k = n;\\
            0, & k\neq 0.
        \end{cases}
    \end{equation}
    Note that $K(G,n)$ only makes sense for abelian $G$ when $n\geq 2$, because $\pi_k(M)$ is abelian for $k\geq 2$. Since $K(G,n)$ admits a CW complex construction, using the natural bijection \cite{HatcherAT} between homotopy class of maps $[M_D, K(G,n)]$ and the cohomology group $H^n(M_D,G)$, we can rewrite the partition function as a sum over cohomology classes:
    \begin{equation}
        \calZ_{\omega}^{BG}[M_D] = \frac{1}{\abs{G}^{b_0}}\sum_{[A_1] \in 
        H^1(M_D, G)} e^{2\pi i\langle \omega(A_1), [M_D]\rangle},
    \end{equation}
    for $G$ abelian, where $\omega(A_1)$ is the evaluation of $\gamma^*\omega$ on $A_1$. This is a convenient representation of the theory as it echoes with the traditional definition of gauge theories in terms of gauge invariant quantities constructed from gauge connections 1-forms. Especially, the 1-form gauge field $A_1$ is a formal analog of the connection 1-form in Yang-Mills theory. In this language, $dA_1=0$ implies that the path-integral measure of Dijkgraaf-Witten theory is defined on the space of flat connections modulo gauge transformations. The generalization to $q$-form gauge theories is straightforward. We simply replace the target space with the $q$-th classifying space $B^qG\equiv K(G,q)$. The $q$-form action is defined by a cohomology class $[\omega]\in H^D(B^qG, \mathrm{U}(1))$ and the partition function reads:
    \begin{equation}
        \calZ_{\omega}^{BG}[M_D] = \frac{1}{\abs{G}^{b}}\sum_{[A_q] \in 
        H^q(M_D, G)} e^{2\pi i\langle \omega(A_q), [M_D]\rangle},
    \end{equation}
    where $b$ is an alternating sum $b = \sum_{i=0}^{q-1}b_{q-i}(M_D)$ of the $i$-th Betti-number $b_i(M_D)$. This partition function can be generalized to the topological sigma model from $M_D$ to a target space $X$ \cite{Delcamp:2019fdp}:
    \begin{equation}
        \calZ_\omega^X[M_D] = \frac{1}{N[X]}\sum_{[\gamma_0]\in \pi_0 \left(\text{Map}(M_D, X)\right)}e^{2\pi i\langle \gamma^*\omega, [M_D]\rangle},
    \end{equation}
    where the target space is $X$, the topological action is $\omega \in C^{D}(X, \mathrm{U}(1))$, $M_D$ is the compact oriented $D$-dimensional physical spacetime, and $N(X)$ is an overall normalization factor dependent on $X$. When the target space is a $k$-stage Postnikov tower, the topological model is a higher group gauge theory \cite{Delcamp:2019fdp}. We will explain them in detail in section \ref{section - type-I_symTFT}. 

    Typically, we choose to model the physical spacetime $M_D$ either on the continuum or on the lattice. By a continuum formulation, we mean that the gauge fields are valued in de Rham cohomologies or \Cech\, cohomologies of $M_D$. By a lattice formulation, we mean the gauge fields are valued in simplicial cohomologies of $M_D$. For simplicity, we choose to work on the continuum with de Rham cohomology variables in this work. These de Rham cohomology variables have $2\pi \mathbb{Z}$ periods and are often referred to as $U(1)$-valued gauge fields. We will use the two terminologies interchangeably in this work and we refer the readers to \cite{Brennan:2023mmt} for further details.

    Note that the simplicial variables and the de Rham variables do not agree with each other if $M_D$ has torsion cycles. Since the $k$-th de Rham cohomology group of $M_D$ is isomorphic to the $k$-th simplicial cohomology group of $M_D$ valued in $\mathbb{R}$, by the universal coefficient theorem, we have:
    \begin{equation}
        0 \rightarrow \text{Ext}^1(H_{k-1}(M_D,\mathbb{Z}),\mathbb{R}) \rightarrow H^k(M_d,\mathbb{R}) \rightarrow \text{Hom}(H_k(M_D, \mathbb{Z}), \mathbb{R}) \rightarrow 0.
    \end{equation}
    Ext vanishes for all torsion elements of $H_{k-1}(M,\mathbb{Z})$, so $H^k(M_d,\mathbb{R})$ does not detect the torsion cycles of $M_D$. Therefore, for simplicity we will restrict to torsionless $M_D$ in this work.

    Finally, we review some details of Dijkgraaf-Witten theories. In (2+1)D, Dijkgraaf-Witten theories are described rigorously by unitary modular tensor category \cite{Etingof}. An \textbf{\textit{untwisted Dijkgraaf-Witten theory}} with gauge group $G$ is described by a trivial action $[0]\in H^3(G,U(1))$. The spectrum of line operators contain 
    \begin{itemize}
        \item   Wilson lines $W_\rho$ labeled by irreducible representations $\rho\in \text{Irr}(G)$ of the gauge group. 
        \item 't Hooft lines $M_g$ labeled by the conjugacy classes of $g$ in $G$. 
        \item Dyon lines labeled by  $([g],\rho)$, where $\rho\in \text{Irr}(Z(g))$ and $Z(g)$ is the centralizer of $g\in G$ in $G$. 
    \end{itemize}

    The gauge invariant data of the untwisted Dijkgraaf-Witten theories admits intuitive physical interpretations in both the path integral picture and the canonical quantization picture. The operator equations and the corresponding correlation functions can be used interchangeably by shrinking. Generically, consider a string of line operators $O_1(\gamma_1)\dots O_n(\gamma_n)$, where $\gamma_i$'s are contractible loops that can have nontrivial mutual linkings. Consider a generic correlation function:
    \begin{equation}
        \expval{O_1(\gamma_1)\dots O_n(\gamma_n)} = \frac{1}{\calZ_0^{BG}(G)}\int \mathcal{D}[\Psi] e^{-S[\Psi]} O_1(\gamma_1)\dots O_n(\gamma_n).
    \end{equation}
     Specifically:
\begin{itemize}
        \item Shrinking a contractible loop produces the quantum dimension of the operator.
            \begin{equation}
                \expval{O_a(\gamma)} = d_a.
            \end{equation}
        When $d_a > 1$, the line is said to be non-invertible. All line operators in a (possibly twisted) Dijkgraaf Witten theory in $(2+1)$D have integer quantum dimensions \cite{Coste:2000tq}.
        \item The Hopf link between two lines defines the entries of the modular $S$-matrix:
        \begin{equation}
            \expval{O_a(\gamma_1) O_b(\gamma_2)} = S_{ab}.
        \end{equation}
        \item The Hopf link of a Wilson and a 't Hooft operator contains information about the character table:
            \begin{equation}\label{eq - character table from Hopf link}
            \expval{W_{\rho}(\gamma)M_g(\gamma')} = \chi_{\rho}(g) d_{[g]},
        \end{equation}
        where $d_{[g]}$ is the quantum dimension of the 't Hooft line. This is equivalent to the operator equation Eq. \eqref{eq - gauge theory EM linking} with the Wilson line shrunk to a point. 
        \item The linking between $O_a(\gamma_1) O_b(\gamma_2)$ is an obstruction to shrinking both operators and it is measured by the complex phase of $\expval{O_a(\gamma_1) O_b(\gamma_2)}$. This reasoning generalizes to $\expval{O_1(\gamma_1)\dots O_n(\gamma_n)}$, where $\gamma_i$'s are contractible loops. 
    \end{itemize}
    In the following, we will also need twisted Dijkgraaf-Witten theories with finite abelian gauge groups. The relevant facts are summarized below.
    \begin{itemize}
        \item For classification purposes, it suffices to consider gauge groups of the form $\mathbb{Z}_N\times\mathbb{Z}_M\times\mathbb{Z}_K$. The relevant cohomology is:
        \begin{equation}
            \begin{split}
                H^3(\mathbb{Z}_N\times\mathbb{Z}_M\times\mathbb{Z}_K,U(1))\simeq & \mathbb{Z}_N \oplus \mathbb{Z}_M \oplus\mathbb{Z}_K\\
                &\oplus\mathbb{Z}_{\gcd(N,M)} \oplus\mathbb{Z}_{\gcd(N,K)}\oplus\mathbb{Z}_{\gcd(M,K)}\\
                & \oplus\mathbb{Z}_{\gcd(N,M,K)}.
            \end{split}
        \end{equation}
        The Dijkgraaf-Witten twists generated by the cohomology generators in the three lines are referred to as type I/II/III twists, respectively. 
        \item The pure Wilson lines are the same as their counterparts in the untwisted case. The dyon lines are associated with projective irreducible representations of the gauge group, which are determined by the twist. The projective representations make the dyon lines generically non-invertible.  
    \end{itemize}

    Note that there exist general formulas for the modular data of a $(2+1)$D Dijkgraaf-Witten theories with any finite gauge group $G$ in terms of the representation theory data of $G$ \cite{Coste:2000tq}. The fusion rules of line operators can be reproduced by applying the Verlinde formula: 
    \begin{equation}
        N^{c}_{ab} = \sum_{x\in \text{Irr}(\mathcal{C})} \frac{S_{ax} S_{bx} S_{cx}^*}{S_{0x}},
    \end{equation}
    where $\text{Irr}(\mathcal{C})$ labels the simple anyons of the UMTC $\mathcal{C}$. In this sense, $(2+1)$D Dijkgraaf-Witten theories are solved\footnote{By solved, we mean given $G$ and $[\omega]\in H^3(G,U(1))$ as an input, we can systematically compute the operator spectrum, modular data, and fusion coefficients, although the calculations in practice can become highly technical. }. 

    In $D$-dimensional spacetime, we are interested in untwisted Dijkgraaf-Witten theories. The relation between shrinking of operators supported on contractible cycles and their correlation functions still makes sense. Eq. \eqref{eq - gauge theory EM linking} still holds, where the 't Hooft operators and Wilson operators are supported on $S^{D-2}$ and $S^1$ that form a Hopf link. Note that Dijkgraaf-Witten theories in its full form should be treated as an extended TQFT, but in this work we are only interested in a truncation where the Wilson lines are supported on path-connected, closed, oriented 1-dimensional submanifolds and the 't Hooft surfaces are supported on path-connected closed oriented codimension-2 submanifolds. In this case, it only makes sense to discuss mutual fusions among Wilson lines and mutual fusions among 't Hooft surfaces, which are uniquely fixed by the group theory $G$. The fusion rules are analogous to their $(2+1)$D counterparts and there exist some dimension-independent behaviors. We will explore these features in Sec. 
    \ref{section - D4 gauge theory}.

    \subsection{Condensation Defects in Untwisted DW Theories}\label{subsection - condensation defects in untwisted DW theories}

    In this subsection, we review the results of higher gauging condensation defects in untwisted abelian DW theories in \cite{Cordova:2024mqg, Cordova:2024jlk} on the lattice. The topological domain walls and their higher codimension generalizations can be cleanly organized in terms of the folding trick and the restriction of the bulk gauge group. We can also construct these objects on the continuum when the gauge group is abelian using the integer lift $(\text{lattice}) = \frac{2\pi}{N}(\text{continuum})$. 
    
    As previously mentioned, a $p$-dimensional defect has its own partition function. Specifically, it has a worldvolume gauge group defined as a restriction $H \triangleleft G\times G$ of the spacetime gauge group $G$, where the $G\times G$ comes from folding trick considerations. We can also associate an appropriate worldvolume Dijkgraaf-Witten action $\alpha\in H^{p}(H,\mathrm{U}(1))$, known as the \textbf{\textit{discrete torsion}}.
 
\begin{figure}[h]
        \centering
        \begin{tikzpicture}[scale=0.7, every node/.style={scale=0.7}]
            \tikzset{
                midarrow/.style={
                  postaction=decorate,
                  decoration={
                    markings,
                    mark=at position 0.5 with {\arrow{>}}
                  }
                }
              }
            
            \coordinate (A) at (0,0);
            \coordinate (B) at (10,0);
            \coordinate (C) at (5,5);
            \coordinate (C2) at (5,2.5);
            \coordinate (D) at (5,0);
            \coordinate (E) at (5,5.5);
            
            \node at ($(E)+(0,0.5)$) [scale=1.6]{$\mathcal{D}_H(\Sigma)$};
            
            \draw [thick, midarrow] (A) -- node[below, scale=1.4]{$g_Lk_L^{-1}$} (D);
            \draw [thick, midarrow] (B) -- node[below, scale=1.4]{$g_Rk_R^{-1}$} (D);
            \draw [thick, midarrow] (B) -- node[above, scale=1.4]{$\; \ \ \ g_Rh_R$} (C);
            \draw [thick, midarrow] (A)-- node[above, scale=1.4]{$g_Lh_L\ \ \ \ $} (C);
            
            \node at ($(D)+(0,-0.35)$) [scale=1.4]{$v_3$};
            \node at ($(A)+(-0.2,0)$) [scale=1.4]{$v_1\ \ $};
            \node at ($(B)+(0.2,0)$) [scale=1.4]{$\ \ v_2$};
            \node at ($(C)+(0,0.25)$) [scale=1.4]{$\quad \ \ \ v_4$};

            \draw [double, ultra thick] (C) -- (E);
            
            \draw [ultra thick, double, midarrow] (D) -- node[right, scale=1.4]{$( k_Lh_L ,k_Rh_R)$} (C2) -- (C);
            
            \filldraw[black] (A) circle (1pt);
            \filldraw[black] (B) circle (1pt);
            \filldraw[black] (C) circle (2pt);
            \filldraw[red] (D) circle (2pt);
        \end{tikzpicture}
        \caption{An equivalent configuration where we performed a defect worldvolume gauge transformation $(k_L, k_R)\in H$ at site $v_3$. }
        \label{fig - holonomy across domain wall 2}
    \end{figure}
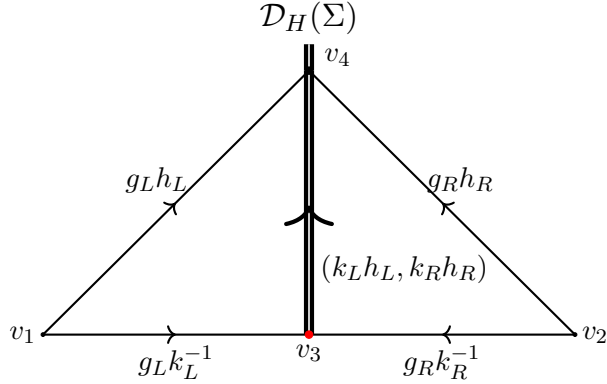

    Let us start from the codimension-1 case, which corresponds to domain walls. The domain walls can be conveniently labeled by the restriction of the spacetime gauge group $G$ to the defect worldvolume. Let $\Sigma_{D-1}$ be a closed oriented connected submanifold, by the folding trick we can define a gauge group $H \triangleleft G\times G$ and associate to $\Sigma_{D-1}$ a twist $\alpha \in H^{(D-1)}(H,\mathrm{U}(1))$. In terms of simplicial variables, a domain wall introduces nontrivial holonomies along a closed loop piercing the wall. See Fig \ref{fig - holonomy across domain wall} and \ref{fig - holonomy across domain wall 2} for an illustration:

    \begin{figure}[h]
        \centering
            
        \begin{tikzpicture}[scale=0.7, every node/.style={scale=0.7}]
          \tikzset{
            midarrow/.style={
              postaction=decorate,
              decoration={
                markings,
                mark=at position 0.5 with {\arrow{>}}
              }
            }
          }
        \coordinate (A) at (0,0);
        \coordinate (B) at (10,0);
        \coordinate (C) at (5,5);
        \coordinate (C2) at (5,2.5);
        \coordinate (D) at (5,0);
        \coordinate (E) at (5,5.5);
        
        \node at ($(E)+(0,0.5)$) [scale=1.6]{$\mathcal{D}_H(\Sigma)$};
        
        \draw [white] ($(A)+(-0.5,-0.2)$) -- node[below,scale=1.4,black]{$g_L$} ($(D)+(0,-0.2)$);
        \draw [white] ($(D)+(0,-0.2)$) -- node[below,scale=1.4,black]{$g_R$} ($(B)+(0.5,-0.2)$);
        
        \draw [thick, midarrow] (A) -- (D);
        \draw [thick, midarrow] (B) -- (D);
        \draw [thick, midarrow] (B) -- node[above, scale=1.4, black]{$\ \ \ \ g_Rh_R$} (C);
        \draw [thick, midarrow] (A)-- node[above, scale=1.4, black]{$ g_L h_L\ \ \ \ $} (C);
        \draw [double,ultra thick, midarrow] (D) -- node[right, scale=1.4]{$(h_L,h_R)$} (C2) -- (C);
        \draw [double,ultra thick] (C) -- (E);
        
        \node at ($(D)+(0,-0.35)$) [scale=1.4]{$v_3$};
        \node at ($(A)+(-0.2,0)$) [scale=1.4]{$v_1\ \ $};
        \node at ($(B)+(0.2,0)$) [scale=1.4]{$\ \ v_2$};
        \node at ($(C)+(0,0.25)$) [scale=1.4]{$\quad \ \ \ v_4$};

        \filldraw[black] (A) circle (1pt);
        \filldraw[black] (B) circle (1pt);
        \filldraw[black] (C) circle (2pt);
        \filldraw[black] (D) circle (2pt);
        
        \end{tikzpicture}
        \caption{Example of holonomies in the presence of a domain wall $\mathcal{D}_H$. The holonomy $(v_1,v_3,v_2,v_4,v_1)$ piercing the wall is nontrivial, while other holonomies remain trivial.}
        \label{fig - holonomy across domain wall}
    \end{figure}
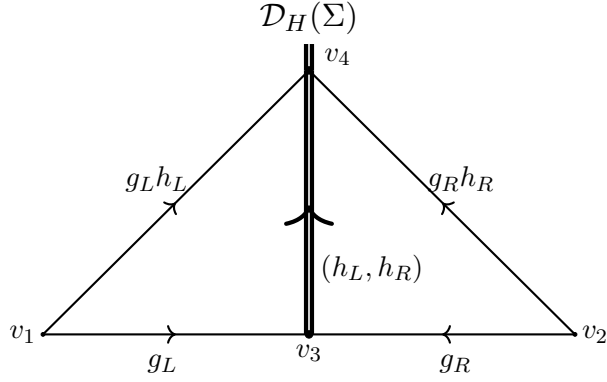

    Special classes of examples include:
    \begin{itemize}
        \item Automorphism domain walls $\mathcal{D}_G^{(\phi)}$: take $H = G$, with $\phi \in \text{Aut}(G)$. As the name suggests, these domain walls implement $\phi \in \text{Aut}(G)$ transformations on the Wilson and 't Hooft operators. The discrete torsion is typically nontrivial. An orientation reversal sends $\mathcal{D}_G^{\phi}$ to $\mathcal{D}_G^{\phi^{-1}}$.
        \item Diagonal domain walls $\mathcal{D}_{K,\alpha}^{(\text{id})}$: take $H = K \triangleleft G$ with discrete torsion $\alpha\in H^{(D-1)}(K,\mathrm{U}(1))$. The diagonal domain walls are invariant under orientation reversals on $\Sigma_{D-1}$.
        \item Magnetic Domain walls $\mathcal{D}_{G\times G}$: take $H = G\times G$ with a trivial discrete torsion. The magnetic domain walls are also invariant under orientation reversal on $\Sigma_{D-1}$. 
    \end{itemize}
    
   These three types of domain walls admit closed fusion rules and a partial list reads\cite{Cordova:2024jlk, Cordova:2024mqg}:
   \begin{align}
       \mathcal{D}_{G^{(\phi)}}\times \mathcal{D}_{G^{(\phi')}} &= \mathcal{D}_{G^{(\phi\circ \phi')}},\\
       \mathcal{D}_{K^{\text{(id)}},\alpha}\times \mathcal{D}_{{K'}^{\text{(id)}},\alpha'} &= \frac{\abs{G}}{\abs{K\cdot K'}}\mathcal{D}_{K\cap K',\alpha\cdot\alpha'}^{(\text{id})},
   \end{align}
   where $K\cdot K'$ denotes the product of subgroup $G$ generated by $K$ and $K'$:
   \begin{equation}
       K\cdot K' = \{kk' \; |\; k\in K, k'\in K'\}.
   \end{equation}
   Since $K, K' \triangleleft G $, $K\cdot K' \triangleleft G$ and it has cardinality $\abs{K}\abs{K'}/\abs{K\cap K'}$. The diagonal walls admit codimension-$n$ generalizations, where the defect worldvolume gauge group is again a normal subgroup $K \triangleleft G\times G$ with a possible discrete torsion $\alpha \in H^{D-n}(K,U(1))$. 
    \subsection{Higher Gauging Condensation Defects as 0-form Symmetry Defects} \label{subsection - higher gauging condensation defects as 0-form symmetry defect}

    In this subsection, we outline a simple procedure for constructing the higher gauging condensation defects generating 0-form group-like symmetries in untwisted DW theories with abelian gauge groups.  

\begin{figure}[htb]
        \centering
        \includegraphics[scale=0.3]{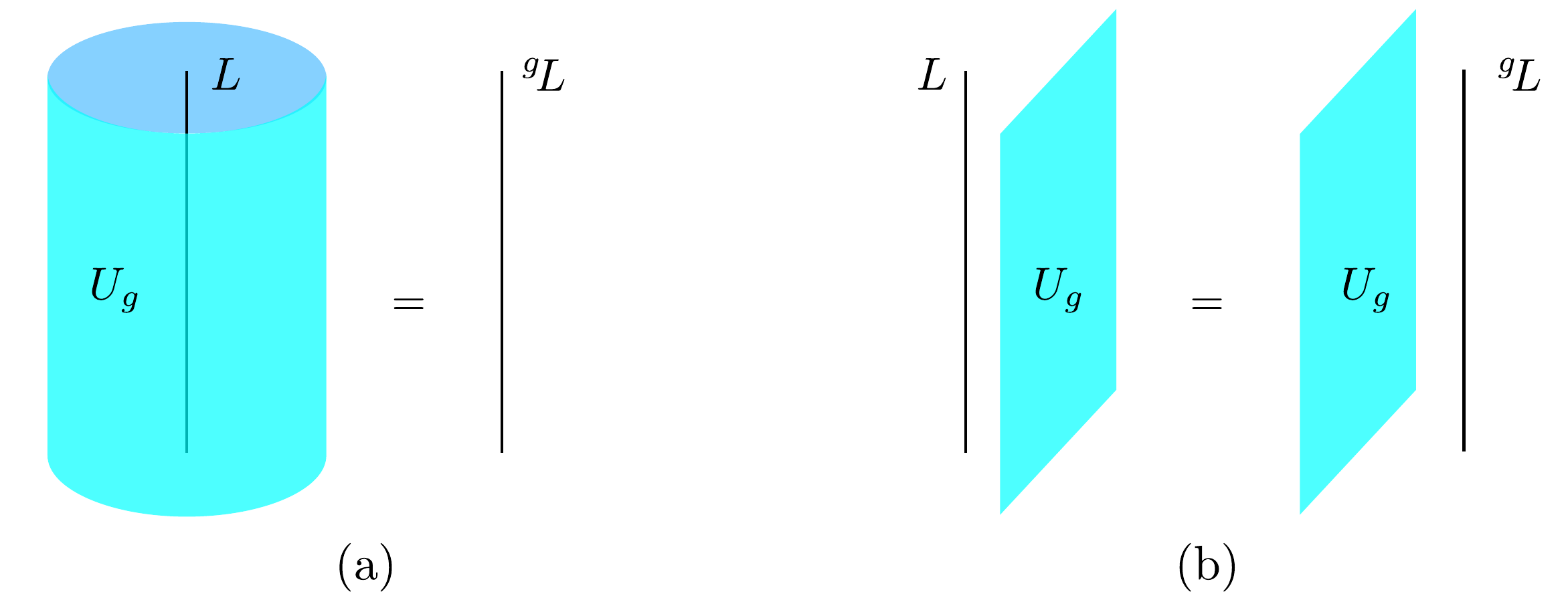}
        \caption{Two equivalent notions of a symmetry transformation on a line operator by a symmetry defect. We associate $g\in 
        G$ to the symmetry defect $U_g$ and denote the line as $L$. The action of the symmetry defect on the line is denoted as $L \mapsto {}^g\!L$.   }
        \label{fig - domain_wall_op}
    \end{figure}

    For an abelian gauge group $G$, a large class of symmetry actions is induced by automorphism actions of $G$. They are generated by the automorphism domain walls \cite{Cordova:2024mqg, Cordova:2024jlk}:
    \begin{equation}
        \label{eq - automorphism wall action}\mathcal{D}_{G^{(\phi)}}\cdot W_{\rho_i} = W_{\rho_i\cdot \phi^{-1}}, \qquad \mathcal{D}_{G^{(\phi)}}\cdot M_g = M_{\phi (g)}.
    \end{equation}
    In this notation, the symmetry action is implemented by wrapping a wall $\mathcal{D}_{G^{(\phi)}}$ around a tubular neighborhood of the line operator. Shrinking the wall to zero implements a $\phi$ transformation on the wrapped operator as Fig. \ref{fig - domain_wall_op}(a). We can also deform the process into the configuration shown in Fig. \ref{fig - domain_wall_op}(b) \cite{Roumpedakis:2022aik}. Namely pushing a charged operator across the wall implements the $\phi$ symmetry transformation. In this work, we take Fig. \ref{fig - domain_wall_op}(b) \cite{Roumpedakis:2022aik} as the canonical move for symmetry transformations and we are only interested in the gauging of $G$-automorphism symmetries. 

    Recall that any finite abelian group admits a unique prime decomposition. The higher gauging condensation defect representations of the elementary automorphisms for prime decomposed gauge groups have been worked out in \cite{Cordova:2024jlk,Cordova:2024mqg}. Here we outline a more heuristic construction for the automophism symmetry defects. 

    Let us start in (2+1)D, where a TQFT with loop excitations is described by a unitary modular tensor category (UMTC). An anyon symmetry transformation is defined by a braided auto-equivalence, which leaves the vacuum line exactly invariant, and the other gauge invariant quantities $N_{ab}^c$, $d_a$, $\theta_a$, $S_{ab}$ invariant up to permutations of the anyon lines \cite{Barkeshli:2014cna}. This also implies that the domain wall implementing the anyon symmetry should not absorb or emit any anyons\footnote{We thank Yi-Zhuang You for a helpful discussion on this intuitive definition. }. In the absence of symmetry fractionalization, the fusion of the symmetry defects follows the usual group multiplication of the 0-form symmetry group. Here we adopt a more heuristic definition in terms of tunneling matrices \cite{Lan:2014uaa}. Let $a$ label a simple anyon and ${}^g\!a$ label the image of the anyon under a $g$-transformation where $g\in G^{(0)}$. The construction follows the simple steps:
    \begin{itemize}
        \item Specify a collection of tunneling matrices $W_{a,{}^g\!a}$ which satisfy the requirement of anyon symmetries. Associate each $W_{a,{}^g\!a}$ to a higher gauging condensation defect. 
        \item Use the folding trick and the tunneling matrix to determine the anyons to be condensed on the symmetry defect.
        \item Since pushing an anyon through a symmetry defect leaves the symmetry defect itself invariant, the symmetry defect must be associated with an appropriate discrete torsion $H^2(G,\mathrm{U}(1))$ so that the nontrivial braiding phase between the anyon line and the condensed line can be absorbed by the symmetry defect. This finishes the construction of individual symmetry defects.
        \item Finally, check that the fusions of symmetry defects follow $G^{(0)}$ group multiplication. 
    \end{itemize}

    Take the untwisted $\mathbb{Z}_3$ gauge theory in $(2+1)$D as an example.  There are nine simple anyons: the vacuum line, two Wilson lines $W, W^2$ corresponding the nontrivial $\mathbb{Z}_3$ irreducible representations, two 't Hooft lines $M, M^2$ corresponding to the two nontrivial conjugacy classes of $\mathbb{Z}_3$, and four extra dyon lines constructed by fusing the Wilson lines and the 't Hooft lines:
        \begin{equation}
            D_{1,1} = W\times M,\quad D_{1,2} = W\times M^2,\quad D_{2,1} = W^2\times M,\quad D_{2,2} = W^2\times M^2.
        \end{equation}
        Let $\{1,\omega,\omega^2\}$ denote the elements in $\mathbb{Z}_3$. The automorphism group Aut$(\mathbb{Z}_3)=\mathbb{Z}_2$ acts on $\mathbb{Z}_3$ by exchanging $\omega$ with $\omega^2$, which induces a charge conjugation symmetry exchanging the two Wilson lines by $W\leftrightarrow W^2$ and the two 't Hooft lines $M\leftrightarrow M^2$. The dyon lines are permuted as follows:
        \begin{equation}
            D_{1,1}\leftrightarrow D_{2,2}, \qquad D_{1,2}\leftrightarrow D_{2,1}.
        \end{equation}
        Since there is only one nontrivial element in $\mathbb{Z}_2$, we only need the following tunneling matrix:
        \begin{equation}
            W_{a,{}^ga} =  \begin{pmatrix}
1 & 0 & 0 & 0 & 0 & 0 & 0 & 0 & 0 \\
0 & 0 & 1 & 0 & 0 & 0 & 0 & 0 & 0 \\
0 & 1 & 0 & 0 & 0 & 0 & 0 & 0 & 0 \\
0 & 0 & 0 & 0 & 1 & 0 & 0 & 0 & 0 \\
0 & 0 & 0 & 1 & 0 & 0 & 0 & 0 & 0 \\
0 & 0 & 0 & 0 & 0 & 0 & 0 & 0 & 1 \\
0 & 0 & 0 & 0 & 0 & 0 & 0 & 1 & 0 \\
0 & 0 & 0 & 0 & 0 & 0 & 1 & 0 & 0 \\
0 & 0 & 0 & 0 & 0 & 1 & 0 & 0 & 0 
\end{pmatrix}  ,
        \end{equation}
        where the tunneling matrix entry is ordered by $\{1,W,W^2,M,M^2,D_{1,1},D_{1,2},D_{2,1}, D_{2,2}\}$.

        Now we construct the higher gauging condensation defect. The folding trick informs us to condense $W$ and $M$ on the wall, so the summand of the condensation defect should contain $W(\gamma)M(\Gamma)$. Consider moving $W(\gamma')$ across the surface $\Sigma$, the tunneling matrix requires $W(2\gamma)$ to be emitted from the right. We have the following interaction between $W(\gamma')$ and the summand of the higher gauging condensation defect:
        \begin{equation}
            \begin{split}
                W(\gamma')\times (W(\gamma)\times M(\Gamma)) &= e^{-\frac{2\pi i}{3}\langle \gamma',\Gamma\rangle } (W(\gamma)\times M(\Gamma))\times W(\gamma')\\
                & = e^{-\frac{4\pi i}{3}\langle \gamma',\Gamma_\rangle } (W(\gamma-\gamma')\times M(\Gamma))\times W(2\gamma')\\
                & = e^{-\frac{4\pi i}{3}\langle \gamma',\Gamma\rangle } (W(\gamma-\gamma')\times M(\Gamma))\times W^2(\gamma').
            \end{split}
        \end{equation}
        Since pushing a bulk line across a symmetry defect does not modify the structure of the symmetry defect itself, the algebraic data on $\Sigma$ must be able to absorb the braiding phase $e^{-\frac{4\pi i}{3}\langle \gamma',\Gamma\rangle}$. This instructs us to stack a discrete torsion term $e^{\frac{4\pi i}{3} \langle\gamma,\Gamma\rangle}$ so that:
        \begin{equation}
            S(\Sigma) \sim \sum_{\gamma,\Gamma\in H_1(\Sigma,\mathbb{Z}_3)} e^{\frac{4\pi i}{3}\langle\gamma-\gamma',\Gamma\rangle} W(\gamma-\gamma')M(\Gamma) = \sum_{\gamma,\Gamma\in H_1(\Sigma,\mathbb{Z}_3)} e^{\frac{4\pi i}{3}\langle\gamma,\Gamma\rangle} W(\gamma)M(\Gamma).
        \end{equation}
        Namely, the braiding factor is absorbed by a redefinition of the homology lattice generator. This choice of discrete torsion indeed produces the correct symmetry action on the 't Hooft lines:
        \begin{equation}
M(\gamma')\times S(\Sigma)
 = S(\Sigma)\times M(2\gamma').
        \end{equation}
        The action on the dyon line proceeds analogously and indeed this higher gauging condensation defect reproduces the correct tunneling matrix. 

        To fix the normalization factor of $S(\Sigma)$, we observe that $S(\Sigma)$ is necessarily an invertible operator as it follows a group-like fusion rule. Shrinking $S(\Sigma)$ produces a multiplicity of $\abs{H_1(\Sigma,\mathbb{Z}_3)}$ because all the lines and the discrete torison are invertible operators. Therefore, the normalization factor is simply the inverse of the volume of the summand:
        \begin{equation}\label{eq - charge conjugation in Z3 gauge theory}
            S(\Sigma) \equiv \frac{1}{\abs{H_1(\Sigma,\mathbb{Z}_3)}}\sum_{\gamma,\Gamma\in H_1(\Sigma,\mathbb{Z}_3)} e^{\frac{4\pi i}{3}\langle\gamma,\Gamma\rangle} W(\gamma)M(\Gamma).
        \end{equation}

        Finally, a quick calculation shows that this condensation defect indeed fuses to the identity operator with itself.
        \begin{equation}
            \begin{split}
                &S(\Sigma)\times S(\Sigma)\\
                =& \frac{1}{\abs{H_1(\Sigma,\mathbb{Z}_3)}^2}\sum_{\gamma,\gamma',\Gamma,\Gamma'}\exp\left( \frac{4\pi i}{3}\left(\langle\gamma,\Gamma\rangle + \langle\gamma', \Gamma'\rangle\right)\right)W(\gamma)M(\Gamma)W(\gamma')M(\Gamma') \\
                = & \frac{1}{\abs{H_1(\Sigma,\mathbb{Z}_3)}^2}\sum_{\gamma,\gamma',\Gamma,\Gamma'}\exp\left( \frac{4\pi i}{3}\left(\langle\gamma,\Gamma\rangle + \langle\gamma', \Gamma'\rangle-\langle\gamma',\Gamma
                \rangle\right)\right)  W(\gamma+\gamma')M(\gamma+\Gamma')\\
                 = & \frac{1}{\abs{H_1(\Sigma,\mathbb{Z}_3)}^2}\sum_{\gamma_{\pm}, \Gamma_{\pm}}\exp\left( \frac{4\pi i}{3} \left( \frac{1}{4}\langle \gamma_+ + \gamma_-, \Gamma_+\rangle + \frac{1}{4}\langle3\gamma_- - \gamma_+,\Gamma_- \rangle   \right) \right)W(\gamma_+) M(\Gamma_+) \\
                 = & \frac{1}{\abs{H_1(\Sigma,\mathbb{Z}_3)}}\sum_{\Gamma_+, \gamma_-}\exp\left( \frac{i\pi}{3}\langle \gamma_-, \Gamma_+\rangle \right)M(\Gamma_+)\\
                 = & 1 ,
            \end{split}
        \end{equation}
        where the summations are over $H_1(\Sigma,\mathbb{Z}_3)$.
        Some explanations are in order. We insert the definition of $S(\Sigma)$ in the second line. Moving $M(\Gamma)$ across $W(\gamma')$ introduces an extra braiding factor. In the fourth line, we define a new homology basis $\gamma_{\pm} = \gamma \pm \gamma'$ and $\Gamma_{\pm} = \Gamma \pm \Gamma'$. In the fifth line, integrating over $\Gamma_-$ implements the constraint $\gamma_+ = 3\gamma_-$, which collapses the $\gamma_+$ sum and eliminates the condensed Wilson line. Finally, integrating over $\gamma_-$ implements the constraint $\Gamma_+ = 0$, which collapses the $\Gamma_+$ sum and eliminates the condensed $M(\Gamma_+)$ line. The multiplicity produced by the remaining sum $\sum_{\gamma_{-}}$ is canceled by the normalization factor, so we obtain the identity operator in the end.

\section{Lagrangian Analysis of \texorpdfstring{$\mathbb{D}_4$}{D4} Gauge Theory} \label{section - D4 gauge theory}

In this section, we review the Lagrangian analysis of the $\mathbb{D}_4$ gauge theory, which can be found in \cite{Cordova:2024mqg,Cordova:2024jlk, He:2016xpi}. The idea is straightforward. One derives the most general off-shell gauge transformations that leave the action invariant on a closed manifold. Operators are defined by restricting the spacetime gauge transformation to closed oriented submanifolds. Operators that require the dressing of higher gauging condensation defects to achieve operator worldvolume gauge invariance are generally non-invertible. Fusion rules and character tables can be correctly reproduced from the character table. Moreover, as pointed out in \cite{He:2016xpi}, the linking invariant calculations in $(2+1)$D require certain lattice regularization procedures, which again reflects the effective field theory nature of the Lagrangian analysis. We review the untwisted $\mathbb{D}_4$ gauge theory action construct in Sec. \ref{subsection - D4 gauge theory} and review the operator spectrum and fusion rule in Sec. \ref{subsection - D4 opertor spectrum}. We review the Hopf link calculation that produces the $\mathbb{D}_4$ character table in Sec. \ref{subsection - D4 character from linking} and the $(2+1)$D lattice regulation in Sec. \ref{subsection - lattice regularization}. The content of this section is not new, but the computational details are complete. We hope this quick review serves as a starting point for future in-depth Lagrangian analysis of Dijkgraaf-Witten theories in arbitrary spacetime dimensions.

\subsection{Lagrangian Formulation of \texorpdfstring{$\mathbb{D}_4$}{D4} DW Theory}\label{subsection - D4 gauge theory}

    In this subsection, we review the construction of the effective action for $\mathbb{D}_4$ gauge theories in arbitrary spacetime dimension by the gauging of a $\mathbb{Z}_2^{(0)}$ 0-form symmetry following \cite{Cordova:2024mqg}. The dihedral group of order 8 can be represented as:
    \begin{equation}
        \mathbb{D}_4 = \{ a,b,c\;|\; a^2=b^2=c^2=(ac)^4=1,ab=ba=cac, bc=cb=aca \}.
    \end{equation}
    This group is isomorphic to $(\mathbb{Z}_2\times\mathbb{Z}_2)\rtimes\mathbb{Z}_2$, where the nontrivial twist is specified by exchanging generators $(1,0)$ and $(1,1)$ of $\mathbb{Z}_2\times\mathbb{Z}_2$. Besides the trivial subgroup and the full group, $\mathbb{D}_4$ has four normal subgroups. The center group $Z(\mathbb{D}_4) = \{1,b\}$ is the only normal subgroup of order 2. There is a cyclic normal subgroup $\mathbb{Z}_4 = \{1,ac,(ac)^2,(ac)^3\}$ and two Klein-four normal subgroups $V_4 = \{1,a,b,ab\}$ and $V_4' = \{1,b,c,bc\}$.

     The original $\mathbb{Z}_2\times\mathbb{Z}_2$ gauge theory contains the Wilson lines $W_1, W_2, W_V$ and 't Hooft surface operators $M_1, M_2, M_V$ generated by the two copies of $\mathbb{Z}_2$'s and the diagonal $\mathbb{Z}_2^{(0)}$ respectively. The  $\mathbb{Z}_2^{(0)}$ symmetry is generated by:
    \begin{equation}
        S(\Sigma) = \frac{1}{\abs{H^1(\Sigma, \mathbb{Z}_2)}}\sum_{\substack{\gamma\in H_1(\Sigma, \mathbb{Z}_2)\\
            \Gamma \in H_{D-2}(\Sigma, \mathbb{Z}_2)}}(-1)^{\langle \gamma , \Gamma\rangle  }W_1(\gamma) M_2(\Gamma).
    \end{equation} 
    Its symmetry action on the lines are given by:
    \begin{equation}\label{eq - D4 Symmetry Action by Condensation Defect}
        \begin{split}
            W_1(\gamma)\times S(\Sigma) &= S(\Sigma)\times W_2(\gamma),   \\
        W_2(\gamma)\times S(\Sigma) &= S(\Sigma) \times W_V(\gamma) ,  \\W_V(\gamma)\times S(\Sigma) &= S(\Sigma)\times W_2(\gamma), \\
        M_1(\gamma)\times S(\Sigma) &= S(\Sigma)\times M_V(\gamma),\\
        M_2(\gamma) \times S(\Sigma) &= S(\Sigma)\times M_2(\gamma), \\
        M_V(\gamma) \times S(\Sigma) &= S(\Sigma)\times M_1(\gamma).
        \end{split}
    \end{equation}

    To gauge this $\mathbb{Z}_2^{(0)}$ symmetry, we first insert the conserved current $*j_1$ of the symmetry to the theory by coupling it to a $\mathbb{Z}_2$ background gauge field. As mentioned in the introduction, in principle there is no Noether current $j_1$ for discrete symmetries. To mimic the current insertion, we invoke \Poincare duality and replace $*j_1$ with the insertion of a mesh of codimension-1 symmetry defects:
    \begin{equation}
        \begin{split}
            S(\Sigma) & = \frac{1}{\abs{H^1(\Sigma, \mathbb{Z}_2)}}\sum_{\substack{\gamma\in H_1(\Sigma, \mathbb{Z}_2)\\
            \Gamma \in H_{D-2}(\Sigma, \mathbb{Z}_2)}}(-1)^{\langle \gamma , \Gamma\rangle  }W_1(\gamma) M_2(\Gamma) \\
            & = \frac{1}{\abs{H^1(\Sigma, \mathbb{Z}_2)}}\sum_{\substack{\gamma\in H_1(\Sigma, \mathbb{Z})\\
            \Gamma \in H_{D-2}(\Sigma, \mathbb{Z})}}e^{\frac{i}{\pi}\int_{\Sigma} A_{D-2}\wedge \tilde{B}_1 + i\oint_{\gamma} a_{1} + i \oint_{\Gamma}\tilde{b}_{D-2}}  \\
            & = \frac{1}{\abs{H^1(\Sigma, \mathbb{Z}_2)}}\sum_{\substack{\tilde{B}_1\in H^1(\Sigma, \mathbb{Z})\\
            A_{D-2} \in H^{D-2}(\Sigma, \mathbb{Z})} }e^{ \frac{i}{\pi}\oint_{\Sigma}(A_{D-2}\wedge\tilde{B}_1  + a_1\wedge A_{D-2} + \tilde{b}_{D-2}\wedge \tilde{B}_{1})}\\
            & = e^{-\frac{i}{\pi}\oint_{\Sigma} a_1\wedge \tilde{b}_{D-2}}.
        \end{split}
    \end{equation}
    Here in the second equation we invoked \Poincare duality and switched over to the differential form notation. $A_{D-2}$ and $\tilde{A}_1$ are constrained to have $\pi\mathbb{Z}$-valued monodromies on $\Sigma$ and the homology cycles $\gamma$ and $\Gamma$ are the \Poincare duals of these gauge fields. In the third equation we integrate out $\tilde{A}_1$, which implements the constraint $\tilde{b}_{D-2} = - A_{D-2}$. Since $a_1\wedge\tilde{b}_{D-2}$ is proportional to the volume form of the closed oriented codimension-1 surface $\Sigma$, it is manifestly a closed form and it formally obeys the conservation law $d*j_1 \sim d(a_1\wedge \tilde{b}_{D_2})=0$. 
    
    Now we insert a mesh of $S(\Sigma)$:
    \begin{equation}
        \sum_{\Sigma\in H_{D-1}(M,\mathbb{Z}_2)}S(\Sigma) = \sum_{\Sigma\in H_{D-1}(M,\mathbb{Z}_2)}e^{-\frac{i}{\pi}\oint_{\Sigma} a_1\wedge \tilde{b}_{D-2}} = \sum_{c_1\in H^1(M,\mathbb{Z}_2)}e^{-\frac{i}{\pi^2}\int_{M} a_1\wedge \tilde{b}_{D-2}\wedge c_1  },
    \end{equation}
    where $c_1$ is the background gauge field of the $\mathbb{Z}_2^{(0)}$ symmetry with monodromy $\oint_{\gamma} c_1 \in \pi\mathbb{Z}$. The insertion contributes a term to the action $-\frac{i}{\pi}\int_{M}a_1\wedge \tilde{b}_{D-2}\wedge c_1$, which can be understood as a discrete analog of the coupling of a conserved current $*j_1$ to the action:
    \begin{equation}
        I[j_1] = I_0  + \int_{M_D}c_1\wedge *j_1 = I_0  -\frac{i}{\pi^2}\int_M a_1\wedge\tilde{b}_{D-2}\wedge c_1,
    \end{equation}
    where $I_0 = \frac{i}{\pi}\int_{M_D}(\tilde{a}_{D-2}\wedge da_1 + \tilde{b}_{D-2}\wedge db_1)$ is the action of the original $\mathbb{Z}_2\times\mathbb{Z}_2$ gauge theory. The equations of motion $da_1 = 0$ and $d\tilde{b}_{D-2}=0$ guarantee that the current $*j_1 \sim a_1\wedge\tilde{b}_{D-2}$ is closed on-shell. Thus it is a good analog of the conserved current.

    The last step is to promote $c_1$ to a dynamical gauge field. This is done by introducing a Lagrange multiplier field $\tilde{c}_{D-2}$. Adding up all contributions, we arrive at the following action:
    \begin{equation} \label{eq - D4 gauge theory action}
        I_{\mathbb{D}_4} = \frac{i}{\pi} \int_M \left(\tilde{a}_{D-2}\wedge da_1 + \tilde{b}_{D-2}\wedge db_1  + \tilde{c}_{D-2}\wedge dc_1 - \frac{1}{\pi}a_1\wedge\tilde{b}_{D-2}\wedge c_1\right).
    \end{equation}
    The equations of motion are:
    \begin{equation}
        \begin{aligned}
        &da_1= 0, \quad d\tilde{b}_{D-2}=0,\quad dc_1=0, \\
        &d \tilde{a}_{D-2} = \frac{1}{\pi}\,\tilde{b}_{D-2}\wedge c_{1},\quad d b_{1} = (-1)^{D-2}\frac{1}{\pi}\,a_{1}\wedge c_{1}, \quad d \tilde{c}_{D-2} = (-1)^{D-1}\frac{1}{\pi}\,a_{1}\wedge \tilde{b}_{D-2},
\end{aligned}
    \end{equation}
    The most general set of gauge transformations that leaves the action invariant off-shell is:
\begin{equation} \label{eqn - D4 gauge transformation}
    \begin{aligned}
    a_1\;&\longmapsto a_1 + d\alpha_0,  \\
    c_1\;&\longmapsto c_1 + d\epsilon_0,  \\
     \tilde{b}_{D-2}\;&\longmapsto \tilde{b}_{D-2}  +d\tilde{\beta}_{D-3
     },  \\
\tilde a_{D-2}\;&\longmapsto\; \tilde a_{D-2}+d\tilde\alpha_{D-3}
  -\frac{1}{\pi }\left(\tilde\beta_{D-3}\wedge c_1
  +(-1)^{D-2}\,\epsilon_{0}\,\tilde b_{D-2}
  +\,\tilde\beta_{D-3}\wedge d\epsilon_{0}\right),\\
b_1\;&\longmapsto\; b_1+d\tilde\beta_0
  -(-1)^{D-2}\frac{1}{\pi }\left(\,\alpha_0 c_1-\,\epsilon_0 a_1
  +\,\alpha_0\,d\epsilon_{0}\right),\\[2pt]
\tilde c_{D-2}\;&\longmapsto\; \tilde c_{D-2}+d\tilde\epsilon_{D-3}
  -(-1)^{D-1}\frac{1}{\pi }\left(\alpha_0\,\tilde b_{D-2}
  +(-1)^{D-2}\,\tilde\beta_{D-3}\wedge a_1
  +\,\alpha_0\,d\tilde\beta_{D-3}\right),
\end{aligned}
\end{equation}
    where $\alpha_0, \beta_{D-3}, \epsilon_0, \tilde{\alpha}_{D-2}, \beta_0, \tilde{\epsilon}_{D-3}$ are gauge transformation parameters with $2\pi$-periodicity. We will demonstrate it derivation in the Sec. \ref{section - type-I action analysis}. 

\subsection{Operator Spectrum of the \texorpdfstring{$\mathbb{D}_4$}{D4} DW Theory}
\label{subsection - D4 opertor spectrum}

In the original $\mathbb{Z}_2\times
\mathbb{Z}_2$ gauge theory, all holonomies of the gauge fields are gauge invariant. In the $\mathbb{D}_4$ gauge theory, the off-shell gauge transformations imply that only the holonomies of $a_1$, $\tilde{b}_{D_2}$ and $c_1$ are gauge invariant. Define:
    \begin{equation}
        U_a(M_1) = e^{i\oint_{M_1}a_1},\qquad U_{\tilde{b}}(M_{D-2}) = e^{i\oint_{M_{D-2}}{\tilde{b}_{D-2}}}, \qquad U_c(M_1) = e^{i\oint_{M_1}c_1},
    \end{equation}
    where $U_{c}(M_1)$ is the dynamical Wilson line from gauging the $\mathbb{Z}_2^{(0)}$ symmetry. Their self-fusions are all of order-2, namely:
    \begin{equation}
        U_a\times U_a =1, \qquad U_{\tilde{b}}\times U_{\tilde{b}} =1, \qquad U_c\times U_c =1.
    \end{equation}
   Meanwhile, $U_a$ can fuse with $U_c$ to form a order-2 Wilson line:
   \begin{equation}
       U_{a+c}(M_1)\equiv U_a(M_1)\times U_c(M_1).
   \end{equation}
    On the other hand, the operator $U_{\tilde{b}}(M_{D-2})$ has different dimensions from $U_a$ and $U_c$ and it only fuses with $U_a$ and $U_c$ in (2+1)D. This concludes the discussion on the invertible sector.

    By the gauge transformations in Eq. \eqref{eqn - D4 gauge transformation}, the holonomies $e^{i\oint_{M_{D-2}} \tilde{a}_{D-2}}$, $e^{i\oint_{M_1}b_1}$ and $e^{i\oint_{M_{D-2}} \tilde{c}_{D-2}}$ are not gauge invariant operators.  For example, $e^{i\oint_{M_1}b_1}$ transforms as:
    \begin{equation}
        e^{i\oint_{M_1}b_1} \mapsto \exp\Bigg(\frac{i(-1)^{D-1}}{\pi} \oint_{M_1} \left(\alpha_0 c_1 - \epsilon_0 a_1 + \alpha_0 d\epsilon_0\right)\Bigg)e^{i\oint_{M_1}b_1}.
    \end{equation}
   The extra $\mathrm{U}(1)$ factor does not vanish for generic values of $\alpha_0$ and $\epsilon_0$, so $e^{i\oint_{M_1}b_1}$ is not gauge invariant. One naive remedy is to sum over all possible gauge transformations, which can be represented by the following object:
   \begin{equation}
       \int \mathcal{D}\alpha_{0}\,\mathcal{D}\epsilon_{0}\;
\exp\!\left(
  i\oint_{M_{1}}\!\left(
    b_{1}\;+\;\frac{(-1)^{D-2}}{\pi}\left(\alpha_{0}\,c_{1}-\epsilon_{0}\,a_{1}-\epsilon_{0}\,d\alpha_{0}\right)
  \right)
\right).
   \end{equation}
    This in fact is an overkill. Let $\iota:M_1 \hookrightarrow M_D$ be the inject that embeds $M_1$ into spacetime $M_D$, then we only need to sum over the $\alpha_0$ and $\epsilon_0$ modes on $M_1$ that lead to a trivial gauge transformation in the pullback configuration $\iota^*a_1$ and $\iota^*c_1$. This constraint can be imposed by adding additional Lagrange multiplier terms $\phi_0,\lambda_0$ on $M_1$:
    \begin{equation}
        \begin{split}
            \hat U_{b}(M_{1})=
\int &\mathcal{D}\alpha_{0}\,\mathcal{D}\epsilon_{0}\,\mathcal{D}\phi_{0}\,\mathcal{D}\lambda_{0}\;
\exp\!\left(
  i\oint_{M_{1}}\! b_{1}
  -\frac{1}{\pi}(-1)^{D-2}\oint_{M_{1}}\! \epsilon_{0}\,d\alpha_{0}
\right)\\
&\;\times\;
\exp\!\left(
  \frac{i}{\pi}(-1)^{D-2}\oint_{M_{1}}
  \left(\phi_{0}(c_{1}-d\epsilon_{0})-\lambda_{0}(a_{1}-d\alpha_{0})\right)
\right).
        \end{split}
    \end{equation}
Integrating out the Lagrange multiplier term gives:
\begin{equation}
    \begin{aligned}
\hat U_{b}(M_{1})
=& \int \mathcal{D}\alpha_{0}\,\mathcal{D}\epsilon_{0}\,\mathcal{D}\phi_{0}\,\mathcal{D}\lambda_{0}\;
   e^{\,i\oint_{M_{1}} b_{1}-\frac{1}{\pi}(-1)^{D-2}\oint_{M_{1}}\epsilon_{0}\,d\alpha_{0}} \\
& \quad \times
   \exp\!\left[\frac{i}{\pi}(-1)^{D-2}\oint_{M_{1}}
     \left(\phi_{0}(c_{1}-d\epsilon_{0})-\lambda_{0}(a_{1}-d\alpha_{0})\right)\right] \\
= &\int \mathcal{D}\alpha_{0}\,\mathcal{D}\epsilon_{0}\;
   e^{\,i\oint_{M_{1}} b_{1}-\frac{1}{\pi}(-1)^{D-2}\oint_{M_{1}}\epsilon_{0}\,d\alpha_{0}}\,
   \delta\!\left(c_{1}-d\epsilon_{0}\right)\,
   \delta\!\left(a_{1}-d\alpha_{0}\right) \\
= & \int \mathcal{D}\alpha_{0}\,\mathcal{D}\epsilon_{0}\;
   e^{\,i\oint_{M_{1}} b_{1}-\frac{1}{\pi}(-1)^{D-2}\oint_{M_{1}}\epsilon_{0}\,d\alpha_{0}}\,
   \delta\!\left(\oint_{M_{1}} c_{1}\right)\,
   \delta\!\left(\oint_{M_{1}} a_{1}\right) \\
\sim &~ e^{\,i\oint_{M_{1}} b_{1}}\,
   \delta\!\left(\oint_{M_{1}} c_{1}\right)\,
   \delta\!\left(\oint_{M_{1}} a_{1}\right). 
\end{aligned}
\end{equation}

    Some explanations are in order for the previous calculation. The second line is obtained by integrating out the Lagrange multiplier terms $\phi_0$ and $\lambda_0$, resulting in Dirac delta functions that impose topological boundary conditions for $a_1$ and $c_1$ on $M_1$ in the third line. There is an overall normalization factor in the fourth line which we will later fix.

    As pointed out in \cite{He:2016xpi}, these constraints can be understood as projection operators of $a_1$ and $c_1$ onto the trivial monodromy sector. This is because the local constraints $c_1 = d\epsilon_0$ and $a_1=d\alpha_0$ imply $\oint_{M_1}a_1=0$ and $\oint_{M_1}c_1=0$ by Stoke's theorem. This projection operator can be conveniently represented in terms of the monodromy operators. Consider the constraint $\delta\left(\oint_{M_1}a_1\right)$. We would like to construct an operator on $M_1$ that evaluates to $0$ when $\oint_{M_1}a_1\neq0$ and evaluates to something proportional to 1 when $\oint_{M_1}a_1\neq0$. An obvious choice is the sum:
    \begin{equation}
        \delta\left(\oint_{M_1}a_1\right)\equiv\frac{1}{2}\left(1 + e^{i\oint_{M_1}a_1}\right).
    \end{equation}
    This is true because $\oint_{M_1}a_1\in\pi\mathbb{Z}$, implying that $\delta\left(\oint_{M_1}a_1\right)$ evaluates to a sum over roots of unity for $\oint_{M_1}a_1\neq0$ and the identity operator when $\oint_{M_1}a_1=0$. Since $\oint_{M_1}c_1\in\pi\mathbb{Z}$, we also have:
    \begin{equation} \label{eq - top degree projection operator}
        \delta\left(\oint_{M_1}c_1\right)\equiv\frac{1}{2}\left(1 + e^{i\oint_{M_1}c_1}\right).
    \end{equation}
    By inspection $\delta\left(\oint_{M_1}c_1\right){}^2 = \delta\left(\oint_{M_1}c_1\right)$, so indeed it's a good projection operator. 

    For higher dimensional operators, to make $e^{i\oint_{M_{D-2}} \tilde{a}_{D-2}}$ gauge invariant, we need to dress it with operators that project $c_1$ and $\tilde{b}_{D-2}$ to the trivial monodromy sector. For $\tilde{b}_{D-2}$, this is done by stacking:
    \begin{equation}
        \delta\left(\oint_{M_{D-2}}\tilde{b}_{D-2}\right) = \frac{1}{2}\left(1+e^{i\oint_{M_{D-2}}\tilde{b}_{D-2}}\right).
    \end{equation}
    For $c_1$, this is done by inserting a mesh of $\delta\left(\oint_{\gamma}c_1\right)$ on all 1-cycles of $M_2$. Define a new projection operator:
    \begin{equation}
        \Delta_c(M_{D-2}) \equiv \prod_{\gamma \in H_1(M_{D-2},\mathbb{Z}_2)}\delta \left(\oint_{\gamma}c_1\right) = \frac{1}{\abs{H_1(M_{D-2},\mathbb{Z}_2)}}\sum_{\gamma\in H_1(M_{D-2},\mathbb{Z}_2)}U_c(\gamma),
    \end{equation}
    which is proportional to a higher gauging condensation defect by condensing $U_c$ on $M_{D-2}$. Similarly, we can define projectors $\Delta_a$ and $\Delta_{a+c}$, which are required by $e^{i\oint_{M_{D-2}}\tilde{c}_{D-2}}$ and $e^{i\oint_{M_{D-2}}(\tilde{a}_{D-2} + \tilde{c}_{D-2})}$, respectively. The non-invertible 'f Hooft surfaces are:
    \begin{equation}
        \begin{split}
            &\hat{U}_{\tilde{a}}\sim e^{i\oint\tilde{a}_{D-2}}\left(\frac{1+ U_{\tilde{b}}}{2}\right)\Delta_c, \\
            &\hat{U}_{\tilde{c}}\sim e^{i\oint\tilde{c}_{D-2}}\left(\frac{1+ U_{\tilde{b}}}{2}\right)\Delta_a, \\
            & \hat{U}_{\tilde{a},\tilde{c}}\sim e^{i\oint (\tilde{a}_{D-2} + \tilde{c}_{D-2})}\left(\frac{1+ U_{\tilde{b}}}{2}\right)\Delta_{a+c}.
        \end{split}
    \end{equation}
    where $\frac{1}{2}(1+U_{\tilde{b}})$ is an $(D-2)$-dimensional projector for $\oint_{M_{D-2}}\tilde{b}_{D-2}$. 
    
    Now we fix the normalization factor. Drawing some intuitions from the anyon theory literature \cite{Simon:2023hdq}, we demand that the fusion of a non-invertible $p$-dimensional operator with its charge conjugation must contain a unique factor of the $p$-dimensional identity operator or an appropriate condensation defect. As previously mentioned,  the charge conjugation of an operator defined on a submanifold $X$ is defined by assigning the same algebraic data to the orientation-reversal $\overline{X}$.

    We start from the non-invertible line $\hat{U}_b$. Since we only consider lines and codimension-2 operators, the fusion between lines can only produce lines. Each 1-dimensional projector contains a factor of $1/2$, so the non-invertible line needs to be scaled by 2 to produce a unique identity line:
    \begin{equation}
        \begin{aligned}
\hat U_{b}(M_{1})\times \hat U_{b}(M_{1})
&= e^{\,2i\oint_{M_{1}} b_{1}}\,
   \left(1+e^{\,i\oint_{M_{1}} c_{1}}\right)\,
   \left(1+e^{\,i\oint_{M_{1}} a_{1}}\right) \\
&= 1 + e^{\,i\oint_{M_{1}} c_{1}}
     + e^{\,i\oint_{M_{1}} a_{1}}
     + e^{\,i\oint_{M_{1}} (a_{1}+c_{1})} \\
&= 1 + U_{c}(M_{1}) + U_{a}(M_{1}) + U_{a+c}(M_{1}).
\end{aligned}
    \end{equation}
    For non-invertible surface operators, we choose to rescale $\Delta_a, \Delta_c$:
    \begin{equation} \label{eq - D4 electric condensation}
        \mathcal{S}_a \equiv 2\Delta_a, \qquad \mathcal{S}_c \equiv 2\Delta_c, \qquad \mathcal{S}_{a+c} = 2\Delta_{a+c}.
    \end{equation}
   so that the objects $\mathcal{S}_a$ and $\mathcal{S}_c$ follow the fusion rules:
   \begin{equation} \label{eq - D4 diagonal wall fusion}
       \mathcal{S}_a \times \mathcal{S}_a = 2\mathcal{S}_a, \qquad \mathcal{S}_c \times \mathcal{S}_c = 2\mathcal{S}_c, \qquad \mathcal{S}_{a+c}\times \mathcal{S}_{a+c} = 2 \mathcal{S}_{a+c}.
   \end{equation}
   Here $\mathcal{S}_a$, $\mathcal{S}_c$, $\mathcal{S}_{a+c}$ are higher codimension analogs of the diagonal walls \cite{Cordova:2024mqg, Cordova:2024jlk}. They corresponds to orientation-reversal invariant condensation defects with worldvolume gauge group: \cite{Cordova:2024mqg, Cordova:2024jlk}:
    \begin{equation}
        \mathcal{S}_a \leftrightarrow \mathcal{D}_{V_4}, \qquad
            \mathcal{S}_c \leftrightarrow \mathcal{D}_{V_4'}, \qquad \mathcal{S}_{a+c} \leftrightarrow \mathcal{D}_{\mathbb{Z}_4}.
    \end{equation}
    Note that the fusion of a non-invertible surface with its charge conjugation has the form:
    \begin{equation}
        \hat{U}_{\tilde{a}}\times \hat{U}_{\tilde{a}} = \mathcal{S}_c + \mathcal{S}_c \times U_{\tilde{b}}, 
    \end{equation}
    where $\mathcal{S}_c$ plays the role of the identity object in the fusion rule. Summarizing, the non-invertible line and surface operators are:
   \begin{align}
       \hat{U}_b(M_1) &= \frac{1}{2}e^{i\oint_{M_1}b_1}(1 + U_a(M_1))(1 + U_c(M_1)),\label{eq - D4 dim-2 irrep} \\
       \hat{U}_{\tilde{a}}(M_{D-2})& = \frac{1}{2}e^{i\oint_{M_{D-2}}\tilde{a}_{D-2}}\mathcal{S}_c(M_{D-2})(1+U_{\tilde{b}}(M_{D-2}))\label{eq - D4 size-2 conjugacy class a},\\
       \hat{U}_{\tilde{c}}(M_{D-2})& = \frac{1}{2}e^{i\oint_{M_{D-2}}\tilde{c}_{D-2}}\mathcal{S}_a(M_{D-2})(1+U_{\tilde{b}}(M_{D-2}))\label{eq - D4 size-2 conjugacy class c},\\
       \hat{U}_{\tilde{a},\tilde{c}}(M_{D-2})& = \frac{1}{2}e^{i\oint_{M_{D-2}}(\tilde{a}_{D-2} + \tilde{c}_{D-2} )}\mathcal{S}_{a+c}(M_{D-2})(1+U_{\tilde{b}}(M_{D-2})).\label{eq - D4 size-2 conjugacy class ac}
   \end{align}
   
   Let us compute the mutual fusion rules between 't Hooft surfaces. Since the product $N_1 N_2$ of two normal subgroups $N_1\triangleleft G$ and $N_1\triangleleft G$ is also a normal subgroup of $G$, the product of any pair of any order-4 normal subgroups of $\mathbb{D}_4$ is the entire $\mathbb{D}_4$. Furthermore, any pair of the order-4 normal subgroups intersect only at the $\mathbb{Z}_2$ center. Therefore, we have the following mutual fusion rules:
   \begin{equation}
       \mathcal{S}_a \times \mathcal{S}_c = \mathcal{S}_a \times \mathcal{S}_{a+c} = \mathcal{S}_c \times \mathcal{S}_{a+c} = \mathcal{D}_{\mathbb{Z}_2}.
   \end{equation}
   Consider the mutual fusion between 't Hooft surfaces, for example:
    \begin{equation}
        \hat{U}_{\tilde{a}} \times \hat{U}_{\tilde{c}} = \frac{1}{2}e^{\oint_{M_{D-2}}(\tilde{a}_{D-2} + \tilde{c}_{D-2})} \mathcal{D}_{\mathbb{Z}_2} (1+ U_{\tilde{b}}(M_{D-2})).
    \end{equation}
    Note that the quantum dimension of $\mathcal{D}_{\mathbb{Z}_2}$ is 4, because it is the product of two condensation defects of dimension-2. By demanding the equality of quantum dimensions of the LHS and RHS of the same operator equation, we have:
    \begin{equation}
        \hat{U}_{\tilde{a}}\times \hat{U}_{\tilde{c}} = 2\hat{U}_{\tilde{a},\tilde{c}}, \qquad \hat{U}_{\tilde{a}}\times \hat{U}_{\tilde{a},\tilde{c}} = 2\hat{U}_{\tilde{c}}, \qquad \hat{U}_{\tilde{c}}\times \hat{U}_{\tilde{a},\tilde{c}} = 2\hat{U}_{\tilde{a}},
    \end{equation}
    where the factor of ``2" should be understood as a TQFT valued coefficient. 
    
    A crucial feature of the condensation defect formalism is that it correctly reproduce the standard UMTC result in $(2+1)$D. For example, consider the mutual fusion $\hat{U}_{\tilde{a}} \times \hat{U}_{\tilde{c}}$ on $S^1$:
    \begin{equation}
        \begin{split}
            \hat{U}_{\tilde{a}}\times \hat{U}_{\tilde{c}}= & \frac{1}{2} e^{i\oint_{S^1}(\tilde{a} + \tilde{c})}\mathcal{S}_a \times\mathcal{S}_c\times (1+ U_{\tilde{b}})\\
           = &  \frac{1}{2} e^{i\oint_{S^1}(\tilde{a} + \tilde{c})}(1+U_a)(1+U_c) (1+ U_{\tilde{b}})\\
           =& \frac{1}{4} e^{i\oint_{S^1}(\tilde{a} + \tilde{c})}(1+U_a)(1+U_c) (1+ U_{\tilde{b}}) + \frac{1}{4} e^{i\oint_{S^1}(\tilde{a} + \tilde{c})}(1+U_a)(1+U_c) (1+ U_{\tilde{b}})U_{a,c}\\
           =& \frac{1}{4} e^{i\oint_{S^1}(\tilde{a} + \tilde{c})}\mathcal{D}_{\mathbb{Z}_2}(1+U_{\tilde{b}} ) + \frac{1}{4} e^{i\oint_{S^1}(\tilde{a} + \tilde{c})}\mathcal{D}_{\mathbb{Z}_2}(1+U_{\tilde{b}} ) U_{a,c}\\
        =& \hat{U}_{\tilde{a},\tilde{c}} + \hat{U}_{\tilde{a},\tilde{c}} \times U_{a,c}.
        \end{split}
    \end{equation}
     Here we have used the fact that $\mathcal{S}_a = 1+U_a$ and  $\mathcal{S}_c = 1+U_c$ in $(2+1)$D. Note that the condensation defect stacked on the non-invertible lines are not the same as the original condensation defects. Nonetheless, the physical operators are labeled by the monodromy factors $e^{i\oint_{S^1}(\tilde{a} + \tilde{c})}$ and one can check that the two different choices of stacked condensation defects both ensure gauge invariance on $S^1$. Therefore this is not an issue. The remaining mutual fusions can be worked out in a similar fashion and we refer the readers to \cite{Cordova:2024jlk,Cordova:2024mqg} for further details.
    \subsection{Character Table and Linking Invariants}\label{subsection - D4 character from linking}

    In this subsection, we show that the expectation values of the Hopf link between Wilson and 't Hooft operators correctly reproduce the $\mathbb{D}_4 $ character table in arbitrary spacetime dimensions, hence justifying Eq. \eqref{eq - D4 Action} as an effective action for the untwisted $\mathbb{D}_4$ gauge theory. This result has been announced in \cite{Cordova:2024mqg} and here we provide the details of the calculation.

   We take $\mathbb{D}_4 \simeq (\mathbb{Z}_2\times\mathbb{Z}_2)\rtimes\mathbb{Z}_2$ as a semi-direct product and denote elements of $\mathbb{D}_4$ as $(x,y,z)$, where $x,y,z$ are integers mod 2. The character table can be easily constructed with Clifford theory, see Appendix \ref{appendix - Mackey's Theory} for an introduction. $\mathbb{D}_4$ has two conjugacy classes of size-1 and three conjugacy classes of size-2:
    \begin{equation}
        \begin{split}
            &[(0,1,0)]=\{(0,1,0),(1,1,0)\}, \\ 
             &[(0,0,1)]=\{(0,0,1),(1,0,1)\}, \\
            &[(0,1,1)]=\{(0,1,1),(1,1,1)\}.
        \end{split}
    \end{equation}
    The first column labels the $\mathbb{D}_4$ irreducible representations and they generically come from induced representations of various subgroups of $\mathbb{D}_4$. See appendix \ref{appendix - Mackey's Theory} for an explanation of the notation.
    
    \begin{table}[h!]
\centering
\begin{tabular}{c|ccccc}

 &$(0,0,0)$ & $(1,0,0)$ & $[(0,1,0)]$ & $[(0,0,1)]$ & $[(0,1,1)]$ \\
\hline
$T$& 1 & 1 & 1 & 1  & 1 \\ 
\hline $T^-$& 1 & 1 & 1 & -1    & -1\\ 
\hline
 $(\chi_0,\chi_1)$& 1 & 1 & -1 & 1   & -1\\ 
 \hline
  $(\chi_0,\chi_1)^-$& 1 & 1 & -1 & -1 & 1\\ 
  \hline
 $[(\chi_0,\chi_1)]$& 2 & -2 & 0 & 0   & 0
\end{tabular}
\caption{$\mathbb{D}_4\simeq (\mathbb{Z}_2\times\mathbb{Z}_2)\rtimes\mathbb{Z}_2$ character table}
\label{table - D4 Characters}
\end{table}

First consider the Hopf link between the a nontrivial operator and an identity operator of the appropriate dimension, which is equivalent to compute the one-point function of the non-trivial operator. Formally, we have:
\begin{equation}
    \expval{O(\Sigma_p)} \equiv \frac{1}{\calZ(M_D)}\int \mathcal{D}[\text{Fields}]e^{-S[\text{Fields}]}O(\Sigma_p),
\end{equation}
where $\Sigma_p$ is an appropriate closed oriented submanifold that can be shrunk to a point. 

As usual, the one point function of the invertible operators $U_a(S^1), U_c(S^1), U_{a,c}(S^1)$, $U_{\tilde{b}}(S^{D-2})$ are equal to 1. Thus, we can identify the invertible Wilson lines $U_a, U_c, U_{a,c}$ as the three nontrivial 1-dimensional irreducible representations $(\chi_0, \chi_1)$, $T^-$, $(\chi_0, \chi_1)^-$, respectively. Similarly, the invertible 't Hooft operator $U_{\tilde{b}}$ can be identified as the nontrivial size-1 conjugacy class $(1,0,0)$. To compute the one point function of the non-invertible $U_b(S^1)$, we use Eq. \eqref{eq - D4 dim-2 irrep}. The shrinking of the monodromy factor gives a trivial contribution, while the shrinking of the stacked condensation defects produces a factor of $2\times2 = 4$, giving:
\begin{equation}
    \expval{\hat{U}_b(S^1)} = \frac{1}{2}\times 4 = 2  = \dim([(\chi_0, \chi_1)]).
\end{equation}
Therefore $\hat{U}_b(S^1)$ should be identified with the only nontrivial 2-dimensional irreducible representation $[(\chi_0,\chi_1)]$.
The one-point functions of the non-invertible surface operators $\hat{U}_{\tilde{a}}(S^{D-2})$, $\hat{U}_{\tilde{c}}(S^{D-2})$, and $\hat{U}_{\tilde{a},\tilde{c}}(S^{D-2})$  can be obtained similarly. Recall Eq. \eqref{eq - D4 size-2 conjugacy class a}, \eqref{eq - D4 size-2 conjugacy class c}, \eqref{eq - D4 size-2 conjugacy class ac}, the shrinking of the magnetic condensation defect produces a factor of 2, meanwhile the shrinking of the electric condensation $\mathcal{S}_a=2\Delta_a$,  $\mathcal{S}_c=2\Delta_c$, $\mathcal{S}_{a+c}$ produces a factor of 2. Therefore, we have:
\begin{align}
    \expval{\hat{U}_{\tilde{a}}(S^{D-2})} =  \expval{\hat{U}_{\tilde{c}}(S^{D-2})} =  \expval{\hat{U}_{\tilde{a},\tilde{c}}(S^{D-2})} = \frac{1}{2}\times 2 \times 2 = 2.
\end{align}
Hence, we should identify $\hat{U}_{\tilde{a}}$, $\hat{U}_{\tilde{c}}$, $\hat{U}_{\tilde{a},\tilde{c}}$ with the size-2 conjugacy classes $[(0,1,0)]$, $[(0,0,1)]$ and $[(0,1,1)]$, respectively.

Now we compute the Hopf links between the nontrivial Wilson and 't Hooft operators. For simplicity, we denote the invertible Wilson lines collectively as $U_{a,c}^{n_a, n_c}$, where $n_a,n_c \in \{0,1\}$. For example, $U_{a,c}^{0,0}=1$, $U_{a,c}^{0,1}=U_c$. Similarly, we denote the non-invertible 't Hooft operators collectively as $\hat{U}_{\Tilde{a},\Tilde{c}}^{\Tilde{n}_a,\Tilde{n}_c}$, where $\Tilde{n}_a,\Tilde{n}_c \in \{0,1\}$.

First consider $\expval{U_{a,c}^{n_a, n_c}(S^1)U_{\Tilde{b}}(S^{D-2})}$. The calculation is similar to the calculation of linkings in untwisted BF theories. The monodromy factor $e^{i\oint_{S^1}(n_a a + n_c c)}$ introduces source terms in the path integral for $\tilde{a}$ and $\tilde{c}$:
\begin{equation}
    I = I_0 + i\int_{S^1}(n_a a_1 + n_c c_1) + i\int_{S^{D-2}}\Tilde{b}_{D-2}. 
\end{equation}
Integrating out $a,c$ leads to the modified equations of motion: 
\begin{equation}
    d\tilde{a} = -\pi n_a \delta (S^1), \qquad d\tilde{c} = -\pi n_c \delta (S^1).
\end{equation}
However, no operators inserted in the path integral are coupled to $\tilde{a}$ or $\tilde{c}$, so there is no obstruction to unlinking $U_{a,c}^{n_a, n_c}(S^1)$ with $U_{\Tilde{b}}(S^{D-2})$. This produces the expectation value:
\begin{equation}
    \expval{U_{a,c}^{n_a, n_c}(S^1)U_{\Tilde{b}}(S^{D-2})} = 1.
\end{equation}
This implies that the characters of the 1-dimensional irreducible representation evaluated on the nontrivial size-1 conjugacy class are 1, which match Table \ref{table - D4 Characters}.

    Now consider $\expval{U_{a,c}^{n_a, n_c}(S^1)\hat{U}_{\Tilde{a},\Tilde{c}}^{\Tilde{n}_a,\Tilde{n}_c}(S^{D-2})}$. The operator insertion modifies the action to:
    \begin{equation}
        I = I_0 + i\int_{S^1}(n_a a_1 + n_c c_1) + i\int_{S^{D-2}}(\Tilde{n}_a\Tilde{a}_{D-2} + \Tilde{n}_c\Tilde{c}_{D-2}).
    \end{equation}
    Integrating out $\Tilde{a}$ and $\Tilde{c}$, we have:
    \begin{equation}
        da = - \pi \Tilde{n}_a\delta(S^{D-2}), \quad dc = -\pi\Tilde{n}_c\delta(S^{D-2}).
    \end{equation}
    This converts the $S^1$-integral into nontrivial linking invariants, which contributes a $\mathrm{U}(1)$ phase $(-1)^{n_a\Tilde{n}_a + n_c\Tilde{n}_c}$. Furthermore, note that $\hat{U}_{\Tilde{a},\Tilde{c}}^{\Tilde{n}_a,\Tilde{n}_c}(S^{D-2})$ contains a condensation defect $(1+U_{\Tilde{b}}(S^{D-2}))/2$. Shrinking this operator does not produce any new factors. However, by Eq. \eqref{eq - D4 electric condensation}, shrinking the electric condensations  produces a factor of 2. Therefore, the expectation value reads:
    \begin{equation}
        \expval{U_{a,c}^{n_a, n_c}(S^1)\hat{U}_{\Tilde{a},\Tilde{c}}^{\Tilde{n}_a,\Tilde{n}_c}(S^{D-2})} = 2\times (-1)^{n_a\Tilde{n}_a + n_c\Tilde{n}_c}.
    \end{equation}
    This implies that the characters of the 1-dimensional irreducible representation evaluated on the size-2 conjugacy classes are $(-1)^{n_a\Tilde{n}_a + n_c\Tilde{n}_c}$, which matches Table \ref{table - D4 Characters}.
    
    The expectation value $\expval{U_b(S^1)U_{\Tilde{b}}(S^{D-2})}$ can be evaluated by integrating out the $b$-field, which enforces $d\Tilde{b} = -\pi\delta(S^1)$. Since $U_{\Tilde{b}}(S^{D-2})$ couples nontrivially to $\Tilde{b}$, now we have nontrivial contributions $-1$ from the linking. Note that the operator $U_b(S^1)$ contains an electric condensation $\frac{1}{2}(1+ U_a(S^1)) (1+ U_{c}(S^{1}))$ and shrinking it away gives a factor of $2$. Therefore, we have:
    \begin{equation}
        \expval{U_b(S^1)U_{\Tilde{b}}(S^{D-2})} = -2.
    \end{equation}
    This implies that the character of the 2-dimensional irreducible representation evaluated on the size-1 conjugacy class is $-2$, which matches Table \ref{table - D4 Characters}.

    Finally, consider the expectation value $\expval{\hat{U}_b(S^1)\hat{U}_{\Tilde{a},\Tilde{c}}^{\Tilde{n}_a,\Tilde{n}_c}(S^{D-2})}$. The operator insertion modifies the action to:
    \begin{equation}
        I = I_0 + i\int_{S^1}b_1 + i\int_{S^{D-2}}(\Tilde{n}_a\Tilde{a}_{D-2} + \Tilde{n}_c\Tilde{c}_{D-2}).
    \end{equation}
    Integrating out the $b$ fields leads to $d\Tilde{b}= -\frac{2\pi}{2}\delta(S^1)$. This converts the condensation defect condensation defect $(1+U_{\Tilde{b}}(S^{D-2}))/2$ on $S^{D-2}$ to a sum over roots of unity, so this expectation value identically vanishes:
    \begin{equation}
    \expval{\hat{U}_b(S^1)\hat{U}_{\Tilde{a},\Tilde{c}}^{\Tilde{n}_a,\Tilde{n}_c}(S^{D-2})} = 0.
    \end{equation}
    This implies that the character of the 2-dimensional irreducible representation equals zero when evaluated on the size-2 conjugacy classes, which matches Table \ref{table - D4 Characters}.

    \subsection{(2+1)D Lattice Regularization}\label{subsection - lattice regularization}

In this subsection, we point out a subtle issue in the linking invariant calculation. It is known that the expectation value of a Hopf link in a (2+1)D DW theory is proportional to the modular S-matrix \cite{Coste:2000tq}. Using the same method, we should be able to reproduce the modular $S$-matrix of $\mathbb{D}_4$ theory in (2+1)D. Our naive path integral calculation suggests that all line operators would have trivial self-linking in (2+1)D, which is generally not true. Therefore an appropriate lattice regularization procedure is required. This effect was originally discovered in \cite{He:2016xpi}, which we quickly review.

First, note that the untwisted $\mathbb{D}_4$ DW theory admits dyonic lines in $(2+1)$D. The fusion between any two invertible Wilson or t' Hooft lines produces an invertible line, giving eight in total: $\{1, U_{a}, U_{\tilde{b}}, U_{c}, U_{a,\tilde{b}}, U_{a,c}, U_{\tilde{b},c}, U_{a,\tilde{b},c}\}$. The mutual fusions of the three non-invertible lines $\hat{U}_{\tilde{a}}, \hat{U}_b,\hat{U}_{\tilde{c}}$ define four more non-invertible lines: $\{\hat{U}_{\tilde{a},b},\hat{U}_{\tilde{a},\tilde{c}},\hat{U}_{b,\tilde{c}},\hat{U}_{\tilde{a},b,\tilde{c}}\}$. We can define more non-invertible dyon lines by fusing the invertible lines with the seven bare non-invertible lines. The topological boundary conditions on the bare non-invertible lines  define equivalence relations among the invertible lines that can be fused with the bare non-invertible lines. For example, fusing the invertible lines $U_{a,\Tilde{b},c}^{n_a, \Tilde{n}_b,n_c}$ with $U_{\Tilde{a},b,\Tilde{c}}$ gives:
\begin{equation}
    U_{\Tilde{a},b,\Tilde{c}}(M_1) = \frac{1}{2}e^{i\oint_{M_1}(\tilde{a}_1 + b_1 + \tilde{c}_1)}(1 + U_{a,\tilde{b}}(M_1))(1+ U_{\tilde{b}+c}(M_1)).
\end{equation}
We use a triple $(n_a, \Tilde{n}_b,n_c)$ to label the invertible lines $U_{a,\Tilde{b},c}^{n_a, \Tilde{n}_b,n_c}$, where the entries take values in integers under addition mod 2. The topological boundary condition implemented by $\mathcal{S}_{a+\Tilde{b}} \mathcal{S}_{\Tilde{b}+c}$ defines an equivalence relation in the following sense. All $U_{a,\Tilde{b},c}^{n_a, \Tilde{n}_b,n_c}$ satisfying $(n_a +\Tilde{n}_b, \Tilde{n}_b+n_c)=(0,0)$ form one equivalence class and the remaining invertible lines form a second equivalence class:
    \begin{equation}
        \{1, U_{a,\Tilde{b},c}^{1,1,0}, U_{a,\Tilde{b},c}^{0,1,1}, U_{a,\Tilde{b},c}^{1,0,1}\}
        \quad \text{and} \quad \{U_{a,\Tilde{b},c}^{0,0,1}, U_{a,\Tilde{b},c}^{0,1,0}, U_{a,\Tilde{b},c}^{1,0,0},U_{a,\Tilde{b},c}^{1,1,1}\}.
    \end{equation}
Note that all lines in the first equivalence class can be absorbed by the higher gauging condensation defect. Meanwhile, all lines in the second equivalence class cannot be completely absorbed by the higher gauging condensation defects and they can be mapped to each other by fusing with an appropriate line factored out of the condensation defect on $U_{\tilde{a},b,\tilde{c}}$. For example:
\begin{equation}
   \begin{split}
        &U_{\Tilde{a},b,\Tilde{c}}(M_1)\times U_{a,\tilde{b},c}(M_1)\\
        =& \frac{1}{2}e^{i\oint_{M_1}(\tilde{a}_1 + b_1 + \tilde{c}_1)}(1 + U_{a,\tilde{b}}(M_1))(1+ U_{\tilde{b},c}(M_1)) \times U_{a,\tilde{b},c}(M_1)\\
        =& \frac{1}{2}e^{i\oint_{M_1}(\tilde{a}_1 + b_1 + \tilde{c}_1)}( U_{a,\tilde{b}}(M_1)+1)(1+ U_{\tilde{b},c}(M_1)) \times U_{a,\tilde{b}}(M_1) \times U_{a,\tilde{b},c}(M_1)\\
        =& \frac{1}{2}e^{i\oint_{M_1}(\tilde{a}_1 + b_1 + \tilde{c}_1)}( 1+U_{a,\tilde{b}}(M_1))(1+ U_{\tilde{b},c}(M_1))  \times U_{c}(M_1)\\
        = & U_{\Tilde{a},b,\Tilde{c}}(M_1)\times U_{c}(M_1).
   \end{split}
\end{equation}
In this sense, $U_{a,\tilde{b},c}^{1,1,1} = U_{a,\tilde{b},c}$ is equivalent to $ U_{a,\tilde{b},c}^{0,0,1}= U_{c} $. As one can check,  the boundary conditions on each bare non-invertible line reduce the set of eight invertible lines into exactly two equivalence classes. Therefore, we have $ 8 + (3+4) \times 2 = 22$ simple lines as expected. 

Before introducing the lattice regularization, let us first compute the $S$-matrix without regularization and see how far the result deviates from the quantum double calculation. Let us adopt the following ordering for the anyon basis:
\begin{equation}
    \begin{split}
        &1, U_{a}, U_{\tilde{b}}, U_{c}, U_{a,\tilde{b}}, U_{a,c}, U_{\tilde{b},c}, U_{a,\tilde{b},c},\\
        &\hat{U}_{\tilde{a}},\hat{U}_{b},\hat{U}_{\tilde{c}},\hat{U}_{\tilde{a}}^{(1,0,0)},\hat{U}_{b}^{(0,1,0)},\hat{U}_{\tilde{c}}^{(0,0,1)},\\
        &\hat{U}_{\tilde{a},b},\hat{U}_{\tilde{a},\tilde{c}},\hat{U}_{b,\tilde{c}},\hat{U}_{\tilde{a},b}^{(1,0,0)},\hat{U}_{\tilde{a},\tilde{c}}^{(0,0,1)},\hat{U}_{b,\tilde{c}}^{(0,0,1)},\hat{U}_{\tilde{a},b,\tilde{c}},\hat{U}_{\tilde{a},b,\tilde{c}}^{(0,0,1)}.
    \end{split}
\end{equation}
With our previous experience in computing linking invariants in $D$-dimensions, the naive $S$-matrix calculation follows these rules:
\begin{enumerate}
    \item The Hopf link of a vacuum line with any operator is its quantum dimension.
    \item The linkings between invertible lines all equal to 1.
    \item Consider correlation functions of the type $\expval{U_a(S^1)\hat{U}_{\tilde{a}}(S^1{}')}$. Integrating out $\tilde{a}$ implements $da = -\frac{2\pi}{2}\delta(S^1)$, so we have a nontrivial contribution from linking. Shrinking the higher gauging condensation defects produces the quantum dimension 2. Therefore:
    \begin{equation}
        \expval{U_a(S^1)\hat{U}_{\tilde{a}}(S^1{}')} = \expval{U_{\tilde{b}}(S^1)\hat{U}_{b}(S^1{}')} = \expval{U_c(S^1)\hat{U}_{\tilde{c}}(S^1{}')}=-2.
    \end{equation}
     Similarly, consider correlation functions of the type $\expval{U_a(S^1)\hat{U}_{b}(S^1{}')}$. $U_a(S^1)$ does not link with the monodromy factor or the condensation defects in $\hat{U}_{b}(S^1{}')$, so the expectation value equals to the product of their quantum dimensions. Therefore:
    \begin{equation}
        \begin{split}
           & \expval{U_a(S^1)\hat{U}_{b}(S^1{}')} = \expval{U_a(S^1)\hat{U}_{c}(S^1{}')} = \expval{U_{\tilde{b}}(S^1)\hat{U}_{\tilde{a}}(S^1{}')}\\
           &= \expval{U_{\tilde{b}}(S^1)\hat{U}_{\tilde{c}}(S^1{}')} = \expval{U_c(S^1)\hat{U}_{a}(S^1{}')} = \expval{U_c(S^1)\hat{U}_{c}(S^1{}')} = 2.
        \end{split}
    \end{equation}
    This observation can be generalized to the Hopf link between any invertible lines and bare non-invertible lines:
    \begin{equation}
        \expval{U_{a,\tilde{b},c}^{n_a, \tilde{n}_b,n_c}(S^1) \hat{U}_{\tilde{a},b,\tilde{c}}^{\tilde{n}_a, n_b, \tilde{n}_c}(S^1{}')} = 2(-1)^{n_a\tilde{n}_a + n_b\tilde{n}_b + n_c\tilde{n}_c},
    \end{equation}
    where $n_a,\tilde{n}_a,n_b,\tilde{n}_b,n_c,\tilde{n}_c\in \{0,1\}$. The same result applies to non-invertible lines stacked with invertible lines. 
    \item Consider the correlation functions between non-invertible lines. Here we only explain three calculations, which can be easily generalized to the remaining entries. Consider $\expval{\hat{U}_{\tilde{a}}(S^1)\hat{U}_{\tilde{a}}(S^1{}')}$, where $\hat{U}_{\tilde{a}}(S^1) = 2e^{i\oint_{S^1}\tilde{a}_1}\delta\left(\oint_{S^1}\tilde{b}_1\right)\delta\left(\oint_{S^1}c_1\right)$. Integrating out $\tilde{a}_1$ implements the constraint $da_1 = -\pi \delta(S^1)$. However, no operators are coupled to $a_1$ in this expectation value, so the linking is trivial and $\expval{\hat{U}_{\tilde{a}}(S^1)\hat{U}_{\tilde{a}}(S^1{}')}=4$, which is the product of their quantum dimensions. Similarly, consider $\expval{\hat{U}_{\tilde{a}}(S^1)\hat{U}_{\tilde{a}}^{(1,0,0)}(S^1{}')}$. In this case, $\hat{U}_{\tilde{a}}^{(1,0,0)}(S^1{}')$ contains a copy of $U_a$, so a nontrivial linking phase is produced by integrating out $\tilde{a}_1$ in $\hat{U}_{\tilde{a}}(S^1)$ and we have $\expval{\hat{U}_{\tilde{a}}(S^1)\hat{U}_{\tilde{a}}^{(1,0,0)}(S^1{}')}=-4$. Finally, consider $\expval{\hat{U}_{\tilde{a}}(S^1)\hat{U}_{b}(S^1{}')}$, where $U_b(S^1) = 2e^{i\oint_{S^1}b_1}\delta\left(\oint_{S^1}a_1\right)\delta\left(\oint_{S^1}c_1\right)$. Integrating out $\tilde{a}_1$ in $\hat{U}_{\tilde{a}}(S^1)$ converts $\delta\left(\oint_{S^1}a_1\right)$ to a sum over roots of unity $\delta\left(\oint_{S^1}a_1\right) = \frac{1}{2}(1-1)=0$, which trivializes this correlation function.
\end{enumerate}

We find that the bottom right $2\times 2$ block of the $S$-matrix obtained this way differs from the quantum double calculation \cite{He:2016xpi} by a sign and the remaining entries agree with the quantum double calculation. To fix this, we need to adopt a lattice regularization scheme \cite{He:2016xpi}. Let us revisit the definition of the operator $\hat{U}_{\tilde{a},b,\tilde{c}}$. In (2+1)D, it is convenient to relabel the gauge field $\tilde{b}_1,a_1, c_1$ as an ordered triple $(A_1^{(1)},A_1^{(2)},A_1^{(3)})$ and $b_1, \tilde{a}_1, c_1$ as an ordered triple $(b_1^{(1)},b_1^{(2)},b_1^{(3)})$. The action Eq. \eqref{eq - D4 gauge theory action} then becomes:
\begin{equation}
    I_{\mathbb{D}_4} = \frac{i}{2\pi}\int_{M_3}\left(2\, b_1^{(i)}\wedge dA_1^{(i)} + \frac{1}{3\pi}\epsilon^{ijk}A_1^{(i)}\wedge A_1^{(j)}\wedge A_1^{(k)}\right).
\end{equation}
The operator $\hat{U}_{\tilde{a},b,\tilde{c}}$ becomes:
\begin{equation}
    \begin{split}
        \hat{U}_{\tilde{a},b,\tilde{c}} &= \int \mathcal{D}\alpha_o^{(i)}\mathcal{D}\lambda_0^{(i)}\exp\left(i\oint_{M_1}\sum_{i=1}^3 b_1^{(i)} + \sum_{i,j,k}i\oint_{M_1} \frac{\epsilon^{ijk}}{\pi}\left(\frac{1}{2} \alpha_0^{(j)}d \alpha_0^{(k)}\right) + (d\alpha_0^{(j)}-A_1^{(j)})\lambda_0^{(k)} \right)\\
        &= 2\exp\left(i\oint_{M_1}\sum_{i=1}^3 b_1^{(i)} + \sum_{i,j,k}i \oint_{M_1} \frac{\epsilon^{ijk}}{2\pi}\omega^{(j)} d\omega^{(k)} \right) \delta(\Bar{\omega}^{(1)}|_{M_1} - \Bar{\omega}^{(2)}|_{M_1}) \delta(\Bar{\omega}^{(2)}|_{M_1} - \Bar{\omega}^{(3)}|_{M_1})\\
        & = 2\exp(i\oint_{M_1}\sum_{i=1}^3 b_1^{(i)} + \sum_{i,j,k}i\oint_{M_1} \frac{\epsilon^{ijk}}{2\pi}\tilde{\omega}^{(j)} d\tilde{\omega}^{(k)} ),
    \end{split}
\end{equation}
where the monodromy $\Bar{\omega}^{(i)} \equiv \oint_{M_1}A_1^{(i)}$ is a formal variable and the objects $\tilde{\omega}$ are the monodromies surviving the projection. The integrals $\oint_{M_1} \frac{\epsilon^{ijk}}{2\pi}\tilde{\omega}^{(j)} d\tilde{\omega}^{(k)}$ are the lattice regulators and they are evaluated by defining a pair of discrete derivative operators:
\begin{equation}
    \begin{split}
        d\tilde{\omega}^{(i)}(r) &= \tilde{\omega}^{(i)}(r+1) - \tilde{\omega}^{(i)}(r),\\
        \Bar{d}\tilde{\omega}^{(i)}(r)&= \tilde{\omega}^{(i)}(r) - \tilde{\omega}^{(i)}(r-1), 
    \end{split}
\end{equation}
where $\Bar{d}$ is the adjoint of $d$, and $r$ labels the lattice sites. It turns out, the only nontrivial regulator is:
\begin{equation}
\oint_{M_1}\left(\tilde{\omega}^{(i)}d\tilde{\omega}^{(j)} - \tilde{\omega}^{(j)}\Bar{d}\tilde{\omega}^{(i)} \right) = - \pi^2 \quad \text{where}\quad i \neq j ,
\end{equation}
where $L$ denotes the total number of lattice sites in the regularization scheme and the factor of $\pi$ comes from the fact that the variables are formally $\mathbb{Z}_2$-valued. Inserting this regulator restores the sign of the bottom right $2\times 2$ block of the linking calculation for the $S$-matrix.

\section{Gauging of Finite Symmetries by Higher Gauging Condensation Defects}
\label{section - various notions of gauging}

In this section, we clarify the idea of gauging of a finite symmetry by higher gauging condensation defects for untwisted Dijkgraaf-Witten theories in arbitrary spacetime dimensions. We saw in the previous section that the fusion rules and linking invariant calculations are independent of spacetime dimensions. Using this universal behavior, we propose a Lagrangian description for an untwisted Dijkgraaf-Witten theories with gauge group $G$ that fits in abelian extensions:
\begin{equation}
    1 \rightarrow A \rightarrow G \rightarrow J \rightarrow 1
\end{equation}
where $A,J$ are both finite abelian groups. Generically, the Lagrangian is that of a Dijkgraaf-Witten theory with gauge group $A\times J$ and a non-trivial $H^{D}(A\times J,U(1))$ twist. In $(2+1)$D, a proper equivalence relation among Dijkgraaf-Witten theory is the braided equivalence of the underlying UMTC. We explain how to use the $(2+1)$D equivalence with the dimension universality to construct effective actions for untwisted Dijkgraaf-Witten theories in arbitrary spacetime dimensions. We will begin with a review of the familiar notions of finite symmetries gauging in $(2+1)$D and outline the gauging procedure and matching criteria in Sec. \ref{subsection - gauging proposal}. In Sec. \ref{subsection - Heisenberg gauge theory}, using this construction we propose a family of effective Lagrangians for untwisted Dijkgraaf-Witten theories whose gauge groups are the Heisenberg groups over $\mathbb{Z}_p$ for $p$-prime.

Before going into the details, we end this introduction with a disclaimer. This construction is a formal analogy of the gauging of continuous symmetries in non-topological Lagrangian QFTs. On the other hand, Noether currents do not exist for finite symmetries since no continuous parameters can parametrize the symmetry group. Furthermore, there can be obstructions to the gauging of finite symmetries. For a $0$-form group like symmetry $J$, the obstruction is quantified by the 't Hooft anomaly $H^{(D+1)}(J,U(1))$, and the anomaly class is determined by the degrees of freedom of the $D$-dimensional Dijkgraaf-Witten theory itself. These obstructions are not the concern of this work. Hence our proposal below should be understood as a gauging procedure in an EFT sense and it should not be treated as a canonical definition. In limited cases, there exists a concrete definition for the 't Hooft anomaly $J$-0-form symmetry in a Dijkgraaf-Witten theories\cite{Thorngren:2015gtw, Mller2018ParallelTO,Muller:2018doa,Kapustin:2014zva, Wang:2021nrp}, which we will summarize at the end of Sec. \ref{subsection - gauging proposal}. We also stress that we are only interested in a \textbf{\textit{truncation}} of Dijkgraaf-Witten theories where the non-trivial operators defined by the gauge transformations of the effective action are defined on closed oriented codimension-2 submanifolds and closed oriented  1-dimensional submanifolds.

\subsection{Effective Action for Untwisted DW Theories with Non-Abelian Gauge Group}\label{subsection - gauging proposal}

    In principle, the gauging of finite symmetries in a Dijkgraaf-Witten theory in arbitrary spacetime dimension should be analyzed with the language of higher categories. This problem is well studied in $(2+1)$D, where the fundamental excitations are line operators and their interactions are described by UMTCs. To discuss anyon symmetries, we need to introduce codimension-1 defects, which enlarges the UMTC to a 2-category. In the generalized symmetry language, these symmetries are 0-form symmetries and their gauging was investigated mathematically in \cite{Cui2015OnGS} and reinterpreted physically in terms of anyon diagrams in \cite{Barkeshli:2014cna}.  It can be understood as the categorification of gauging by minimal coupling. This is a two-step procedure: one couples background gauge fields of the symmetry to the theory, then promotes the background gauge fields to dynamical gauge fields. Alternatively, the first step corresponds to the insertion of symmetry defects and non-genuine defect sector states\footnote{Hence the name defectification}, while the second step corresponds to the projection onto the gauge-invariant states. Mathematically, the first step corresponds to the extension of a UMTC $\mathcal{C}$ to a $J$-crossed braided tensor category $\mathcal{C}_J^{\times}$ and the second step corresponds to the $J$-equivariantization of $\mathcal{C}_J^{\times}$.  The inverse of gauging is anyon condensation\cite{Kong2013AnyonCA} and the inverse of defectification is the confinement of defect sector states. The relations between them are summarized in the following diagram \cite{Barkeshli:2014cna}:
    \begin{center}
        \begin{tikzpicture}[baseline=(current bounding box.center),>=stealth,thick]
  \node (C)   at (0,0) {$\mathcal{C}$};
  \node (CG)  at (4,0) {$\mathcal{C}^{\times}_{J}$};
  \node (CGG) at (8,0) {$(\mathcal{C}^{\times}_{J})^{J}$};

  \draw[->] (C)  to[bend left=20]  node[above] {Defectification} (CG);
  \draw[<-] (C)  to[bend right=20] node[below] {Confinement}     (CG);

  \draw[->] (CG)  to[bend left=20]  node[above] {Gauging}        (CGG);
  \draw[<-] (CG)  to[bend right=20] node[below] {Condensation}   (CGG);
\end{tikzpicture}
    \end{center}

    Just like the gauging of group-like symmetries in usual non-topological QFTs, there are also obstructions to the gauging of anyon symmetries in $(2+1)$D. However, the obstruction happens only at the defectification step \cite{Cui2015OnGS}. Let the 0-form symmetry be $J^{(0)}$ and pick an action $\rho: J\rightarrow \text{Aut}(\mathcal{C})$:
    \begin{itemize}
        \item Symmetry fractionalization is possible only if a specific $[O_3]\in H_{\rho}^3(J, \mathcal{A})$ vanishes, where $\mathcal{A}$ is the subset of abelian topological lines in $\mathcal{C}$. $[O_3]$ is known as the fractionalization obstruction class.
        \item When $[O_3]$ vanishes, the possible fusion rules are parameterized by $[\alpha]\in H^2(J, \mathcal{A})$. Given the pair $(\rho, \alpha)$, the fusion rules are associative only if $[O_4]\in H^4(J,\mathrm{U}(1))$ vanishes. $[O_4]$ is known as the defectification obstruction class and it can be understood as an analog of the 't Hooft anomaly for the 0-form symmetry. 
    \end{itemize}
    When these two obstructions vanish, there is no additional obstruction to the equivariantization  step \cite{Cui2015OnGS}. 

    Now we return to the Lagrangian description of 0-form symmetry gauging by higher gauging condensation defects in untwisted Dijkgraaf-Witten theories. We only consider the gauging of finite symmetries in untwisted Dijkgraaf-Witten theories with operators supported on closed oriented submanifolds. As reviewed in Sec. \ref{subsection - D4 gauge theory}, we see that the following data of an untwisted Dijkgraaf-Witten theory does not depend on the spacetime dimension $D$ for $D\geq 3$:
    \begin{itemize}
        \item Operator spectrum of pure Wilson and t' Hooft operators. 
        \item Mutual fusion among Wilson and t' Hooft operators. Especially in $D=3$, we ``forget" fusions between Wilson lines and t' Hooft lines so that the fusion ring is projected down to a direct sum of mutual fusions among Wilson lines and mutual fusions among t' Hooft lines.
        \item Expectation values of Wilson lines on $S^1$ and t' Hooft surfaces on $S^{(D-2)}$ on a local patch isomorphic to $\mathbb{R}^D$.
        \item Expectation values of Hopf links between Wilson lines and t' Hooft surfaces in a local patch isomorphic to $\mathbb{R}^D$. 
    \end{itemize}
    These data are dimension-independent because they only depend on the gauge group. We can trust the EFT descriptions for these moves on contractible local patches of $M_D$. To probe the precision of the EFT Lagrangian in terms of twisted abelian gauge theories, we need to compute more partition functions and correlation functions from the homotopy theory perspective. We leave the exploration for future investigations.
    
    In $(2+1)$D, we start with an untwisted Dijkgraaf-Witten theory $\text{Rep}(\calD(A))$ with abelian gauge group $A$. Let $J$ be a 0-form global symmetry that fits in a short exact sequence:
    \begin{equation} \label{eq - anyon symmetry short exact sequence}
        1 \rightarrow A \rightarrow G \rightarrow J \rightarrow 1.
    \end{equation}
    In the vanishing of the $[O_3]$ and $[O_4]$ obstruction, it is known that gauging the $J^{(0)}$ symmetry leads to the untwisted Dijkgraaf-Witten theory $G$ \cite{Maier2011EquivariantMC}. On the other hand, the Lagrangian description in general produces a twisted Dijkgraaf-Witten theory $\text{Rep}(\calD^{\text{III}}(A\times J))$ with gauge group $A\times J$, where the discrete torsion term corresponds to a type-III twist. To declare the action a good effective action for the $G$-gauge theory, we need to ensure the equivalence of the two theories in $(2+1)$D, where we choose the equivalence relation to be braided equivalence between braided tensor categories. This leads us to the following proposal:
    \begin{tcolorbox}[colback=green!10!white,colframe=blue!75!black,
                    ]
    Consider an untwisted Dijkgraaf-Witten theory with an abelian gauge group $A$ and an abelian $J^{(0)}$ symmetry that fits into a short exact sequence in Eq. \eqref{eq - anyon symmetry short exact sequence}. One can construct an effective action $I_{eff}$ by summing over higher gauging condensation defects and promoting the $J^{(0)}$ background gauge fields to dynamical gauge fields. Suppose there is a braided equivalence between $\text{Rep}(\calD^{\text{III}}(A\times J))$ and $\text{Rep}(\calD(G))$ in (2+1)D,  then $I_{eff}$ can be treated as an effective action for the untwisted Dijkgraaf-Witten theory with gauge group $G$ in arbitrary spacetime dimensions. 
\end{tcolorbox}
    
    A few explanations on the extension in Eq. \eqref{eq - anyon symmetry short exact sequence} are in order. This extension determines a \textbf{\textit{weak $J$-action}} on $A$ \cite{Maier2011EquivariantMC}, which we will discuss in Appendix \ref{appendix - gauging finite symmetry in DW}. For a fixed pair of $A$ and $J$, isomorphism classes of weak $J$-actions correspond to isomorphism classes of group extensions \cite{Maier2011EquivariantMC}.
    For each $j\in J$, we specify a collection of $A$-automorphisms so that $\rho_i\circ \rho_j =  \rho_{ij} $. For each pair $i,j \in J$, we specify a collection of $c_{i,j}\in A$ so that:
    \begin{equation}
        \rho_i(c_{j,l})\cdot c_{i,jk} = c_{i,j}\cdot c_{ij,k} \quad \text{and} \quad c_{1,1}=1.
    \end{equation}
    When $c_{i,j}=1$ for all $i,j\in A$, the weak action reduces to a \textbf{\textit{strict $J$-action}} on $A$ and the group extension splits. This was systematically studied in \cite{Maier2011EquivariantMC}, where the group $A$ and $J$ both need not be abelian. In those cases, the weak $J$-action on $A$ is associative up to inner $A$-automorphisms labeled by $c_{i,j}$. We will explain this in detail in Appendix \ref{appendix - gauging finite symmetry in DW}. For strict J-actions, our proposal becomes:
    \begin{tcolorbox}[colback=green!10!white,colframe=blue!75!black,
                    ]
    Consider an untwisted Dijkgraaf-Witten theory with an abelian gauge group $A$ and an abelian $J^{(0)}$ symmetry that acts on $A$ by group automorphisms. One can construct an effective action $I_{eff}$ by summing over higher gauging condensation defects and promoting the $J^{(0)}$ background gauge fields to dynamical gauge fields. Suppose there is a braided equivalence between $\text{Rep}(\calD^{\text{III}}(A\times J))$ and $\text{Rep}(\calD(A\rtimes J))$ in (2+1)D,  then $I_{eff}$ can be treated as an effective action for the untwisted Dijkgraaf-Witten theory with gauge group $A\rtimes J$ in arbitrary spacetime dimensions. 
\end{tcolorbox}

    Finally, we mention that there exists a rigorous definition of 't Hooft anomalies for Dijkgraaf-Witten theories in arbitrary spacetime dimensions for the global symmetries induced by the automorphisms of the gauge group. Note that Dijkgraaf-Witten theories in (2+1)D admit a much richer set of symmetries than Dijkgraaf-Witten theories in other spacetime dimensions. In the language of \cite{Etingof2009FusionCA, Cui2015OnGS}, the symmetry of $\text{Rep}(\mathcal{D}^{\omega}(H))$ is the group $\text{Aut}_{\otimes}^{\mathfrak{br}}( \mathcal{D}^{\omega}(H))$ of equivalence classes of braided tensor auto-equivalences of $\text{Rep}(\mathcal{D}^{\omega}(H))$. For example, when $A$ is abelian and $\omega$ is the trivial class, $\text{Aut}_{\otimes}^{\mathfrak{br}}( \mathcal{D}^{\omega}(H))$ has a $\mathbb{Z}_2$ EM duality subgroup exchanging the Wilson lines and the 't Hooft lines. This symmetry is absent in higher dimensions because the Wilson and 't Hooft operators are defects in different dimensions. On the other hand, global symmetries induced by the automorphisms of the gauge group are well defined in all spacetime dimension, and they are the symmetries of interest in our proposal. 

    Consider a Dijkgraaf-Witten theory in arbitrary spacetime dimension with gauge group $H$ and a topological action $\omega \in H^{D+1}(H,U(1))$. Here $H$ need not be abelian. Consider a $J^{(0)}$ global symmetry described by the group extension:
    \begin{equation}
        1 \rightarrow H \xrightarrow{\iota} G \rightarrow J \rightarrow 1
    \end{equation}
    In \cite{Muller:2018doa}'s language, gauging the $J$ symmetry should be understood as finding an appropriate $\widehat{\omega}\in H^{D+1}(G,U(1))$ so that its pull back $\iota^*\widehat{\omega}$ to $BH$ by the embedding map $\iota$ is cohomologous to the original action $\omega$. The 't Hooft anomaly of this $J^{(0)}$ global symmetry is an obstruction to such $\widehat{\omega}$ and the anomaly can be computed from the Lyndon-Hochschild-Serre spectral sequence associated to the group extension. We refer the readers to \cite{Wang:2021nrp, Kapustin:2014zva} for a physical derivation when $H$ is finite abelian. When this 't Hooft anomaly vanishes, our gauging procedure by summing over higher gauging condensation defects can be understood as a precise formulation of $J$-gauging.

    \subsection{Example - Heisenberg Gauge Theory}\label{subsection - Heisenberg gauge theory}

    In this subsection, we demonstrate our proposal by constructing the effective action for untwisted Dijkgraaf-Witten theory with the Heisenberg gauge group $H_3(\mathbb{Z}_p)$. We study in detail the operator spectrum for $p=3$ and show that the $H_3(\mathbb{Z}_p)$ character table can be reproduced exactly from Hopf links in arbitrary spacetime dimensions.

    The Heisenberg group $H_3(\mathbb{Z}_p)$ is one of the simplest extraspecial groups. This group is well-studied in the math literature. Consider the Heisenberg group $H_3(\bbZ_p)$ with prime $p$. Define 
\begin{equation}
    x = \begin{pmatrix}
        1 & 0 & 0 \\ 0 & 1 & 1 \\ 0 & 0 & 1
    \end{pmatrix},\quad 
    y = \begin{pmatrix}
        1 & 0 & 1 \\ 0 & 1 & 0 \\ 0 & 0 & 1
    \end{pmatrix},\quad
    z = \begin{pmatrix}
        1 & 1 & 0 \\ 0 & 1 & 0 \\ 0 & 0 & 1
    \end{pmatrix},
\end{equation}
with the relations $y = zx z^{-1}x^{-1} $,  $xy = yx $,  $yz= zy$, and $x^p = y^p = z^p = 1$. $H_3(\bbZ_p)$ is a group of $3\times 3$ upper triangular matrices of the form
\begin{equation}
    \begin{pmatrix}
        1 & c & b \\ 0 & 1 & a \\ 0 & 0 & 1
    \end{pmatrix} = (a,b, c),
\end{equation}
with the multiplication $(a,b,c)\cdot (a',b',c') = (a+a', a'c+b+b', c+c') $. The Heisenberg group has $p+1$ order-$p^2$ normal subgroups of the form
\begin{equation}
    H_\infty = \{(a, b, 0) \},\quad H_\lambda = \{(\lambda c, b, c) \},
\end{equation}
where $\lambda \in \bbZ_p$ and $a,b,c \in \bbZ_p$.

    To verify the effective action, we need the character table for $H_3(\mathbb{Z}_p)$. If a finite group $G$ is isomorphic to a semi-direct product of two finite abelian groups, then the irreducible representations and linear characters of $G$ can be constructed from the representation theory data of $A$ and $B$ by Clifford's theory, which we review in Appendix \ref{appendix - Mackey's Theory}. The conjugacy classes of $H_3(\mathbb{Z}_p)$ can be obtained from the group multiplication law. The size-1 conjugacy classes are the identity elements and $(0,y,0)$ for $y\in \{1,2,d\dots, p-1\}$. The size-$p$ conjugacy classes are $[(x,0,0)]$ and $[(x,0,z)]$, where $x\in \{0,1,\dots,p-1\}$ and $z\in \{1,\dots, p-1\}$. The 1 dimensional irreducible representations of $H_3(\mathbb{Z}_p)$ are labeled by $(\chi_0,\omega_n)$ for $n\in\{1,2,\dots, p-1\}$, and $(\chi_a, \chi_0)^{\omega_n}$ for $n\in \{0,1,\dots p-1\}$ with $a\in \{0,1,\dots, p-1\}$. The remaining irreducible representations $(\chi_a, \chi_b)$ are all $p$-dimensional, where $b\in \{ 1, \dots, p-1\}$ labels the irreducible representations. The label $a$ in $(\chi_a,\chi_b)$ is a representative of a $\mathbb{Z}_p$ orbit on $\widehat{\mathbb{Z}}_p\times \widehat{\mathbb{Z}}_p$ and different values of ``$a$" under the same ``$b$" corresponds to the same $H_3(\mathbb{Z}_p)$ irreducible representation. We postpone the explanation of the representation theory detail to Appendix \ref{appendix - Mackey's Theory}. The full character table is summarized in Table \ref{table - Heisenberg Group Character Table}.
\begin{table}[h!]
\centering
\begin{tabular}{c|c|c|c|c}

$H_3(\mathbb{Z}_p)$ & e & $[(0,y,0)]$ where $y\neq 0$ & $[(x,0,0)]$ where $x\neq 0$ & $[(x,0,z)]$ where $z\neq 0$ \\ 
\hline
$T$  & 1 &1  &1 & 1 \\ 
$T^{\omega_n}$ where $n\neq0$ & 1 & 1 & 1 & $\xi^{nz}$ \\ 
$(\chi_a, \chi_0)^{\omega_n}$ & 1 &  1 & $\xi^{ax}$ & $\xi^{ax + nz}$ \\ 
$(\chi_a,\chi_b)$ & $p$ & $p\xi^{by}$ & 0 &  0
\end{tabular}
\caption{$H_3(\mathbb{Z}_p)$ Character Table, where $\xi$ is the p-th Primitive Root of Unity.}
\label{table - Heisenberg Group Character Table}
\end{table}
    
Now we construct the Lagrangian for the $H_3(\mathbb{Z}_p)$ gauge theory. The untwisted $\mathbb{Z}_p\times\mathbb{Z}_p$ DW theory in arbitrary spacetime dimension:
\begin{equation}
    I_{\mathbb{Z}_p^2} = \frac{ip}{2\pi}\int_{M_D}\left(\tilde{a}_{D-2}\wedge da_1 + \tilde{b}_{D-2}\wedge db_1\right).
\end{equation}
Using the procedure in Sec. \ref{subsection - higher gauging condensation defects as 0-form symmetry defect}, we find the following higher gauging condensation defect:
\begin{equation}
    S(\Sigma) = \frac{1}{\abs{H_1(\Sigma,\mathbb{Z}_p)}}\sum_{\gamma,\Gamma }\exp(\frac{2\pi i}{p}\langle \gamma, \Gamma \rangle)W_1(\gamma)M_2(\Gamma),
\end{equation}
where $\langle,\rangle$ is a $\mathbb{Z}_p$-valued intersection form defined on a closed oriented codimension-1 submanifold $\Sigma$. It satisfies the fusion rule:
\begin{equation}
    S(\Sigma)\times S(\Sigma) = \frac{1}{\abs{H_1(\Sigma,\mathbb{Z}_p)}}\sum_{\gamma,\Gamma }\exp(\frac{2\pi i}{p}\langle \gamma, \Gamma \rangle)W_1(\gamma)M_2(2\Gamma),
\end{equation}
which implies its $p$-th power is the identity operator. It leaves $W_1$, $M_2$ invariant and acts on the remaining Wilson and 't Hooft operators as:
\begin{equation}\label{eq - Heisenberg group condensation defect action}
    \begin{split}
         M_1(M_{D-2}) \times S(\Sigma) &= S(\Sigma)\times M_1(M_{D-2})M_2(M_{D-2}),\\
         W_2(l)\times S(\Sigma) &= S(\Sigma)\times W_1(l)W_2(l).
    \end{split}
\end{equation}
Therefore, $S(\Sigma)$ is the symmetry defect generating the $\mathbb{Z}_p^{(0)}$ symmetry\footnote{We mention that the automorphism twist in Eq. (\ref{eq - Heisenberg group condensation defect action}) actually induces an action $(\chi_0, \chi_1)\mapsto (\chi_{p-1}, \chi_1)$ on irreducible representations. However, since $p$ is prime, the two actions are equivalent up to a relabeling of the $H_3(\mathbb{Z}_p)$ irreducible representations.}. 

To gauge the symmetry, we first convert $S(\Sigma)$ to a U$(1)$ phase by carrying out the sum. Expressing the condensed Wilson and 't Hooft operator as $W_1(M_1) = \exp(i\oint_{M_1} a_1)$ and $M_2(M_{D-2}) = \exp(i\oint_{M_{D-2}}\tilde{b}_{D-2})$, we have:
\begin{equation}
    S(\Sigma) = \exp\left(-\frac{ip}{2\pi}\int_{\Sigma} a_1\wedge \tilde{b}_{D-2}\right).
\end{equation}
Coupling the conserved current to the action, we have:
\begin{equation}
    I_{\mathbb{Z}_p^2}[S] = I_{\mathbb{Z}_p^2} - \frac{ip^2}{(2\pi)^2}\int_{M_D}a_1\wedge \tilde{b}_{D-2}\wedge c_1,
\end{equation}
where $c_1$ is the background gauge field for the $\mathbb{Z}_p^{(0)}$ symmetry. Finally, promoting $c_1$ to a dynamical flat gauge field, we have:
\begin{equation}\label{eq - Heisenberg gauge theory action}
    I_{H_3(\mathbb{Z}_p)} =  \frac{i}{2\pi}\int_{M_D}\left(p\,\tilde{a}_{D-2}\wedge da_1 + p\,\tilde{b}_{D-2}\wedge db_1 + p\, \tilde{c}_{D-2}\wedge dc_1 - \frac{p^2}{2\pi} a_1\wedge \tilde{b}_{D-2}\wedge c_1\right).
\end{equation}
Setting $D=3$, we find that the action is exactly
the action for the $\mathbb{Z}_p^3$ Dijkgraaf-Witten theory with a unit type-III twist \cite{He:2016xpi}.
Furthermore, the braided equivalence between this theory and $\text{Rep}(\calD(H_3(\mathbb{Z}_p)))$ was shown in \cite{Lu:2024lzf}. Therefore, we propose Eq. \eqref{eq - Heisenberg gauge theory action} as an effective action for $H_3(\mathbb{Z}_p)$ gauge theories in arbitrary spacetime dimension. 

For concreteness, let's examine the operator spectrum and their fusion for $p=3$ on closed oriented $(D-2)$-dimensional and $1$-dimensional submanifolds. The generalization to other odd prime $p$ is straightforward. The most general gauge transformation that leaves the action invariant off-shell is:
\begin{equation}\label{eq - Heisenberg Z3 gauge transformation}
    \begin{aligned}
    a_1\;&\longmapsto a_1 + d\alpha_0,  \\
    c_1\;&\longmapsto c_1 + d\epsilon_0,  \\
     \tilde{b}_{D-2}\;&\longmapsto \tilde{b}_{D-2} + d\tilde{\beta}_{D-3
     },  \\
\tilde a_{D-2}\;&\longmapsto\; \tilde a_{D-2}+d\tilde\alpha_{D-3}
  -\frac{3}{2\pi}\left(\tilde\beta_{D-3}\wedge c_1
  +(-1)^{D-2}\,\tilde\epsilon_{0}\,\tilde b_{D-2}
  +\tilde\beta_{D-3}\wedge d\epsilon_{0}\right), \\[2pt]
b_1\;&\longmapsto\; b_1+d\beta_0
  -(-1)^{D-2}\frac{3}{2\pi }\left(\alpha_0 c_1-\epsilon_0 a_1
  +\alpha_0\,d\epsilon_{0}\right),\\[2pt]
\tilde c_{D-2}\;&\longmapsto\; \tilde c_{D-2}+d\tilde\epsilon_{D-3}
  -(-1)^{D-1}\frac{3}{2\pi }\left(\alpha_0\,\tilde b_{D-2}
  +(-1)^{D-2}\,\tilde\beta_{D-3}\wedge a_1
  +\alpha_0\,d\tilde\beta_{D-3}\right).
\end{aligned}
\end{equation}
Let $n_a, n_c\in \{0,1,2\}$. Define the invertible Wilson lines:
\begin{equation}
    U_{a,c}^{n_a, n_c}(M_1) = e^{i\oint_{M_1}(n_a a_1 + n_c c_1)}, 
\end{equation}
which corresponds to the nine 1 dimensional irreducible representations. There are two non-invertible Wilson lines:
\begin{equation}
    \hat{U}_b(M_1) = \frac{1}{3} e^{i\oint n_b b}(1+U_a(M_1) + U_a^2(M_1))(1+U_c(M_1) + U_c^2(M_1)),
\end{equation}
where $n_b\in\{1,2\}$, which corresponds to the two 3-dimensional irreducible representations.

The spectrum of codimension-2 t' Hooft operators is slightly involved. We have three invertible surfaces:
\begin{equation}
    U_{\tilde{b}}^{\tilde{n}_b}(M_{D-2}) = e^{i\oint_{M_{D-2}}\tilde{n}_b\tilde{b}},
\end{equation}
where $\tilde{n}_b\in \{0,1,2\}$. The non-invertible surfaces have quantum dimension $3$ and are defined by dressing appropriate electric condensation defects to the monodromy integrals $e^{i\oint_{M_{D-2}}(\tilde{n}_a\tilde{a} + \tilde{n}_c \tilde{c})}$. Therefore, the higher gauging condensation defects to be dressed to the monodromy integrals must be associated to a defect worldvolume gauge group of order 9.

Let us first consider those integrals involving $\tilde{a}$ and $\tilde{c}$ only:
\begin{equation}
    \begin{split}
        \hat{U}_{\tilde{a}}(M_{D-2}) &= \frac{1}{3}e^{i\tilde{n}_a\oint_{M_{D-2}}\tilde{a}_{D-2}}\mathcal{S}_{c}(M_{D-2})(1 + U_{\tilde{b}}(M_{D-2}) +  U_{\tilde{b}}^2(M_{D-2})),\\
        \hat{U}_{\tilde{c}}(M_{D-2}) &= \frac{1}{3}e^{i\tilde{n}_c\oint_{M_{D-2}}\tilde{c}_{D-2}}\mathcal{S}_{a}(M_{D-2})(1 + U_{\tilde{b}}(M_{D-2}) +  U_{\tilde{b}}^2(M_{D-2})),
    \end{split}
\end{equation}
where $\mathcal{S}_a$ and $S_c$ are the following condensation defects:
\begin{equation}
    \begin{split}
        \mathcal{S}_{a}(M_{D-2}) &= \frac{3}{\abs{H_1(M_{D-2},\mathbb{Z}_3)}}\sum_{\gamma\in H_1(M_{D-2},\mathbb{Z}_3)}U_a(\gamma), \\
        \mathcal{S}_{c}(M_{D-2}) &= \frac{3}{\abs{H_1(M_{D-2},\mathbb{Z}_3)}}\sum_{\gamma\in H_1(M_{D-2},\mathbb{Z}_3)}U_c(\gamma).
    \end{split}
\end{equation}

Now we fix the remaining monodromy integrals $e^{i\oint_{M_{D-2}}(\tilde{n}_a\tilde{a} + \tilde{n}_c\tilde{c})}$ 
where $\tilde{n}_a,\tilde{n}_c\in\{0,1,2\}$. They fall into two types, the diagonal ones where $\tilde{n}_a = \tilde{n}_c$ and the off-diagonal ones where $\tilde{n}_a\neq\tilde{n}_c$. By inspection, the diagonal ones take the following form:
\begin{equation}
    \begin{split}
        \hat{U}_{\tilde{a},\tilde{c}}^{(1,1)}(M_{D-2}) &= \frac{1}{3}e^{i\oint_{M_{D-2}}(\tilde{a}_{D-2} + \tilde{c}_{D-2})}\mathcal{S}_{a+c}(M_{D-2})(1+ U_{\tilde{b}} + U_{\tilde{b}}^2)\\
        \hat{U}_{\tilde{a},\tilde{c}}^{(2,2)}(M_{D-2}) &= \frac{1}{3}e^{i\oint_{M_{D-2}}(2\tilde{a}_{D-2} + 2\tilde{c}_{D-2})}\mathcal{S}_{a+c}(M_{D-2})(1+ U_{\tilde{b}} + U_{\tilde{b}}^2)
    \end{split}
\end{equation}
 Similarly the off-diagonal ones are:
\begin{equation}
    \begin{split}
        \hat{U}_{\tilde{a},\tilde{c}}^{(1,2)}(M_{D-2}) &= \frac{1}{3}e^{i\oint_{M_{D-2}}(\tilde{a}_{D-2} + \tilde{c}_{D-2})}\mathcal{S}_{a-c}(M_{D-2})(1+ U_{\tilde{b}}(M_{D-2}) + U_{\tilde{b}}^2(M_{D-2})),\\
        \hat{U}_{\tilde{a},\tilde{c}}^{(2,1)}(M_{D-2}) &= \frac{1}{3}e^{i\oint_{M_{D-2}}(2\tilde{a}_{D-2} + 2\tilde{c}_{D-2})}\mathcal{S}_{a-c}(M_{D-2})(1+ U_{\tilde{b}}(M_{D-2}) + U_{\tilde{b}}^2(M_{D-2})).
    \end{split}
\end{equation}
$\mathcal{S}_{a-c}$ and $\mathcal{S}_{a+c}$ have defect world-volume gauge group $H_1$ and $H_2$, respectively. Expanding them out in terms of the condensed lines, we have:
\begin{equation}
    \begin{split}
        \mathcal{S}_{a+c}(M_{D-2}) &= \frac{3}{\abs{H_1(M_{D-2},\mathbb{Z}_3)}}\sum_{\gamma\in H_1(M_{D-2},\mathbb{Z}_3)}U_{a,c}^{(1,1)}(\gamma), \\
        \mathcal{S}_{a-c}(M_{D-2}) &= \frac{3}{\abs{H_1(M_{D-2},\mathbb{Z}_3)}}\sum_{\gamma\in H_1(M_{D-2},\mathbb{Z}_3)}U_{a,c}^{(1,2)}(\gamma).
    \end{split}
\end{equation}

One can check that there is a one-to-one correspondence between the order-$9$ normal subgroups of $H_3(\mathbb{Z}_3)$ and the higher gauging condensation defects $\mathcal{S}_a, \mathcal{S}_c, \mathcal{S}_{a+c}, \mathcal{S}_{a-c}$. For example, $\mathcal{S}_a$ and $\mathcal{S}_c$ are associated to the normal subgroups $H_0$ and $H_{\infty}$ respectively. They satisfy the fusion
\begin{equation}
    \mathcal{D}_{H_0} \times \mathcal{D}_{H_\infty} = \frac{|H_3(\mathbb{Z}_3)|}{|H_0\cdot H_\infty|} \mathcal{D}_{H_0 \cap H_\infty}.
\end{equation}
where $H_0 \cdot H_{\infty} = H_3(\mathbb{Z}_3)$ and $H_0 \cap H_{\infty} = \mathbb{Z}_3 = Z(H_3(\mathbb{Z}_3))$. Hence, 
\begin{equation}
    \mathcal{S}_a \times  \mathcal{S}_c =  \mathcal{D}_{H_0} \times \mathcal{D}_{H_\infty} = \mathcal{D}_{H_0\cap H_\infty} = \mathcal{D}_{\bbZ_3}.
\end{equation}
Similar to the $\mathbb{D}_4$ case, the condensation defect $\mathcal{D}_{\mathbb{Z}_3}$ has quantum dimension $9$. Therefore, mutual fusions between t' Hooft defects contain TQFT valued coefficients. For example:
\begin{equation}
    \hat{U}_{\tilde{a}}(S_{D-2})\times \hat{U}_{\tilde{c}}(S_{D-2}) = \frac{1}{3} e^{i\oint_{M_{D-2}}(\tilde{a} + \tilde{c})} \mathcal{D}_{\mathbb{Z}_3}(1 + U_{\tilde{b}}(M_{D-2}) +  U_{\tilde{b}}^2(M_{D-2})) \equiv 3 \hat{U}_{\tilde{a},\tilde{c}}^{(1,1)}(M_{D-2}).
\end{equation}

Finally, we show that the $H_3(\mathbb{Z}_p)$ character table can be reproduced from Hopf links between Wilson and t' Hooft operators in arbitrary spacetime dimension. The Wilson lines for general odd-prime $p$ are:
\begin{equation}
    \begin{split}
        U_{a,c}^{(n_a, n_c)}(M_1) &= e^{i\oint_{M_1}(n_a a_1 + n_c c_1)}, \\
        \hat{U}_b^{(n_b)}(M_1) &= \frac{1}{p}e^{in_b\oint_{M_1}b_1}\left(\sum_{l=0}^{p-1}U_a^{(l)}(M_1)\right)\left(\sum_{l=0}^{p-1}U_c^{(l)}(M_1)\right),
    \end{split}
\end{equation}
and t' Hooft surfaces are:
\begin{equation}
    \begin{split}
        U_{\tilde{b}}^{(\tilde{n}_b)}(M_{D-2}) &= e^{i\oint_{M_{D-2}} \tilde{n}_b\tilde{b}_{D-2} }, \\
        U_{\tilde{a},\tilde{c}}^{\tilde{n}_a,\tilde{n}_c}(M_{D-2}) &=    \frac{1}{p}e^{i\oint_{M_{D-2}}(\tilde{n}_a\tilde{a} + \tilde{n}_c\tilde{c})}\mathcal{S}_{\tilde{n}_a\tilde{a}, \tilde{n}_c\tilde{c}}(M_{D-2})\left(\sum_{l=0}^{p-1}U_{\tilde{b}}^{(l)}(M_{D-2})\right),
    \end{split}
\end{equation}
where $\mathcal{S}_{\tilde{n}_a\tilde{a}, \tilde{n}_c\tilde{c}}$ correspond to condensation defects associated to an order-$p^2$ normal subgroup. Note that all $\mathcal{S}_{\tilde{n}_a\tilde{a}, \tilde{n}_c\tilde{c}}$ have quantum dimension $p$. The evaluation is exactly the same as the $\mathbb{D}_4$ example. The relevant Hopf links are:
\begin{align}
    \expval{U_{a,c}^{n_a, n_c}(S^1)U_{\Tilde{b}}(S^{D-2})} &= 1, \\
    \expval{U_{a,c}^{n_a, n_c}(S^1)\hat{U}_{\Tilde{a},\Tilde{c}}^{\Tilde{n}_a,\Tilde{n}_c}(S^{D-2})} &= p\xi^{n_a\Tilde{n}_a + n_c\Tilde{n}_c}, \\
    \expval{U_b^{n_b}(S^1)U_{\Tilde{b}}^{\tilde{n}_b}(S^{D-2})} &= p\xi^{n_b \tilde{n}_b}, \\
    \expval{\hat{U}_b^{n_b}(S^1)\hat{U}_{\Tilde{a},\Tilde{c}}^{\tilde{n}_a,\tilde{n}_c}(S^{D-2})} &= 0,
\end{align}
which corresponds exactly to the entries of Table \ref{table - Heisenberg Group Character Table}.  

\section{Generalized Type-I Action}
\label{section - type-I action analysis}

In this section, we study the general features of type-I actions. They are obtained by gauging $\mathbb{Z}_K^{(0)}$ symmetries by summing over higher gauging condensation defects. We show that their off-shell local gauge transformations and the on-shell deformations agree with each other up to a sign. In this sense, these effective topological actions are well-behaved. In Sec. \ref{subsection - type-I general analysis}, we study the structure of gauge transformations for the type-I actions. In Sec. \ref{subsection - p-form type-I general analysis}, we study a $q$-form generalizations of type-I action.

\subsection{General Analysis of Type-I Action \label{subsection - type-I general analysis}}

    Consider a general type-I action: 
    \begin{equation} \label{eq - 1-form type-I action}
        I = \frac{i}{2\pi}\int_{M_D}\left( N\tilde{a}_{D-2}\wedge da_1 + M\tilde{b}_{D-2}\wedge db_1 + K\tilde{c}_{D-2}\wedge dc_1 - \frac{p}{2\pi} a_1\wedge \tilde{b}_{D-2} \wedge c_1 \right),
    \end{equation}
    with equations of motions:
    \begin{equation} \label{eq - type-I Action EOM}
        \begin{aligned}
N\,d \tilde{a}_{D-2} &= \frac{p}{2\pi}\,\tilde{b}_{D-2}\wedge c_{1},
&\qquad
N\,d a_{1} &= 0,\\[4pt]
M\,d b_{1} &= (-1)^{D-2}\frac{p}{2\pi}\,a_{1}\wedge c_{1},
&\qquad
M\,d \tilde{b}_{D-2} &= 0,\\[4pt]
K\,d \tilde{c}_{D-2} &= (-1)^{D-1}\frac{p}{2\pi}\,a_{1}\wedge \tilde{b}_{D-2},
&\qquad
K\,d c_{1} &= 0.
\end{aligned}
    \end{equation}
    Actions of this type arise naturally when gauging a $\mathbb{Z}_K^{(0)}$ symmetry in an untwisted $\mathbb{Z}_N\times\mathbb{Z}_M$ gauge theory by higher gauging condensation defect summation. As we have seen, the integer $p$ in general contains information of the higher gauging condensation defects that generate the gauged 0-form symmetry. Since this action is constructed out of 1-form gauge fields, the action must be invariant under large gauge transformations. By Appendix \ref{appendix - large gauge transformation}, this imposes the following constraint:
    \begin{equation}
        p\in \text{lcm}(NM, MK, NK) \mathbb{Z}, \qquad p\sim p + NMK.
    \end{equation}

    Let us derive off-shell gauge transformations. Notice that the kinetic terms are invariant under local gauge transformation deformations $a_1 \mapsto a_1 + d\alpha_0$, $\tilde{b}_{D-2} \mapsto \tilde{b}_{D-2} + d\tilde{\beta}_{D-3}$ and $c_1\mapsto c_1 + d\epsilon_0$, but the discrete torsion term is not. In fact, since we are considering discrete gauge theories, we should consider gauge transformation of the actions to all orders. Instead of directly performing an ad hoc derivation of the local gauge transformations of the action off-shell, we can first examine the on-shell physics and determine appropriate deformations of the equations of motion, then use them as an ansatz to motivate the off-shell gauge transformations.

    Now we examine the equations in the left column of Eq. \eqref{eq - type-I Action EOM}. By definition, the field strengths are representatives of some de Rham cohomology classes whose local expressions are given by the equations of motions. Consider deformations $a_1 \mapsto a_1 + d\alpha_0$, $\tilde{b}_{D-2} \mapsto \tilde{b}_{D-2} + d\tilde{\beta}_{D-3}$ and $c_1\mapsto c_1 + d\epsilon_0$, which are trivial deformations that leave the field strengths $d a_1, d\tilde{b}_{D-2}, d c_1$ exactly invariant. By Eq. \eqref{eq - type-I Action EOM}, they induce nontrivial deformations of field strengths $d\tilde{a}_{D-2}$, $d b_1$, $d \tilde{c}_{D-2}$: 
    \begin{equation}\label{eq - okay action field strengths deformation}
        \begin{aligned}
\delta\!\left(d\tilde a_{D-2}\right)
&= \frac{p}{2\pi N}\Bigl(
   d\tilde\beta_{D-3}\wedge c_{1}
   + \tilde b_{D-2}\wedge d\epsilon_{0}
   + d\tilde\beta_{D-3}\wedge d\epsilon_{0}
\Bigr),\\[4pt]
\delta\!\left(d b_{1}\right)
&= (-1)^{D-2}\frac{p}{2\pi M}\Bigl(
   d\alpha_{0}\wedge c_{1}
   + a_{1}\wedge d\epsilon_{0}
   + d\alpha_{0}\wedge d\epsilon_{0}
\Bigr),\\[4pt]
\delta\!\left(d\tilde c_{D-2}\right)
&= (-1)^{D-1}\frac{p}{2\pi K}\Bigl(
   d\alpha_{0}\wedge \tilde b_{D-2}
   + a_{1}\wedge d\tilde\beta_{D-3}
   + d\alpha_{0}\wedge d\tilde\beta_{D-3}
\Bigr),
\end{aligned}
    \end{equation}
    which descend to the gauge fields as:
    \begin{equation}
    \begin{aligned}
\tilde a_{D-2}\;&\longmapsto\; \tilde a_{D-2}+d\tilde\alpha_{D-3}
  +\frac{p}{2\pi N}\left(\tilde\beta_{D-3}\wedge c_1
  +(-1)^{D-2}\,\tilde\epsilon_{0}\,\tilde b_{D-2}
  +\tilde\beta_{D-3}\wedge d\epsilon_{0}\right),\\[2pt]
b_1\;&\longmapsto\; b_1+d\beta_0
  +(-1)^{D-2}\frac{p}{2\pi M}\left(\alpha_0 c_1-\epsilon_0 a_1
  +\alpha_0\,d\epsilon_{0}\right),\\[2pt]
\tilde c_{D-2}\;&\longmapsto\; \tilde c_{D-2}+d\tilde\epsilon_{D-3}
  +(-1)^{D-1}\frac{p}{2\pi K}\left(\alpha_0\,\tilde b_{D-2}
  +(-1)^{D-2}\,\tilde\beta_{D-3}\wedge a_1
  +\alpha_0\,d\tilde\beta_{D-3}\right).
\end{aligned}
\end{equation}
Namely, the induced deformations of the field strengths Eq. \eqref{eq - okay action field strengths deformation} shifts the field strengths by $d$-exact terms. Therefore, these deformations do not change the cohomology classes of the field strengths and only change their representatives.

To deduce the off-shell gauge transformations that leave the action invariant, we use the above transformations as an ansatz and postulate:
\begin{equation}\label{eq - type-I action on shell transformation}
    \begin{aligned}
    a_1\;&\longmapsto a_1 + d\alpha_0, \qquad  c_1\;\longmapsto c_1 + d\epsilon_0, \qquad
     \tilde{b}_{D-2}\;\longmapsto \tilde{b}_{D-2} + d\tilde{\beta}_{D-3
     },  \\
\tilde a_{D-2}\;&\longmapsto\; \tilde a_{D-2}+d\tilde\alpha_{D-3}
  +\frac{p}{2\pi N}\left(\xi\tilde\beta_{D-3}\wedge c_1
  +(-1)^{D-2}\,\xi\tilde\epsilon_{0}\,\tilde b_{D-2}
  +\lambda\tilde\beta_{D-3}\wedge d\epsilon_{0}\right),\\[2pt]
b_1\;&\longmapsto\; b_1+d\beta_0
  +(-1)^{D-2}\frac{p}{2\pi M}\left(\xi\alpha_0 c_1-\xi\epsilon_0 a_1
  +\lambda\alpha_0\,d\epsilon_{0}\right),\\[2pt]
\tilde c_{D-2}\;&\longmapsto\; \tilde c_{D-2}+d\tilde\epsilon_{D-3}
  +(-1)^{D-1}\frac{p}{2\pi K}\left(\xi\alpha_0\,\tilde b_{D-2}
  +(-1)^{D-2}\,\xi\tilde\beta_{D-3}\wedge a_1
  +\lambda\alpha_0\,d\tilde\beta_{D-3}\right),
\end{aligned}
\end{equation}
for some constants $\xi$ and $\lambda$. The parameters are fixed by demanding  gauge invariance for the action on a closed manifold to all orders. We find that setting $\xi=\lambda=-1$ and we will outline this calculation in Appendix \ref{appendix - small gauge transformations}. 

The specific details of the operator spectrum depend on number-theoretic properties of $N,M,K$ and $p$. There are at least $NK$ invertible Wilson lines and $M$ 't Hooft surfaces:
\begin{equation}
    U_{a,c}^{n_a,n_c}(M_1) = e^{i\oint_{M_1}(n_a a_1 + n_c c_1)},\qquad U_{\tilde{b}}(M_{D-2})=e^{i\tilde{n}_b\oint_{M_{D-2}}\tilde{b}_{D-2}},
\end{equation}
where $n_a \in \{0,1,\dots, N\}$, $\tilde{n}_b \in  \{0,1,\dots, M\}$, $n_c \in  \{0,1,\dots, K\}$. In general, the monodromy integrals $e^{i\oint M_{D-2} \tilde{a}_{D-2}}$, $e^{i\oint_{M_{D-2}} \tilde{c}_{D-2}}$ and $e^{i\oint_{M_1} b_1}$ define non-invertible surface/line operators and need to be dressed with appropriate higher gauging condensation defects to ensure gauge invariance. However, there are exceptions.

For concreteness, set $D=3$ and consider an untwisted $\mathbb{Z}_2\times\mathbb{Z}_4$ Dijkgraaf-Witten theory:
\begin{equation}
    I = \frac{i}{2\pi}\int_{M_3}\left( 2\tilde{a}_1\wedge da_1 + 4\tilde{b}_1\wedge db_1 \right).
\end{equation}
Let $W_1, M_1$ be the Wilson lines and 't Hooft lines of $\mathbb{Z}_2$ and $W_2,M_2$ be the Wilson line of $\mathbb{Z}_4$. The group $\mathbb{Z}_2\times\mathbb{Z}_4$ has a order-2 automorphism symmetry acting on the generators as $(1,0)\mapsto (1,2)$. This action is realized by the higher gauging condensation defect:
\begin{equation}
    S(\Sigma) = \frac{1}{\abs{H_1(\Sigma,\mathbb{Z}_2)}}\sum_{\gamma,\Gamma\in H_1(\Sigma,\mathbb{Z}_2) }e^{\frac{2\pi i}{2}\langle \gamma, \Gamma \rangle} W_1(\gamma) M_2^2(\Gamma),
\end{equation}
which leaves $M_2, W_1$ invariant and act on $M_1, W_2$ as:
\begin{equation}
    M_1(\gamma) \times S(\Sigma) = S(\Sigma) \times M_1(\gamma)M_2^2(\gamma), \qquad W_2(\gamma)\times S(\Sigma) = S(\Sigma)\times W_1(\gamma)W_2(\gamma).
\end{equation}
Gauging this $\mathbb{Z}_2^{(0)}$ global symmetry leads to the following action:
\begin{equation}
    I = \frac{i}{\pi}\int_{M_3}\left( \tilde{a}_1\wedge da_1 + 2\tilde{b}_1\wedge db_1 + \tilde{c}_1\wedge dc_1 - \frac{1}{\pi}a_1\wedge\tilde{b}_1\wedge c_1\right).
\end{equation}
The off-shell gauge transformation for $b_1$ is:
\begin{equation}
    \delta b_1 \sim d\beta_0 + \frac{1}{2\pi}\left( \alpha_0 c_1 - \epsilon_0 a_1 + \alpha_0 d\epsilon_0 \right)
\end{equation}
 Consider the operator defined by the monodromy factor $e^{2i\oint_{M_1} b_1}$. To maintain gauge invariance, we need to stack higher gauging condensation defects proportional to these projectors:
\begin{equation}
    \Delta_a (M_1) = \frac{1}{2}\left(1 + U_a(2M_1)\right), \qquad \Delta_c (M_1) = \frac{1}{2}\left(1 + U_c(2M_1)\right) 
\end{equation}
 on $e^{2i\oint_{M_1} b_1}$. However, since $U_a$, $U_c$ themselves square to identity, the projectors are actually trivial. Therefore, $U_b(M_1) \equiv e^{i\oint_{M_1} 2b_1}$ itself is gauge invariant and it is of order two, namely $U_b(M_1)\times 
U_b(M_1) = 1$.

    \subsection{\texorpdfstring{$q$}{q}-Form Type-I Action}
    \label{subsection - p-form type-I general analysis}

    Let us generalize the above analysis to $q$-form BF theories. For example, consider the $q$-form BF theory:
    \begin{equation}
    I = \frac{i}{\pi} \int_{M_D} \left( \Tilde{a}_{D-q-1} \wedge d a_q +  \Tilde{b}_{D-q-1} \wedge d b_q  \right),
\end{equation}
which contains Wilson and 't Hooft surfaces:
\begin{equation}
    U^{(n_a,n_b)}_{a,b}(\gamma_q) = e^{i\oint_{\gamma_q}(n_a a_{q} + n_b b_q)},\qquad U^{(\tilde{n}_a,\tilde{n}_b)}_{\tilde{a},\tilde{b}}(\Gamma_{D-q-1}) = e^{i\oint_{\Gamma_{D-q-1}}(\tilde{n}_a \tilde{a}_{q} + \tilde{n}_b \tilde{b}_q)}, 
\end{equation}
where $n_a, n_b, \tilde{n}_a,\tilde{n}_b \in \{0,1\} $. Consider the higher gauging condensation defect:
\begin{equation}
    S(\Sigma_{D-1}) = \frac{1}{\abs{H_{q}(\Sigma_{D-1}, \mathbb{Z}_2)}}\sum_{\substack{\gamma_q\in H_q(\Sigma, \mathbb{Z}_2)\\
            \Gamma_{D-q-1} \in H_{D-q-1}(\Sigma, \mathbb{Z}_2)}}e^{i\pi\langle \gamma_q,\Gamma_{D-q-1}\rangle}W_1(\gamma_q)M_2(\Gamma_{D-q-1}),
\end{equation}
which follows $S(\Sigma_{D-1})\times S(\Sigma_{D-1}) = 1$ and it transforms the Wilson and 't Hooft surfaces as Eq. \eqref{eq - D4 Symmetry Action by Condensation Defect}. Since the operator is supported on a codimension-1 submanifold, it generates a $\mathbb{Z}_2^{(0)}$ symmetry. Gauging $\mathbb{Z}_2^{(0)}$ by summing over higher gauging condensation defects leads to the action:
\begin{equation}
    I = \frac{i}{\pi} \int_{M_D} \left( \Tilde{a}_{D-q-1} \wedge d a_q +  \Tilde{b}_{D-q-1} \wedge d b_q  + \tilde{c}_{D-2}\wedge dc_1 - \frac{1}{\pi}a_q\wedge\tilde{b}_{D-q-1}\wedge c_1 \right), 
\end{equation}
which describes a topological sigma model from $M_D$ to $B^{q}\mathbb{Z}_2\times B^q\mathbb{Z}_2\times B\mathbb{Z}_2$ with a nontrivial topological action valued in $H^{D}(B^{q}\mathbb{Z}_2\times B^q\mathbb{Z}_2\times B\mathbb{Z}_2, \mathrm{U}(1))$. 

 Let us examine the operator spectrum. We have three nontrivial order-2 invertible operators:
\begin{equation}
    U_c(M_1)= e^{i\oint_{M_1}c_1},\qquad U_{\tilde{b}}(M_{D-q-1})=e^{i\oint_{M_{D-q-1}}\tilde{b}_{D-q-1}}, \qquad U_a(M_q)= e^{i\oint_{M_q}a_q}. 
\end{equation}
There are three non-invertible operators:
\begin{equation}
    \begin{split}
        \hat{U}_{\tilde{a}}(M_{D-q-1}) & \sim e^{i\oint_{M_{D-q-1}}\tilde{a}_{D-q-1}}\left( \frac{1+U_{\tilde{b}}(M_{D-q-1})}{2} \right)\Delta_{c}(M_{D-q-1}), \\
        \hat{U}_{\tilde{c}}(M_{D-2}) & \sim e^{i\oint_{M_{D-2}}\tilde{c}_{D-2}}\Delta_{\tilde{b}}(M_{D-2})\Delta_{a}(M_{D-2}), \\
        \hat{U}_b(M_q) & \sim e^{i\oint_{M_q}b_q}\left( \frac{1 + U_a(M_q)}{2} \right)\Delta_{c}(M_q),
    \end{split}
\end{equation}
where $\Delta_{a}(M_{D-2})$, $\Delta_{\tilde{b}}(M_{D-2})$ ,$\Delta_{c}(M_{D-p-1})$ are the following projectors:
\begin{equation}
    \begin{split}
        \Delta_{a}(M_{D-2}) & \equiv \frac{1}{\abs{H_q(M_{D-2},\mathbb{Z}_2)}}\sum_{\Gamma_{q}\in H_{q}(M_{D-2},\mathbb{Z}_2)}U_a(\Gamma_q), \\
        \Delta_{b}(M_{q}) & \equiv \frac{1}{\abs{H_{D-q-1}(M_{D-2}, \mathbb{Z}_2)}}\sum_{\Gamma_{D-q-1}\in H_{D-q-1}(M_q,\mathbb{Z}_2)}U_{\tilde{b}}(M_{D-q-1}),\\
        \Delta_{c}(M_{q}) & \equiv \frac{1}{\abs{H_1(M_{q},\mathbb{Z}_2)}}\sum_{\Gamma_{1}\in H_1(M_{D-2},\mathbb{Z}_2)}U_c(\Gamma_1),
    \end{split}
\end{equation}
where the normalization factors can be fixed by homotopy theory calculations of the topological sigma model.

Now let us consider the general case. The $q$-form Type-I action reads:
\begin{equation}\label{eq - q-form type-I action}
    I = \frac{i}{2\pi} \int_{M_D} \left( N \Tilde{a}_{D-q-1} \wedge d a_q + M \Tilde{b}_{D-q-1} \wedge d b_q + K \Tilde{c}_{D-2} \wedge  d c_1 - \frac{p}{2\pi} a_q \wedge \Tilde{b}_{D-q-1}\wedge c_1 \right).
\end{equation}
The equations of motion read:
\begin{equation}
    \begin{aligned}
        d \Tilde{a}_{D-q-1} &=  (-1)^{(D-q)(1+q)} \frac{p}{2\pi N} \Tilde{b}_{D-q-1}\wedge c_1, & \quad  d a_q &= 0,\\
        d b_q  &= (-1)^{(D-q-1)q}\frac{p}{2\pi M} a_q \wedge c_1,  & \quad d \Tilde{b}_{D-q-1} &= 0,\\
        d \Tilde{c}_{D-2} &  =  (-1)^{D-1} \frac{p}{2\pi K} a_q\wedge\Tilde{b}_{D-q-1},& \quad  d c_1 &=0.\\
    \end{aligned}
\end{equation}
Following the logic for the 1-form type-I action, it is straightforward to derive the deformations of the equations of motion and use them as an ansatz for off-shell gauge transformation. We find:
\begin{align}
    a_q\longmapsto & a_q +  d\alpha_{q-1}, \\
    \Tilde{b}_{D-q-1} \longmapsto & \Tilde{b}_{D-q-1} +  d\Tilde{\beta}_{D-q-2}, \\
    c_1 \longmapsto & c_1 +  d \epsilon_0, \\
    \Tilde{a}_{D-q-1} \longmapsto & \Tilde{a}_{D-q-1} +  d \Tilde{\alpha}_{D-q-2} + (-1)^{(D-q)(1+q)} \frac{p}{2\pi N} \left((-1)^{D-q-1} \Tilde{b}_{D-q-1}\wedge\epsilon_0 + \right. \nonumber \\
    & \left. \Tilde{\beta}_{D-q-2}\wedge c_1 + \Tilde{\beta}_{D-q-2}\wedge d \epsilon_0)\right), \\
    b_q \longmapsto & b_q +  d\Tilde{\beta}_{q-1} + (-1)^{(D-q-1)q} \frac{p}{2\pi M}\left( (-1)^q a_q\wedge \epsilon_0  + \alpha_{q-1}\wedge c_1 + \alpha_{q-1}\wedge  d\epsilon_0 \right), \\
    \Tilde{c}_{D-2} \longmapsto & \Tilde{c}_{D-2} +  d \tilde{\epsilon}_{D-3} + (-1)^{D-1}\frac{p}{2\pi K} \left((-1)^q a_q\wedge \Tilde{\beta}_{D-q-2} + \alpha_{q-1}\wedge\Tilde{b}_{D-q-1}\right.\nonumber \\
    & \left.+ \alpha_{q-1}\wedge  d\Tilde{\beta}_{D-q-2}\right).
\end{align}

\section{Generalized Type-I Action as a SymTFT \label{section - type-I_symTFT}}

So far, we have been examining type-I actions and their generalizations over closed oriented spacetime manifolds. The next step is to examine the general behavior of their edge modes/boundary conditions. Since the on-shell deformations and off-shell gauge transformations of type-I actions agree with each other up to a sign, type-I actions are good continuum models of topological sigma models. Therefore, we can treat them as the symTFT for a wide class of finite categorical global symmetries \cite{Freed:2022qnc}. Especially, since the gauge transformation is reminiscent of the transformation of higher group global symmetries as in \cite{Cordova:2018cvg}, one might naturally suspect that topological boundary conditions of generalized type-I action can realize higher-group global symmetry. In fact this is not true and a proper justification requires further concepts in higher group structures. In Sec. \ref{subsection - Higher Group Gauge Theory}, we review higher group gauge theories as topological sigma models. In Sec. \ref{subsection - symTFT overview}, we review some elementary structures of symTFT and point out some subtleties that tend to be ignored in naive Lagrangian analysis. In Sec. \ref{subsection - type-I action as higher group symTFT}, we establish a few no-go theorems for higher group global symmetries based on topological sigma model and Postnikov tower considerations. We demonstrate that type-I actions when treated as a symTFTs cannot realize any higher group global symmetries.

\subsection{Higher Group Gauge Theories as Topological Sigma Models} \label{subsection - Higher Group Gauge Theory}

As previously mentioned, higher group gauge theories can be defined as topological sigma model. To motivate the bundle structure, consider a direct sum of a collection of $p_i$-form abelian gauge theory, where $p_i$ is an ordered set of positive integers with $p_i>p_{i-1}$. We can define a topological sigma model from $M_D$ to the target space:
\begin{equation}
    B\tilde{G} = B^{p_1}G_1 \times B^{p_2}G_2\times\dots\times B^{p_{k-1}}G_{k-1}
\end{equation}
with topological action $\omega\in H^D(B\tilde{G},U(1))$ and a partition function:
\begin{equation}
    \calZ_{\omega}^{B\tilde{G}}(M_D) \sim \mathlarger{\sum}_{ \tilde{A} = (A_{p_1}, \dots A_{p_i},\dots, A_{p_k}) }e^{\langle \gamma^*\omega, [M_D] \rangle  },
\end{equation}
where $A_{p_i}\in H^{p_i}(M_{D},G_i)$. The classification of topological actions of this theory can be done by computing $H^{D}(B\tilde{G},U(1))\simeq H^{D+1}(B\tilde{G},\mathbb{Z})$ by repeated application of the K\"{u}nneth formula. 

This theory can be generalized by introducing ``twists" in the space $B\tilde{G}$. Loosely speaking, given a pair of smooth manifolds $X$ and $Y$, one can construct a twisted space by consider the following short exact sequence of topological spaces:
\begin{equation}
     Y \rightarrow E \rightarrow X 
\end{equation}
so that $E$ locally looks like $X \times Y$. Such twists are often, but not always, classified by a cohomology invariant. This leads to the notion of a (smooth) fiber bundle. We can generalize this idea as follows. Consider a contractible local path $U$ of $X$, we can try to find a smooth manifold $E$ with a projection $\pi: E\rightarrow X$ so that the following diagram commutes up to homotopy:
\begin{equation}
    \begin{tikzcd}
f^*E = U\times Y \arrow[r] \arrow[d,"\pi'"'] & E \arrow[d,"\pi"] \\
U \arrow[r,"f"'] & X
\end{tikzcd}.
\end{equation}
In this case, $Y$ is called a \textit{\textbf{homotopy fiber}}. A further generalization of this construction to a generic pair of topological spaces $X$ and $Y$ (often taken to be CW complexes in physics applications) is called a \textbf{\textit{homotopy fibration}}.

Applying this construction to $B\tilde{G}$ inductively defines the ``bundle" structure of a higher group gauge theory. The first stage is the classifying space $B^{p_1}G_1$. The second stage introduces a twist so that $B^{p_2}G_2$ non-trivially fibers over $B^{p_1}G_1$ with the possible twists classified by a Postnikov class $[\Omega_2]\in H^{p_2 + 1}(B^{p_1}G_1,G_2)$:
\begin{equation}
    B^{p_2}G_2 \rightarrow E_2 \rightarrow B^{p_1}G_1
\end{equation}
We can continue with a third stage where we fiber $B^{p_3}G_3$ over $E_2$:
\begin{equation}
     B^{p_3}G_3 \rightarrow E_3 \rightarrow E_2 
\end{equation}
with the possible twists classified by another Postnikov class $[\Omega_3]\in H^{p_3+1}(B^{p_3}E_2,G_3)$. This iteration terminates at the $k$-th step:
\begin{equation}
    B^{p_k}G_k \rightarrow E_k \rightarrow E_{k-1}
\end{equation}
which is classified by $[\Omega_k]\in H^{p_k+1}(E_{k-1},G_k)$. The space $E_k $ is the total space of the higher group bundle $B\mathbb{G}\equiv E_k$ and this inductive fibration construction is known as the \textit{\textbf{Postnikov tower}} \cite{HatcherAT}. 

To define a topological sigma model, we again need to classify homotopy classes of maps from $M_D$ to $\mathbb{E}_k$. This is classified by a $k$-tuple:
\begin{equation}
    \mathbb{A}_k = \{ (A_1,A_2, \dots, A_k) \in C^{p_1}(M_D, G_1)\times C^{p_2}(M_D, G_2)\times \dots\times  C^{p_k}(M_D, G_k)  \},
\end{equation}
satisfying the generalized cocycle condition:
\begin{equation} \label{eq - generalized cocycle condition}
    \begin{split}
        dA_1 &= 0, \\
        dA_2 & = \Omega_2(A_1), \\
        dA_3 & = \Omega_3(A_2, A_1), \\
        &\vdots\\
        dA_k & = \Omega_k(A_1, A_2, \dots, A_{k-1}).
    \end{split}
\end{equation}
where the RHS are precisely the Postnikov classes classifying the sequence of fibration.  The null homotopies correspond to gauge redundancies:
\begin{equation}\label{eq - higher group gauge transformation}
    \begin{split}
        A_1 &= A_1 + d\phi_1, \\
        A_2 & = A_2 + d\phi_2 + \zeta_2(A_1,\phi_1), \\
        A_3 & = A_3 + d\phi_3 + \zeta_3(A_1,\phi_1; A_2,\phi_2), \\
        &\vdots\\
        A_k & = A_k + d\phi_k + \zeta_k(A_1,\phi_1; A_2,\phi_2; \dots; A_{k-1},\phi_{k-1}). 
    \end{split}
\end{equation}
where the functions $\zeta_j$ are descendants of the $j$-th Postnikov class $\Omega_j$:
\begin{equation}
    d\zeta_{j}(A_1,\phi_1; A_2,\phi_2; \dots; A_{j-1},\phi_{j-1}) = \beta_j(A_1 + d\phi_1; \dots; A_{j-1} + d\phi_j) -  \beta_j(A_1 ; \dots; A_{j-1} ).
\end{equation}

    This gauge redundancy can be conveniently packaged into a compact notation. We can define a generalized coboundary operator $D_{E_k}$ so that the generalized cocycle condition Eq. \eqref{eq - generalized cocycle condition} gets translated to:
    \begin{equation}
        D_{E_k}\mathbb{A}_k = 0, \qquad \text{i.e.}\quad \mathbb{A}_k\in \ker(D_{E_k}).
    \end{equation}
    Similarly, the gauge redundancy gets translated to:
    \begin{equation}
        \mathbb{A}_k \mapsto \mathbb{A}_k + D_{E_k}^{\flat}\Phi_k,
    \end{equation}
    where $\Phi_k$ is a $k$-tuple:
    \begin{equation}
        \Phi_k = \{(\phi_1, \phi_2, \dots, \phi_k)\in C^{p_1-1}(M_D, G_1)\times C^{p_2-1}(M_D, G_2)\times \dots \times C^{p_k-1}(M_D, G_k) \}.
    \end{equation}
    The generalized coboundary operators satisfy $D_{E_k}\circ D_{E_k}^\flat = 0$, so they define a generalized cohomology of $M_D$:
    \begin{equation}
        H_{E_k}^{\Vec{p}}(M_D) \equiv \frac{\ker (D_{E_k})}{\text{im}(D_{E_k}^\flat)},
    \end{equation}
    where $\Vec{p} = (p_1, p_2, \dots, p_k)$.
    The object $\mathbb{A}_k$ is \textbf{\textit{a flat $p_k$-connection}} on the higher group bundle and the higher group gauge theory is a topological sigma model with action $[\omega]\in H^{d+1}(E_k, \text{U}(1))$ with partition function \cite{Delcamp:2019fdp}:
    \begin{equation}
        \mathcal{Z}_{\omega}^{E_k}[M_D] = \frac{1}{\prod_{j=1}^k\abs{G}_j^{b_0 \rightarrow q_j -1}}\sum_{[\mathbb{A}_k \in H_{E_k}^{\vec{p}}(M_D)]}e^{ 2\pi i \langle \omega(\mathbb{A}_k),[M_D]\rangle }. 
    \end{equation}

    There is one possible generalization. Note that the Postnikov classes take values in $H^{p_j+1}(B^{p_{j-1}}G_{j-1}, G_{j})$, where we used the $G_{j}$-valued simplicial cohomology of the space $B^{p_{j-1}}G_{j-1}$. There is an equivalent algebraic definition of group cohomologies by treating $G_{j-1}$ as a $G_{j}$ module \cite{Brown1982CohomologyOG}. In this picture, one needs to specify a $G_{j}$-action on $G_{j-1}$. When the action is nontrivial, we have twisted group cohomologies\footnote{Twisted group cohomology also admits a topological definition, see \cite{HatcherAT}}. Therefore, the inductive fibration can be generalized by turning on this \textbf{\textit{algebraic twist}} $\alpha_j \in \text{Hom}(G_j , \text{Aut}(G_{j-1}))$ at each stage. For example, in a 2-stage fibration where both $G_1$ and $G_2$ are finite abelian groups:
    \begin{equation}
         B^2G_2 \rightarrow B\mathbb{G}\rightarrow BG_1 
    \end{equation}
    the 2-group data is specified by a quadruple $(G_1, G_2, \beta_3, \alpha)$, where $\alpha\in \text{Hom}(G_1,\text{Aut}(G_2))$. See \cite{Kapustin:2013uxa} for an in-depth discussion of this fibration. Later we will see that the type-II actions are in fact higher group gauge theories with a nontrivial algebraic twist.

\subsection{SymTFT and 't Hooft Anomalies}\label{subsection - symTFT overview}

\begin{figure}[h]
        \centering\includegraphics[width=0.6\textwidth]{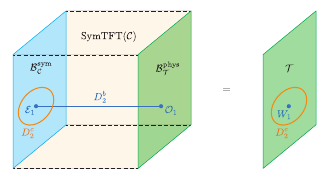}
        \caption{ An illustration of a symTFT sandwich.
        The operator $
        D_2^c$ restricted to the topological boundary $\mathcal{B}^{\text{Sym} }_{\mathcal{C} }$ produces a Hilbert space twisted by the symmetry defect $D_2^c$ upon interval compactification. }
        \label{fig - symTFT sandwich}
    \end{figure}

As motivated in the introduction, a symTFT for a QFT on $M_D$ is a TQFT defined on the cylinder $M_D\times I$, where the interval $I$ is parameterized by $t\in \{0,1\}$. At $t=0$, we have a topological boundary where a subsector of the bulk topological operator are realized as symmetry defects. At $t=1$ we have a dynamical boundary where the dynamical degrees of freedom of the QFT live. Compactifying the interval produces a partition function $Z[A]$, which is the partition function of the QFT coupled to a class of background gauge fields $A$ of its global symmetry. See Fig \ref{fig - symTFT sandwich} for an demonstration.

For example, a pure $\mathrm{SU}(N)$ Yang-Mills theory in $D$ spacetime dimension has a center 1-form symmetry. It's symTFT is the BF theory:
\begin{equation}
    I = \frac{iN}{2\pi}\int_{M_D\times I} b_{D-2}\wedge da_2.
\end{equation}
The boundary conditions can be analyzed by simple field theoretic reasoning. Here we adopt the method in \cite{Brennan:2024fgj}. All $q$-form gauge fields on the cylinder $M_D \times I$ decompose as $\omega_q = \underline{\omega} + dt \wedge \omega_t$, where $\underline{\omega}$ is a $q$-form and $\omega_t$ is a $(q-1)$-form that only has $M_D$ dependence. Performing a bulk $\delta a_2$ variation produces the boundary term:
\begin{equation}
    \delta I_\text{bdy} \sim \frac{iN}{2\pi}\int_{X_d} b_{D-2}\wedge \delta a_2.
\end{equation}
The vanishing of this boundary term implies that either $\delta
 a_2 = 0$ or $b_{D-2}|_\text{bdy} = 0$. The former is the Dirichlet boundary condition for $a_2$ while the latter is the Neumann boundary condition for $a_2$\footnote{There is ambiguity in labeling these boundary conditions. The Dirichlet boundary condition for $a_2$ is equivalent to the Neumann boundary condition for $b_{D-2}$. Similarly, the Neumann boundary condition for $a_2$ is equivalent to the Dirichlet boundary condition for $b_{D-2}$. In practice, one needs to coherently work in one of the two conventions, which is equivalent to choosing a \textbf{\textit{polarization}} for the BF theory. }.

In the sandwich construction, choosing a Dirichlet boundary condition for $a_2$ realizes the center 1-form symmetry, which can be understood as picking a basis that diagonalizes the $U_a(M_2)$ operators:
\begin{equation}
    U_a^{(n)}(M_2) \ket{D_a} = e^{in\oint_{M_2} a_2} \ket{D_a}.
\end{equation}
One can also pick the Neumann boundary condition for $a_2$, which can be understood as picking a basis that diagonalizes the $U_{\tilde{b}}^{(n)}(M_{D-2})$ operators:
\begin{equation}
    U_{\tilde{b}}^n\ket{N_b} = e^{in\oint_{M_{D-2}}\tilde{b}_{D-2}}\ket{N_b}.
\end{equation}
A change of basis from $\ket{A}$ to $\ket{B}$ can be understood as gauging the $\mathbb{Z}_N^{(1)}$ symmetry.

The two types of boundary conditions are formally related by a Fourier transform\footnote{Here we temporarily switch to the cocycle notation, which is customary in the literature.}:
\begin{equation}
    \begin{split}
        \ket{N_b} &= \frac{1}{\sqrt{\abs{H^2(M_D,\mathbb{Z}_N)}}}\sum_{a\in H^2(M_D; \frac{2\pi}{N} \mathbb{Z}_N)}e^{\frac{iN}{2\pi}\int a\cup b} \ket{D_a}, \\
        \ket{D_a} &= \frac{1}{\sqrt{\abs{H^{D-2}(M_D,\mathbb{Z}_N)}}}\sum_{b\in H^{D-2}(M_D; \frac{2\pi}{N} \mathbb{Z}_N)}e^{-\frac{iN}{2\pi}\int a\cup b} \ket{N_b}.
    \end{split}
\end{equation}
In the first line, the change of basis corresponds to the gauging of $\mathbb{Z}_N^{(1)}$-symmetry. It is done by summing over all flat $a_2$ modes modulo background gauge transformations, and the \Poincare dual statement is the condensation of all symmetry operators $U_{\tilde{b}}(M_{D-2})$. Similarly, in the second line, the change of basis corresponds to the gauging of $\mathbb{Z}_N^{(D-2)}$-symmetry. It is done by summing over all flat $b_{D-2}$ modes modulo background gauge transformations, and the \Poincare dual statement is the condensation of all symmetry operators $U_{a}(M_{2})$.

Finally, we point out a subtlety that tends to be ignored in Lagrangian analysis of symTFT. Consider $(1+1)D$ theories with a non-anomalous $G^{(0)}$ symmetry, whose topological defects are described by a fusion category $\text{Vec}_G$. The symTFT for $G^{(0)}$ is a $(2+1)$D untwisted Dijkgraaf-Witten theory with gauge group $G$, which is described by the Drinfeld center $Z(\text{Vec}_G)$ of $\text{Vec}_G$. This category is equivalent to the representation category of the Hopf algebra $\mathcal{D}(G)$ \cite{Etingof}. The symmetry is realized by a choosing a canonical Dirichlet boundary condition on the topological boundary \cite{Freed:2022qnc}. In the naive Lagrangian analysis,  we say that the symmetry defects are parallelly projected down to the topological boundary. However, notice that the defects follow non-commutative fusions on the topological boundary when $G$ is non-abelian, but all line operators in the bulk have commutative fusion. This suggests that the fusion rules of the symmetry defects on the topological boundary are not obtained by a naive restriction of the bulk fusion rules. In fact, when the symTFT can be treated as a topological sigma model with target space $BG$, the fusion rule of the boundary defects for a Dirichlet boundary condition realizing a $G^{(0)}$ symmetry can be directly computed by evaluating the so-called pair-of-chaps configuration, see \cite{Freed:2022qnc} for further detail.

\subsection{Generalized Type-I Actions as a SymTFT} \label{subsection - type-I action as higher group symTFT}

Let us consider a Type-I action on cylinder $M_D\times I$:
\begin{equation}
        I = \frac{i}{2\pi}\int_{M_D\times I}\left( N\tilde{a}_{D-1}\wedge da_1 + M\tilde{b}_{D-1}\wedge db_1 + K\tilde{c}_{D-1}\wedge dc_1 - \frac{p}{2\pi} a_1\wedge \tilde{b}_{D-1} \wedge c_1 \right).
    \end{equation}
We will only discuss 1-form type-I actions and the generalization to $q$-form type-I action is straightforward. Depending on the dynamical boundary condition, there are two possibilities for interpreting the Eq. \eqref{eq - type-I action as a symTFT} as a symTFT:
\begin{enumerate}
    \item By picking an unconventional polarization for the conjugate variable pair $(b_1, \tilde{b}_{D-2})$, it can be understood as a symTFT for  a $\mathbb{Z}_N^{(0)}\times\mathbb{Z}_M^{(D-1)}\times \mathbb{Z}_K^{(0)}$ symmetry with a triple mixed anomaly. 
    \item If the action can be identified as the effective action of an untwisted Dijkgraaf-Witten theory with a non-abelian gauge group, then it can also be understood as the symTFT for a non-abelian 0-form global symmetry. 
\end{enumerate}
The second case has been discussed in \cite{Bergman:2024its}, which gives a holographic derivation of symTFTs for 3D ABJM theories with orthosympletctic gauge groups. In this case, we stress again that the action constructed with abelian gauge fields is only an EFT for the original untwisted Dijkgraaf-Witten theory with non-abelian gauge group. Therefore, from first principles it is not guaranteed that the Lagrangian analysis can reproduce the full set of boundary conditions of the untwisted Dijkgraaf-Witten theory.

Let us go back to the first case. The symTFT for $\mathbb{Z}_N^{(0)}\times\mathbb{Z}_M^{(D-2)}\times \mathbb{Z}_K^{(0)}$ with an $a_1\wedge \tilde{b}_{D-1} \wedge c_1 $ mixed anomaly is\footnote{Here we used the conventional polarization, so the bulk action should be compensated by an integration by parts.}:
\begin{equation} \label{eq - type-I action as a symTFT}
        I = \frac{i}{2\pi}\int_{M_D\times I}\left( N\tilde{a}_{D-1}\wedge da_1 + Mb_1\wedge d\tilde{b}_{D-1}  + K\tilde{c}_{D-1}\wedge dc_1 - \frac{p}{2\pi} a_1\wedge \tilde{b}_{D-1} \wedge c_1 \right),
    \end{equation}
with Dirichlet boundary condition for $(a_1, \tilde{b}_{D-1},c_1)$. Let us place the topological boundary of the symTFT at $t=0$. By truncating the dynamical gauge transformation at $t=0$, we can extract qualitative features of the global symmetries upon interval compactification. The truncation is performed by identifying the gauge fields with Dirichlet boundary conditions at $t=0$ as background gauge fields and freezing the remaining dynamical gauge field transformation parameters. 

For simplicity, consider the following gauging options:
\begin{enumerate}
    \item Gauging $\mathbb{Z}_N^{(0)}$, which is equivalent to assigning $a_1$ a Neumann boundary condition. The truncated transformation reads:
    \begin{equation}
        \begin{aligned}
\tilde a_{D-1}&\longmapsto \tilde a_{D-1}+d\tilde\alpha_{D-2}
  -\frac{p}{2\pi N}\left(\tilde\beta_{D-2}\wedge c_1
  +(-1)^{D-1}\,\tilde\epsilon_{0}\wedge\tilde b_{D-1}
  +\tilde\beta_{D-2}\wedge d\epsilon_{0}\right),\\
     \tilde{b}_{D-1}&\longmapsto \tilde{b}_{D-1} + d\tilde{\beta}_{D-2 
     },  \\
           c_1&\longmapsto c_1 + d\epsilon_0.  
\end{aligned}
    \end{equation}
    We stress that this is not a higher group global symmetry, although the transformation resembles one. The Green-Schwarz type shift should really be interpreted as a consequence of the mixed anomaly. 
    \item Gauging $\mathbb{Z}_M^{(D-2)}$, which is equivalent to assigning $\tilde{b}_{D-1}$ a Neumann boundary condition. The truncated transformation reads:
    \begin{equation}
        \begin{aligned}
    a_1&\longmapsto a_1 + d\alpha_0,  \\
b_1&\longmapsto b_1+d\beta_0
  -(-1)^{D-1}\frac{p}{2\pi M}\left(\alpha_0 c_1-\epsilon_0 a_1
  +\alpha_0\,d\epsilon_{0}\right),\\
      c_1&\longmapsto c_1 + d\epsilon_0  .
\end{aligned}
    \end{equation}
    This is a $0$-form symmetry, which is necessarily group-like.
    \item Gauging $\mathbb{Z}_N^{(0)} \times \mathbb{Z}_K^{(0)}$, which is equivalent to assigning $a_1$ and $c_1$ Neumann boundary conditions. The truncated gauge transformation reads:
        \begin{equation}
        \begin{aligned}
\tilde a_{D-1}&\longmapsto \tilde a_{D-1}+d\tilde\alpha_{D-2}
  -\frac{p}{2\pi N}\left(\tilde\beta_{D-2}\wedge c_1
  \right),\\
       \tilde{b}_{D-1}&\longmapsto \tilde{b}_{D-1} + d\tilde{\beta}_{D-2 
     },  \\
\tilde c_{D-1}&\longmapsto \tilde c_{D-1}+d\tilde\epsilon_{D-2}
  +\frac{p}{2\pi K}\left(\,\tilde\beta_{D-2}\wedge a_1
  \right).
\end{aligned}
    \end{equation}
    Since dynamical gauge fields $a_1$ and $c_1$ appear in background gauge field transformations, this is a generic higher fusion categorical symmetry.
    \item Gauging $\mathbb{Z}_N^{(0)}\times\mathbb{Z}_M^{(D-2)}$, which is equivalent to assigning $a_1$ and $\tilde{b}_{D-2}$ Neumann boundary conditions. The truncated gauge transformation reads: 
    \begin{equation}
        \begin{aligned}
\tilde a_{D-1}&\longmapsto \tilde a_{D-1}+d\tilde\alpha_{D-2}
  -\frac{p}{2\pi N}
  (-1)^{D-1}\,(\epsilon_{0}\,\tilde b_{D-1}),
  \\[2pt]
b_1 &\longmapsto b_1+d\beta_0
  +(-1)^{D-1}\frac{p}{2\pi M}(\epsilon_0 a_1),\\
      c_1&\longmapsto c_1 + d\epsilon_0. 
\end{aligned}
    \end{equation}
    Similar to the previous case, dynamical gauge fields $\tilde{b}_{D-2}$ and $c_1$ appear in a background gauge transformation. This signals a generic higher fusion categorical symmetries with possible interactions between symmetry operators of different dimensions. 
\end{enumerate}

Now we clarify the absence of higher group global symmetries upon gauging a $\mathbb{Z}_N^{(0)}$ symmetry. This can be understood as the consequence of some simple no-go theorems based on symTFT and Postnikov tower considerations. For concreteness, let us start from 2-group symmetries.  Let $M_D$ be the physical spacetime. A 2-group global symmetry consists of a $G_{0}$ 0-form symmetry and a $G_{1}$ 1-form symmetry. Let $B_2$ be a background gauge field for $G_{1}$ and $B_{1}$ be a background gauge field for $G_{0}$. The signature of the 2-group global symmetry is the non-closure of $B_2$:
\begin{equation}
    dB_2 = B_1^*O_3,
\end{equation}
where $[O_3] \in H^3(G_{0}, G_{1})$ is the Postnikov class classifying the twist in the 2-group bundle:
\begin{equation}
     B^2G_1 \rightarrow \mathbb{BG} \rightarrow BG_0 ,
\end{equation}
Here we assume a trivial action of $G_0$ on $G_1$. The symTFT is a topological sigma model from $M_D\times I$ to $\mathbb{BG}$ \cite{Freed:2022qnc}.

On the other hand, if in a theory we see that the background gauge field of a $G$ $p$-form symmetry is non-closed and it satisfies:
\begin{equation}
    dB_{p+1} = f(B_p, B_{p-1},\dots),
\end{equation}
namely the non-closure of $dB_{p+1}$ is measured by the background gauge field of $k$-form global symmetries for $k\leq p$, we cannot directly conclude that this theory has a higher group global symmetry. As we have seen in the previous example, the non-closure of $B_{p+1}$ can also arise from mixed 't Hooft anomalies between various higher form symmetries. 

One concrete conclusion we can draw is the following. The Postnikov class $[O_3] \in H^3(G_0, G_1)$ takes value in the $G_1$ simplicial cohomology of $BG_0=K(G_0,1)$. If a 2-group bundle is trivial, then $[O_3] = 0$ and its pullback $B_1^{*}O_3$ to the simplicial cohomology of $M_D$ must also be a trivial $G^{(1)}$-valued simplicial cohomology class of $M_D$. Now consider a family of theories $\Sigma \equiv\{\mathcal{T}_1,\mathcal{T}_2, \dots, \mathcal{T}_n\}$ related to each other by discrete gauging, then
\begin{tcolorbox}[colback=green!10!white,colframe=blue!75!black,
                    ]
    If the symTFT for $\Sigma$ is described by a topological sigma model from $M_D\times I$ onto $B^2 G_1 \times B G_0$ with a possible twist $[\omega_{D+1}]\in H^{D+1}(G_1\times G_0,U(1))$
    and a trivial $G_0$ action on $G_1$, then none of the theories in $\Sigma $ can have a nontrivial 2-group global symmetry.
\end{tcolorbox}

There is a type of nontrivial 2-group global symmetries characterized by a vanishing Postnikov class. In this case, the nontriviality comes from a twist. Namely the 2-group bundle is characterized by $[O_3] = [0] \in H_{\rho}^3(G_0, G_1)$,  where $\rho: G_0 \rightarrow \text{Aut}(G_1)$ is a nontrivial action of $G_0$ on $G_1$. In this case, the above statement generalizes:
\begin{tcolorbox}[colback=green!10!white,colframe=blue!75!black,
                    ]
    Let $\Sigma \equiv\{\mathcal{T}_1,\mathcal{T}_2, \dots, \mathcal{T}_n\}$ be a family of QFTs on $M_D$ related to each other by discrete gauging. Consider a 2-stage Postnikov tower:
        $$ B^2G_1 \rightarrow \mathbb{BG} \rightarrow BG_0 $$ 
        specified by $[0] \in H_{\rho}^3(G_0, G_1)$ for a nontrivial twist $\rho: G_0 \rightarrow \text{Aut}(G_1)$. If their symTFT is described by a topological sigma model from $M_D\times I$ onto $\mathbb{BG}$  with a possible twist$[\omega_{D+1}]\in H^{D+1}(\mathbb{BG},U(1))$, then the only type of nontrivial 2-group symmetry that any of the theories in $\Sigma$ can have is a split 2-group global symmetry.
\end{tcolorbox}
For a family of theories $\Sigma$ whose symTFT is described by a topological sigma model from $M_D\times I$ to a $k$-stage Postnikov tower, the above statements can be applied to each stage of the Postnikov tower construction. 

Let's apply these observations to the symTFT in Eq. \eqref{eq - type-I action as a symTFT}, where the target space of the sigma model is a trivial three-stage fibration. Since the symTFT is a topological sigma model from $M_D\times I$ to $B^{(D-1)}\mathbb{Z}_M \times B\mathbb{Z}_N\times B\mathbb{Z}_K$ with the discrete torsion measuring a non-trivial twist valued in $H^{(D+1)}(B^{(D-1)}\mathbb{Z}_M \times B\mathbb{Z}_N\times B\mathbb{Z}_K, U(1))$, none of the topological boundary conditions can realize a none-trivial higher group global symmetry. 

Our conclusion seems to contradict some known results in the literature on the nose. As mentioned previously, the type-I action for $\mathbb{D}_4$ Dijkgraaf-Witten theory appeared as a leading order topological truncation of 3D ABJM theories with orthosympletctic gauge groups\cite{Bergman:2024its}. However, the symmetry web described by this symTFT contains split 2-group global symmetries like $(\mathbb{Z}_2^{(1)} \times \mathbb{Z}_2^{(1)})\rtimes \mathbb{Z}_2^{(0)}$ and $\mathbb{Z}_4^{(1)}\rtimes \mathbb{Z}_2^{(0)}$. This contradiction is resolved by noticing that the untwisted $\mathbb{D}_4$ Dijkgraaf-Witten theories is secretly a dualized 2-group gauge theory. Moreover, there are a few equivalent representations of the 2-group bundle given by group extensions that $\mathbb{D}_4$ can fit in. This observation generalizes to higher group gauge theories that can be dualized into Dijkgraaf-Witten theories. When this happens, the Dijkgraaf-Witten theory can serve as the symTFT of higher group global symmetries.

Finally, we point out an alternative approach to treat Dijkgraaf-Witten theories and split 2-group gauge theories as symTFTs of higher group global symmetries by coupling higher group bundles to the bulk boundary system. For example, \cite{Barkeshli:2022edm} pointed out that $(D-1)$-group symmetries can be realized in $D$-dimensional Dijkgraaf-Witten theories with a generic gauge group $G$. Topologically, this means that we can couple the theory to a $(D-1)$-group bundle as well as a $G$-bundle and perform quantization by only summing over homotopy classes of maps from the physical spacetime to $BG$. This echoes with the observation in \cite{Thorngren:2015gtw} that symmetries and anomalies of Dijkgraaf-Witten theories can be understood by appropriate extensions to some higher symmetries. If the Dijkgraaf-Witten theory is defined on a manifold with boundary and the coupling of $(D-1)$-group bundle can be consistently extended to the boundary, then we have a legal realization of the symTFT of a higher group symmetries by Dijkgraaf-Witten theories.

\section{An Example of Type-II Action} \label{section - type-II action analysis}

In this section, we study a specific example of type-II actions by gauging the charge conjugation symmetry of an untwisted $\mathbb{Z}_4$ Dijkgraaf-Witten gauge theory by summing over appropriate higher gauging condensation defects. Unlike type-I actions, a higher group structure emerges in these actions. For concreteness we will work in $(3+1)D$ where the discrete 2-group gauge theory actions have been classified \cite{Kapustin:2013uxa}. We give a quick review of 2-group gauge theory action in Sec. \ref{subsection - dualization} following \cite{Kapustin:2013uxa}. In Sec. \ref{subsection - on shell evidence}, we investigate the $(3+1)$D example and show that the on-shell constraints of the type-II action resembles a split 2-group gauge theory whose dualized action is that of an untwisted $\mathbb{D}_4$ Dijkgraaf-Witten gauge theory. We also demonstrate a mismatch between the on-shell deformations and off-shell gauge transformations of the action. 

\subsection{Dualization of 4D 2-Group Gauge Theories} \label{subsection - dualization}

    In this subsection, we quickly review the dualization of a certain class of 2-group gauge theories in $(3+1)$D following \cite{Kapustin:2013uxa}. Consider an abelian 2-group bundle:
    \begin{equation}
         B^2G_2 \rightarrow B\mathbb{G}\rightarrow BG_1 
    \end{equation}
    classified by a Postnikov class $[\beta] \in H^3(BG_1, G_2)$, where both $G_1$ and $G_2$ are finite abelian groups. In general, there can be an $G_1$ action on $G_2$, so the 2-group bundle is classified by the quadruple $(G_1,G_2, \alpha, [\beta])$, where $[\beta]_\alpha \in H^3_{\alpha}(BG_1, G_2)$ takes value in a local coefficient system. 
    
    The topological sigma model action sums over homotopy classes of maps $[\gamma]$ from $M_4$ to $B\mathbb{G}$ and the topological actions are classified by $H^4(B\mathbb{G}, \mathrm{U}(1))$. The cohomology of the total space of a fibration can be deduced from the cohomologies of the fiber and the base space by the Serre spectral sequence. In $(3+1)$D, let $(A_1,B_2)$ denote a pair gauge fields, which are equivalent to a simplicial map $\gamma: M_4 \rightarrow X_4$, where $A_1$ locally maps onto the $BG_1$ sector of $B\mathbb{G}$ and $B_2$ locally maps onto the $B^2G_2$ sector of $B\mathbb{G}$. As usual, gauge equivalence classes of $(A_1,B_2)$ are equivalent to homotopy classes of $\gamma$ and $H^4(B\mathbb{G}, \mathrm{U}(1))$ can be computed with the Serre spectral sequence. The most general action reads:
    \begin{equation}
        \begin{split}
            I(A_1,B_2)  &\equiv 2\pi i \langle [\gamma], [M_4] \rangle\\
            & = 2\pi i \int_{M_4} q_*(\mathfrak{B}B_2) + 2\pi i\int_{M_4}\langle \mathcal{A}^*\lambda_2, \cup B_2\rangle + 2\pi i \int_{M_4} \mathcal{A}^* \omega,
        \end{split}
    \end{equation}
    where:
    \begin{itemize}
        \item $q$ is the group of quadratic functions  $q: G_2 \rightarrow \mathrm{U}(1)$ isomorphic to $H^4(B^2G_2, \mathrm{U}(1))$. See \cite{Kapustin:2013qsa} for more on detail on $q$ and $q_*$.
        \item $\mathfrak{B}$ is the Pontryagin square, which is symmetric and bilinear in $B_2$. 
        \item $\mathcal{A}^*$ denotes the pull-back map of cohomological quantities from $BG_1$ to $M_4$.
        \item $[\omega]\in H^4(BG_1, \mathrm{U}(1))$ is a Dijkgraaf-Witten action of $G_1$
        \item $\lambda \in H^2(BG_1, \hat{G}_2)$.
    \end{itemize}
    
    When the quadratic term in $B_2$ vanishes, the action can be dualized into \cite{Kapustin:2013uxa}:
    \begin{equation} \label{eq - dualized 4D 2-group action}
        I(A_1, C_1) = 2\pi i \int_{M_4}\mathcal{A}^*\omega + 2\pi i \int_{M_4}\langle C_1, \cup \mathcal{A}^* \beta_3 \rangle \equiv 2\pi i \int_{M_4}\Omega,
    \end{equation}
    where $A_1$ is still a $G_1$ valued 1-cocyle, and $C_1$ is a $G_2$-valued 1-cochain subject to the constraint $\delta_A C = \lambda(A)$. Since the dualized action only depends on a pair of 1-cochain a 1-cocyle, it describes an ordinary Dijkgraaf-Witten theory with gauge group $G$ fixed by the extension
    \begin{equation}
        1 \rightarrow G_2 \rightarrow G \rightarrow G_1  \rightarrow 1
    \end{equation}
    which is specified by the pair $(\alpha, \lambda)$ and the sigma model action $[\Omega] \in H^4(BG, \mathrm{U}(1))$.

\subsection{On Shell Evidence of a \texorpdfstring{$\mathbb{D}_4$}{D4} Gauge Group}
\label{subsection - on shell evidence}

    Consider gauging the charge conjugation symmetry of an untwisted $(3+1)$D $\mathbb{Z}_4$ Dijkgraaf-Witten theory, which acts on the elementary Wilson lines and 't Hooft lines as:
    \begin{equation}
        W\leftrightarrow W^3 \qquad M\leftrightarrow M^3
    \end{equation}
    and it leaves $W^2$ and $M^2$ invariant. The higher gauging condensation defect generating the charge conjugation symmetry is given by:
    \begin{equation}\label{eq - Z4 charge conjugation}
        S(\Sigma) = \frac{1}{\abs{H_1(\Sigma, \mathbb{Z}_4)}} \sum_{
             \substack{
             \gamma\in H_1(\Sigma, \mathbb{Z}_4)\\
            \Gamma \in H_{1}(\Sigma, \mathbb{Z}_4)
            }} e^{\frac{2\pi i}{4}\cdot2\cdot\langle\gamma,\Gamma \rangle}W^2(\gamma) M^2(\Gamma).
    \end{equation}
    Since $\mathbb{D}_4 \simeq \mathbb{Z}_4\rtimes \mathbb{Z}_2$, where the twist is specified by the charge conjugation action, we expect that gauging the charge conjugation symmetry produces an alternative Lagrangian description of $\mathbb{D}_4$.

    First, let us construct the action. Coupling the conserved current to the $\mathbb{Z}_2$ charge conjugation symmetry introduces the following discrete torsion term:
    \begin{equation}
        I_{\text{torsion}} = -\frac{4i}{\pi^2}\int_{M_3}a_1\wedge \tilde{a}_{2}\wedge c_1. 
    \end{equation}
    Promoting $c_1$ to a dynamical background gauge field produces the following action:
    \begin{equation}\label{eq - type-II D4 action}
        I = \frac{i}{\pi}\int_{M_4}\left(2\tilde{a}_{2}\wedge d a_1 + \tilde{c}_{2}\wedge dc_1 - \frac{4}{\pi} a_1\wedge \tilde{a}_2\wedge c_1\right),
    \end{equation}
    with equations of motion:
    \begin{equation}\label{eq - type-II D4 OG EOM}
        \begin{aligned}
            & dc_1 = 0, && d\tilde{c}_2 = -\frac{4}{\pi}a_1\wedge \tilde{a}_2, \\
            & d a_1 = - \frac{2}{\pi}c_1\wedge a_1, && d \tilde{a}_2 =  \frac{2}{\pi}c_1\wedge \tilde{a}_2.
        \end{aligned}
    \end{equation}

    Now we show that the on-shell constraints from the action action Eq. \eqref{eq - type-II D4 action} matches the on-shell constraints of the dualized action. First notice that $\mathbb{D}_4 \simeq \mathbb{Z}_4\rtimes\mathbb{Z}_2$ fits into the split extension:
    \begin{equation}
        1\rightarrow \mathbb{Z}_4 \rightarrow \mathbb{Z}_4\rtimes\mathbb{Z}_2\rightarrow \mathbb{Z}_2 \rightarrow 1
    \end{equation}
    where $\mathbb{Z}_2$ acts on $\mathbb{Z}_4$ by $g\mapsto g^{-1}$. To see the 2-group structure, let us rewrite the equations of motion as:
    \begin{equation}
        \begin{aligned}
             &dc_1 = 0, &&d_{-\gamma c_1}\tilde{a}_2 = 0, \\
             &d_{\gamma c_1}a_1 = 0, &&d\tilde{c}_2 =- 2\gamma\, a_1\wedge \tilde{a}_2, 
        \end{aligned}
    \end{equation}
    where $\gamma = \frac{2}{\pi}$.
    Therefore, $(c_1, a_1)$ should be identified with $(A_1, C_1)$ in the sigma model definition and we indeed have an trivial group extension class. $\tilde{a}_2$, $\tilde{c}_2$ are Lagrangian multipliers enforcing the correct constraints. The twisted coboundary operator is $d_{\gamma c_1} = d + \gamma\, c_1\wedge$. It is more suggestive to rewrite the continuum action as:
    \begin{equation} \label{eq - type-II D4 action rewritten}
        S = \frac{i}{\pi}\int_{M_4}\tilde{c_2}\wedge dc_1 + \frac{2i}{\pi}\int_{M_4} \tilde{a}_2\wedge d_{\gamma c_1}a_1,
    \end{equation}
    Inserting the equations of motion back to the action annihilates the second term because it contains $a_1\wedge a_1 = 0$. Meanwhile the first term corresponds to $[0] \in H^4(B\mathbb{Z}_2, \text{U}(1))$ in the sigma model notation. Therefore, the entire action Eq. \eqref{eq - type-II D4 action rewritten} is classically equivalent to the trivial action $[0]\in H^4(B\mathbb{D}_4, \mathrm{U}(1))$ in the sigma model notation and we conclude that the on-shell physics of Eq. \eqref{eq - type-II D4 action} and Eq. \eqref{eq - type-II D4 action rewritten} correctly reproduce an untwisted $\mathbb{D}_4$ Dijkgraaf-Witten theory in $(3+1)$D. 

    Finally, we comment on the off-shell gauge transformation. When going off-shell, we must turn on the gauge transformation $c_1\mapsto c_1 + d\epsilon_0$. Here it is instructive to consider a general type-II action:
    \begin{equation}
        I_{\text{Type-II}} = \frac{i}{2\pi}\int_{M_D}\Big(N\tilde{a}_{D-2}\wedge d a_1 + M\tilde{c}_{D-2}\wedge dc_1 + \frac{p}{2\pi} a_1\wedge \tilde{a}_{D-2}\wedge c_1\Big)
    \end{equation}
    where $M_D$ is a closed oriented $D$-dimensional manifold. Define the constant $\tilde{\gamma} = (-1)^{(D-2)}\frac{p}{2\pi N}$. One can still find a set of gauge transformations that leave the action invariant off-shell on $M_D$:
    \begin{equation}
    \begin{aligned}
a_1 &\longmapsto e^{-\tilde{\gamma}\epsilon_0+c}\bigl(a_1+d_{-c_1}\alpha_0\bigr)\\
\tilde a_{D-2} &\longmapsto e^{\tilde{\gamma}\epsilon_0+c}\bigl(\tilde a_{D-2}+d_{c_1}\tilde\alpha_{D-2}\bigr)\\
c_1 &\longmapsto c_1+d\epsilon_0\\
\tilde c_{D-2}
&\longmapsto \tilde c_{D-2}+d\tilde\epsilon_{D-3}
 +\frac{N\tilde{\gamma}}{M}\Big(-\,a_1\wedge\tilde\alpha_{D-3}
   +\tilde a_{D-2}\wedge\alpha_0+\alpha_0\wedge d\tilde\alpha_{D-3}
   +\tilde{\gamma}\,\alpha_0 \wedge c_1\wedge\tilde\alpha_{D-3}\Big)
\end{aligned}
\end{equation}
    Note that $\epsilon_0$ is a periodic scalar, so we have $\epsilon \sim \epsilon_0 + 2\pi$. However, this shift ruins the periodicity of the $a_1$ gauge field\footnote{We thank Zhengdi Sun for pointing out this subtlety}. Therefore, the off-shell gauge transformation is intrinsically incompatible with the $U(1)$-variables and the off-shell physics of type-II actions cannot be trusted.

\section{Conclusion and Discussion}\label{section - conclusion and discussion}

Summarizing, we showed that one can construct the effective Lagrangians for a large families of discrete gauge theories from gauging 0-form symmetries in abelian discrete gauge theories using U$(1)$ gauge fields. We performed the gauging by formally identifying higher gauging condensation defects as the \Poincare dual of the conserved currents for the 0-form symmetries. When the result is a type-I action, we gave a criteria for when the effective action can be trusted. We also gave a general analysis of the gauge transformations and operator spectrum for type-I actions. When the result is a type-II action, we showed in an concrete $(3+1)$D example that the equations of motion produce correct constraints on-shell, but the off-shell gauge transformations are inconsistent with the on-shell constraints. We also studied the physical implications of type-I action as a symTFT and proposed a few no-go theorems for higher group global symmetries.

Here we point out a few limitations of our analysis. Since we are using U$(1)$-valued gauge fields, the resulting actions are subject to constraints due to large gauge transformations. Although the gauging procedure is rather algorithmic, we must discard those actions that are not invariant under large gauge transformations. This is an artifact unique to $U(1)$ variables. For example, consider gauging the charge conjugation symmetry in a (2+1)D $\mathbb{Z}_3$ gauge theory. The higher gauging condensation defect was constructed in Sec. \ref{subsection - higher gauging condensation defects as 0-form symmetry defect}:
\begin{equation}
            S(\Sigma) = \frac{1}{\abs{H_1(\Sigma,\mathbb{Z}_3)}}\sum_{\gamma,\Gamma} e^{\frac{4\pi i}{3}\langle\gamma,\Gamma\rangle} W(\gamma)M(\Gamma).
\end{equation}
Gauging by summing over higher gauging condensation defects produces the following action:
\begin{equation}
    S = \frac{i}{2\pi}\int_{M_3}\left( 3\tilde{a}_1\wedge d a_1 +  2\tilde{c}_1\wedge d c_1  + \frac{3}{2\pi}a_1\wedge\tilde{a}_1\wedge c_1\right),
\end{equation}
where $\oint a_1\in\frac{2\pi\mathbb{Z}}{3}$ and $\oint c_1\in \pi\mathbb{Z}$. In the standard form for type-II actions, we have $p=3$. By Appendix \ref{appendix - large gauge transformation}, we see that the BF kinetic terms demands that:
\begin{equation}
    p \in 3\times \text{lcm}(3,2)\mathbb{Z}=18\mathbb{Z}, \qquad\text{with } p\sim p +18,
\end{equation}
which does not include $p=3$. Therefore, this action is not invariant under large gauge transformations and should not be considered at the beginning. However, it is a well-known fact that gauging the charge conjugation symmetry in $\text{Rep}(\mathcal{D}(\mathbb{Z}_3))$ produces the theory $\text{Rep}(\mathcal{D}(\mathbb{Z}_3\rtimes \mathbb{Z}_2)) = \text{Rep}(\mathcal{D}(S_3))$ in the condensed matter physics literature\cite{Barkeshli:2014cna}. We expect similar examples to exist for type-I actions.

There is another limitation with the U$(1)$ gauge fields. When the higher gauging condensation defect has trivial discrete torsion term, our procedure produces a trivial current insertion term in the action. For example, in (2+1)D the electric-magnetic duality symmetry is generated by the following higher gauging condensation defect:
    \begin{equation}
        I_\text{EM}(\Sigma) \sim \sum_{\substack{\gamma\in H_1(\Sigma, \mathbb{Z}_N)\\
            \Gamma \in H_{1}(\Sigma, \mathbb{Z}_N)}}W(\gamma)M(\Gamma) \sim \sum_{\substack{A_1\in H^1(\Sigma, \mathbb{Z}_N)\\
            \tilde{A}_1 \in H^{1}(\Sigma, \mathbb{Z}_N)}}e^{\frac{iN}{2\pi}\int_{\Sigma}(a_1\wedge A_1 + \tilde{a}_1\wedge \tilde{A}_1  )}.
    \end{equation}
    Integrating out the defect world-volume $A_1$ and $\tilde{A}_1$ trivializes $a_1$ and $\tilde{a}_1$, which is equivalent to imposing a Dirichlet boundary condition on $\Sigma$. Inserting a mesh of this defects into the 3-dimensional spacetime introduces a trivial contribution to the action, so formally after gauging the $\mathbb{Z}_2$ EM duality the action in terms of U(1) variables looks like:
    \begin{equation}
        I = \frac{iN}{2\pi}\int \tilde{a}_1\wedge a_1 + \frac{i}{\pi}\int \tilde{c}_1\wedge c_1,
    \end{equation}
    This is the action for an untwisted $\mathbb{Z}_N\times\mathbb{Z}_2$ gauge theory, so this manipulation does not make sense. Therefore, the Lagrangian manipulation in terms of differential form variables are incapable of carrying out this type of gauging. 
    
\begin{figure}[h]
        \centering
        \includegraphics[scale=0.8]{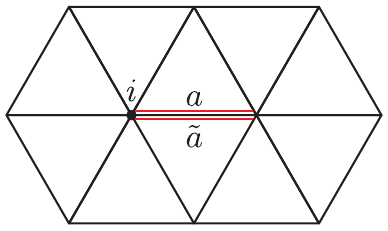}
        \caption{A lattice configuration with a nontrivial 2-cochain coupled to the red link.}
        \label{fig - lattice}
    \end{figure}

     Let's consider again the type-II actions actions using cocycle descriptions have also appeared previously in the literature. A Lagrangian description for type-II action is still possible, but we need to use a different cohomology theory. For example, consider the action in the K-matrix formulation:
    \begin{equation}\label{eq - symTFT action}
        I = \frac{\pi}{N}\int_{M_3}\vb{a}^T \cup_c K \delta_c \vb{a} + \pi\int_{M_3} x\cup\delta c,
    \end{equation}
    where $c$ and $x$ are $\mathbb{Z}_2$ valued 1-cocycles, and  $\vb{a} = (a, \tilde{a})$ is a vector of $\mathbb{Z}_N$ valued 1-cochains twisted by $c$, and $\delta_c$ is a twisted coboundary operator. Here the $K$-matrix is $K = \begin{pmatrix}
        0 & 1\\ 1 & 0
    \end{pmatrix}$.  Similar to the twisted exterior derivative defined in the main text, $\delta_c^2=0$ only when $\delta c=0$. This theory appeared as a cocycle description for the symTFT of (1+1)D QFTs with $\text{TY}(\mathbb{Z}_N)$ categorical symmetries \cite{Kaidi:2022cpf}. Similar to the type-II actions in the main text, Eq. \eqref{eq - symTFT action} is  only flat on-shell. The off-shell gauge transformation leaving the action invariant contains the following shift on $\vb{a}$:
    \begin{equation}
        \vb{a}_{ij}\mapsto K^{-\omega_i}(\vb{a}_{ij} + (\delta_c \vb{g})_{ij} ),
    \end{equation}
    where $c_{ij} \mapsto c_{ij} + (\delta \omega)_{ij}$ is the off-shell gauge transformation on $c$. Apparently this is a discrete analog of $a_1\mapsto e^{-\epsilon_0-C}(a_1 + d_{c_1}\alpha_0)$. However, on a triangulation, the non-linear scaling actually can be canceled by demanding $\vb{n}^T K = \vb{n}^T$, where $\vb{n}$ is the charge vector for $\vb{a}$. Consider a lattice configuration in Fig. \ref{fig - lattice}.
   The non-linear scaling happens only at junctions of links, so one can define gauge invariant operators:
    \begin{equation}
        U_{\vb{n}}(\gamma) = e^{i\frac{2\pi}{N}\oint_{\gamma}\vb{n}^T\cdot\vb{a}},\qquad \vb{n}\in\mathbb{Z}_N\times \mathbb{Z}_N.
    \end{equation}
    On the other hand, on the continuum the non-linear scaling happens at every point on the loop $\gamma$. Therefore, no such configurations can be made gauge invariant on the continuum. This difference implies more possibilities for cocycle effective Lagrangian descriptions using simplicial cohomology variables. It would be interesting to systematically extend the higher gauging condensation defect analysis of gauging onto the lattice. 
    
\section*{Acknowledgement}

     We thank Ken Intriligator, Zhengdi Sun, Yi-Zhuang You, John McGreevy, Tarun Grover, Daniel E. Parker, Oren Bergman, Aiden Sheckler, and Linhao Li for helpful discussions.
    E.Y.Y. and Z.Z. are supported by Simons Foundation award 568420 (Simons Investigator) and award
888994 (The Simons Collaboration on Global Categorical Symmetries). Y.X. thanks the hospitality of Kavli Institute of Theoretical Physics (KITP), where part of this work was completed during the program: Learning the Fine Structure of Quantum Dynamics in Programmable Quantum Matter. This research was supported in part by grant NSF PHY-2309135 to the KITP.

    \begin{appendix}

    \section{Large Gauge Transformations} \label{appendix - large gauge transformation}

Since we are explicitly working with U$(1)$ gauge fields in this work, we need to check the invariance of the action under large gauge transformations. Specifically, let $a_p, \tilde{a}_{D-p-1}$ be a pair of $p$-form gauge fields. The corresponding Wilson and 't Hooft surface operators:
\begin{equation}
W(\gamma_p)=e^{i\oint_{\gamma_p}a_p } \qquad M(\Gamma_{D-p-1})=e^{i\oint_{\Gamma_{D-p-1}} \tilde{a}_{D-p-1} }
\end{equation}
are invariant under the shift \begin{equation}
    \oint_{\gamma_p} a_p \mapsto \oint a_{p}+2\pi \qquad \oint_{\Gamma_{D-p-1}}\tilde{\alpha}_{D-p-1}\mapsto \oint_{\Gamma_{D-p-1}}\tilde{\alpha}_{D-p-1} + 2\pi
\end{equation}
It then must be that the action is also invariant under this shift. The BF kinetic term is automatically invariant under this transformation. Consider $I_{BF} = \frac{iN}{2\pi}\int \tilde{a}_{D-p-1}\wedge da_p$, which transforms under the large gauge transformation as:
\begin{equation}
    I_{BF}\mapsto I_{BF} + \frac{iN}{2\pi}\cdot(2\pi)^2 = I_{BF} + 2\pi iN
\end{equation}
where the transformation of $\tilde{a}_{D-p-1}$ contributes a factor of $2\pi$ and $da_p$ contributes $2\pi$ by flux quantization. 

The nontrivial part comes from the discrete torsion term. The form degree is irrelevant to the large gauge transformation analysis, so here we temporarily suppress them. The relevant discrete torsion term is:
\begin{equation}
    I_{\text{torsion}} = \frac{ip}{(2\pi)^2}\int a\wedge b \wedge c
\end{equation}
where the holonomies are given by $\oint a \in \frac{2\pi}{N_1}$, $\oint b \in \frac{2\pi}{N_2}$, and $\oint c \in \frac{2\pi}{N_3}$. Under three separate large discrete gauge transformations for $a, b, c$, the transformed term contributes a factor of $2\pi$ while the rest contribute a factor proportional to their holonomy. The three different gauge transformations give:
\begin{equation}
    \begin{split}
        \delta \oint a&: \delta I = \frac{ip}{(2\pi)^2}(2\pi) \left(\frac{2\pi}{N_2}\right) \left(\frac{2\pi}{N_3}\right)= \frac{2\pi i p}{N_1 N_2}  \\
        \delta \oint b&: \delta I = \frac{ip}{(2\pi)^2}\left(\frac{2\pi}{N_1}\right)(2\pi)  \left(\frac{2\pi}{N_3}\right) = \frac{2\pi i p}{N_1 N_3} \\
        \delta \oint c&: \delta I = \frac{ip}{(2\pi)^2}\left(\frac{2\pi}{N_1}\right) \left(\frac{2\pi}{N_2}\right)(2\pi) = \frac{2\pi i p}{N_1 N_2} 
    \end{split}
\end{equation}
Gauge invariance requires each all of them to be valued in $2\pi \mathbb{Z}$, which requires $p\in \text{lcm}(N_1N_2, N_2N_3,N_1N_3)\mathbb{Z}$.
One also needs to fix the periodicity of $p$. Note that the discrete torsion term is proportional to:
\begin{equation}
    I_{\text{torsion}} \sim \frac{ip}{(2\pi)^2} \left( \frac{(2\pi)^3}{N_1 N_2 N_3}\right) = \frac{2\pi ip }{N_1 N_2 N_3}
\end{equation}
The smallest shift in $p$ that leaves this action invariant $p$ is $p \mapsto p + N_1 N_2 N_3$. Summarizing, the large gauge transformation fixes $p$ to be:
\begin{equation}
    p\in \text{lcm}(N_1N_2, N_2N_3,N_1N_3)\mathbb{Z} \quad\text{and}\; p\sim p+N_1 N_2 N_3
\end{equation}

This constraint simplifies for type-II actions where the discrete torsion is of the form:
\begin{equation}
    I_{\text{torsion}} = \frac{ip}{(2\pi)^2}\int a\wedge \tilde{a} \wedge c
\end{equation}
Here $a$ and $\tilde{a}$ have the same holonomy. Using the fact that:
\begin{equation}
    \operatorname{lcm}(N^{2},\, N\,N_{3},\, N\,N_{3}) = N\,\operatorname{lcm}(N, N_{3})
\end{equation}
we find that $p$ takes value in $p\in N\,\operatorname{lcm}(N, N_{3}) $ with $ p\sim p N^2N_3$.

As a simple example, we see that the action for the $\mathbb{D}_4$ gauge theory is invariant under large gauge transformations. The discrete torsion term reads:
\begin{equation}
        I_{\mathbb{D}_4 \text{ torsion}} =- \frac{i}{\pi^2} \int_M a_1\wedge\tilde{b}_{D-2}\wedge c_1 = - \frac{4i}{4\pi^2} \int_M a_1\wedge\tilde{b}_{D-2}\wedge c_1 
    \end{equation}
so $p=-4$. Setting $N_1=N_2=N_3=2$, we see that:
\begin{equation}
    p\in \text{lcm}(2^2,2^2,2^2)\mathbb{Z} = 4\mathbb{Z} \;\;\qq{\text{and}}\;\; p\sim p+8
\end{equation}
which means $p\in \{0,4,8\}$. Therefore, this action is invariant under large gauge transformations and by the flux identification requirement we can also write $p=4$ in the action. 
\section{Gauge Transformations Derivation}\label{appendix - small gauge transformations}

    Consider the following type-I action:
    \begin{equation}
        I = I_{\text{kinetic}} + I_{\text{torsion}}
    \end{equation}
    where:
    \begin{equation}
        \begin{split}
            I_{\text{kinetic}} &=  \frac{i}{2\pi}\int_{M _D}\left(N\tilde{a}_{D-2}\wedge da_1 + M\tilde{b}_{D-2}\wedge db_1 + K\tilde{c}_{D-2}\wedge dc_1  \right) \\
        I_{\text{torsion}}&=\frac{p}{2\pi}\int_{M_D}a_1\wedge\tilde{b}_{D-2}\wedge c_1
        \end{split} 
    \end{equation}
    Using the ansatz Eq. \eqref{eq - type-I action on shell transformation} motivated by the on-shell deformation of field strengths, we find the following full gauge transformation of the discrete torsion term:
\begin{equation}
        \begin{split}
            \delta  I_{\text{torsion}} =& \frac{ip}{4\pi^2}\int_{M_D} (a_1 + d\alpha_0)\wedge (\tilde{b}_{D-2}+ d\tilde{\beta}_{D-3}) \wedge (c_1+d\epsilon_0) - \frac{ip}{4\pi^2}\int a_1\wedge \tilde{b}_{D-2} \wedge c_1\\
            =& \frac{ip}{4\pi^2}\int_{M_D} \left( a_1\wedge \tilde{b}_{D-2} \wedge d\epsilon_0 + a_1 \wedge d\tilde{\beta}_{D-3} \wedge c_1 + a_1 \wedge d\tilde{\beta}_{D-3} \wedge d\epsilon_0 \right. \\
            & \left.+ d\alpha_0\wedge \tilde{b}_{D-2} \wedge c_1 + d\alpha_0 \wedge \tilde{b}_{D-2} \wedge d\epsilon_0 + d\alpha_0 \wedge d\tilde{\beta}_{D-3} \wedge c_1 + d\alpha_0 \wedge d\tilde{\beta}_{D-3} \wedge d\epsilon_0\right)
        \end{split}
    \end{equation}

    The last term is a total derivative and it can be dropped. Now we compute the gauge transformation of the kinetic term. It is useful to rewrite $I_{\text{kinetic}}$ as follows:
    \begin{equation}
       I_{\text{kinetic}} = \frac{i}{2\pi}\int_{M_D}\left( N\tilde{a}_{D-2}\wedge d a_1 + M b_1 \wedge d\tilde{b}_{D-2} + K\tilde{c}_{D-2}\wedge d c_1\right)
    \end{equation}
    Schematically, the gauge transformation of each term looks like:
    \begin{equation}
        \delta I \sim \int\left( A\wedge d(\delta B) + \delta A\wedge dB + \delta A\wedge d(\delta B)\right)
    \end{equation}
    where $\delta A$'s are the transformations that contain compensating shifts. Therefore, all terms except $A\wedge d\delta B$ are total derivatives. Inserting ansatz Eq. \eqref{eq - type-I action on shell transformation}, we have:
    \begin{equation}
        \begin{split}
             \delta I_{\text{kinetic}} =& \frac{i}{2\pi}\int_{M_D}\Bigg(\frac{p}{2\pi} \left( \xi\tilde{\beta}_{D-3}\wedge c_1 + (-1)^{D-2}\xi\epsilon_0\wedge\tilde{b}_{D-2}+\lambda\tilde{\beta}_{D-3}\wedge d\epsilon_0   \right)\wedge d a_1 \\
             & +(-1)^{D-2}\frac{p}{2\pi}\left(\xi\alpha_0 \wedge c_1 - \xi\epsilon_0 \wedge a_1 +\lambda\alpha_0\wedge d\epsilon_0  \right)\wedge d\tilde{b}_{D-2}\\
             &+ (-1)^{D-1}\frac{p}{2\pi}\left( \xi\alpha_0\wedge\tilde{b}_{D-2}  + (-1)^{D-2}\xi\tilde{\beta}_{D-3} \wedge a_1 + \lambda\alpha_0 \wedge d\tilde{\beta}_{D-3} \right)\wedge d c_1 \Bigg)
        \end{split} 
    \end{equation}
    Integrating by parts and reorganizing the terms, we get:
    \begin{equation}
        \begin{split}
            \delta I_{\text{kinetic}} =&\frac{i p}{(2\pi)^2}\int_{M_D}\left(
\xi\,a_1\wedge d\tilde{\beta}_{D-3}\wedge c_1
+\xi\,a_1\wedge\tilde{b}_{D-2}\wedge d\epsilon_0
+\xi\,d\alpha_0\wedge\tilde{b}_{D-2}\wedge c_1\right. \\
&+\lambda\left(
a_1\wedge d\tilde{\beta}_{D-3}\wedge d\epsilon_0
+d\alpha_0\wedge d\tilde{\beta}_{D-3}\wedge c_1
+d\alpha_0\wedge\tilde{b}_{D-2}\wedge d\epsilon_0
\right)\\
&-\xi\left(
d a_1\wedge\tilde{\beta}_{D-3}\wedge c_1
-a_1\wedge d\tilde{\beta}_{D-3}\wedge c_1
+(-1)^{D-2}a_1\wedge\tilde{\beta}_{D-3}\wedge d c_1
\right)\\
&
+(-1)^{D-3}\xi\left(
d a_1\wedge\tilde{b}_{D-2}\wedge\epsilon_0
-a_1\wedge d\tilde{b}_{D-2}\wedge\epsilon_0
+(-1)^{D-3}a_1\wedge\tilde{b}_{D-2}\wedge d\epsilon_0
\right)\\
&
+\xi\left(
d\alpha_0\wedge\tilde{b}_{D-2}\wedge c_1
+\alpha_0\wedge d\tilde{b}_{D-2}\wedge c_1
+(-1)^{D-2}\,\alpha_0\wedge\tilde{b}_{D-2}\wedge d c_1
\right)
        \end{split}
    \end{equation}
    where the last three lines are total derivatives. To cancel the $\delta I_{\text{torison}}$ contribution, we need to set $\xi=\lambda=-1$.
    \begin{equation}
    \begin{aligned}
    a_1\;&\longmapsto a_1 + d\alpha_0  \\
    c_1\;&\longmapsto c_1 + d\epsilon_0  \\
     \tilde{b}_{D-2}\;&\longmapsto \tilde{b}_{D-2} + d\tilde{\beta}_{D-3
     }  \\
\tilde a_{D-2}\;&\longmapsto\; \tilde a_{D-2}+d\tilde\alpha_{D-3}
  -\frac{p}{2\pi N}\left(\tilde\beta_{D-3}\wedge c_1
  +(-1)^{D-2}\,\tilde\epsilon_{0}\wedge\tilde b_{D-2}
  +\tilde\beta_{D-3}\wedge d\epsilon_{0}\right)\\[2pt]
b_1\;&\longmapsto\; b_1+d\tilde\beta_0
  -(-1)^{D-2}\frac{p}{2\pi M}\left(\alpha_0\wedge c_1-\epsilon_0 \wedge a_1
  +\alpha_0\wedge d\epsilon_{0}\right)\\[2pt]
\tilde c_{D-2}\;&\longmapsto\; \tilde c_{D-2}+d\tilde\epsilon_{D-3}
  -(-1)^{D-1}\frac{p}{2\pi K}\left(\alpha_0\wedge\tilde b_{D-2}
  +(-1)^{D-2}\,\tilde\beta_{D-3}\wedge a_1
  +\alpha_0\wedge d\tilde\beta_{D-3}\right)
\end{aligned}
\end{equation}
Since no constraints from the equations of motion were used in this derivation, we conclude that the result leaves the action invariant off-shell.

\section{Gauging Finite Symmetries in \texorpdfstring{$(2+1)$}{(2+1)}D Untwisted DW Theories} \label{appendix - gauging finite symmetry in DW}

In this appendix, we summarize a result in  \cite{Maier2011EquivariantMC} on gauging a specific type of $J^{(0)}$ finite symmetries of $(2+1)$D untwisted Dijkgraaf Witten theory with gauge group $G$, where $J$ has a weak action on $G$. Here $J$ and $G$ need not to be abelian. Intuitively, a weak $J$-action on $G$ induces an automorphism of the Hopf algebra $\mathcal{D}(G)$ as well as an automorphism of the representation category $\text{Rep}(\calD(G))$. As usual, gauging the $J^{(0)}$ symmetry is a two-step procedure: performing $J$-extension and  $J$-equivariantization. For a weak $J$-action, we can define similar operations on the Hopf algebra $\mathcal{D}(G)$. Namely, the Hopf algebra $\mathcal{D}(G)$ is first promoted to a $J$-Hopf algebra $\mathcal{D}^J(G)$, then we construct the orbifold algebra $\widehat{\mathcal{D}^J(G)}{}^J$. The representation categories of these two algebras correspond to the $J$-cross extended category $\text{Rep}(\calD(G))_J^{\times}$ and its $J$-equivariantization $\left( \text{Rep}(\calD(G))_J^{\times} \right){}^{J}$.

Unfortunately, in \cite{Maier2011EquivariantMC} the $J$-extension step was referred to as $J$-equivariantization and $J$-equivariantization was referred to as $J$-orbifolding. In this appendix we will adopt the terminologies of \cite{Maier2011EquivariantMC}. There is an alternative description of this gauging in terms of principal bundles and we refer the readers to \cite{Maier2011EquivariantMC} for further details. Since the representation category side of the story is by now rather standard in the physics literature, we will focus more on the Hopf algebra perspective. In Sec. \ref{subsection - weak J actions}, we will define weak $J$-actions on the Hopf algebra and their representation categories. In Sec. \ref{subsection - J-gauging}, we will briefly summarize the main theorems and state their relation with the SET construction in \cite{Barkeshli:2014cna, Cui2015OnGS}. We will spend the rest of this introduction on defining weak $J$-actions.

 Let $J$ be a finite group. Consider a collection of group automorphisms
$\rho_j : G \to G$ labeled by $j \in J$ and a collection of group elements $c_{i,j} \in G$ labeled by a pair of elements $i,j \in J$. For all $i,j,k \in J$, let $\mathrm{Inn}_g$ denote the $G$ inner automorphism associated to an element $g \in G$. We have a \textbf{\textit{weak $J$-action on $G$}} when:
\begin{equation}
    \rho_i \circ \rho_j = \mathrm{Inn}_{c_{i,j}} \circ \rho_{ij}
\qquad
\rho_i(c_{j,k}) \cdot c_{i,jk} = c_{i,j} \cdot c_{ij,k}
\qquad
c_{1,1} = 1
\end{equation}
In this sense, a weak $J$-action on $G$ is a $J$ action by $G$-automorphism that is associative up to $G$ inner automorphisms. 
Two weak actions $(\rho_j,c_{i,j})$ and $(\rho_j',c_{i,j}')$ of $J$ on $G$ are \textbf{\textit{isomorphic}} if there is a collection of group elements $h_j \in G$ labeled by $j \in J$ such that
\begin{equation}
    \rho_j' = \mathrm{Inn}_{h_j} \circ \rho_j
\qquad
c_{ij}'\cdot h_{ij} = h_i \cdot \rho_i(h_j) \cdot c_{ij}
\end{equation}

It is a highly nontrivial fact that weak $J$ actions on $G$ are related to the group extension problem \cite{Schreier1926berDE}. Let $(\rho_i,c_{i,j})$ be a weak $J$-action on $G$. On the set $H = G\times J$, define a multiplication:
\begin{equation}
    (g,i)\cdot (g',j') \equiv (g\cdot \rho_i(g')\cdot c_{i,j}, ij)
\end{equation}
which turns $H$ into a group fitting in the short exact sequence:
\begin{equation}
    1\rightarrow G\rightarrow H \xrightarrow{\pi} J \rightarrow1
\end{equation}
On the other hand, the weak $J$-action can be reconstructed from the above group extension by choosing a set-theoretic sections $s:J\rightarrow H$ and $\pi: H\rightarrow J$ with $s(1)=1$. Since $H\triangleleft G$, the conjugation $s(j)\,g\, s^{-1}(j)$ leaves $G$ invariant, hence $j\in J$ defines an automorphism on $G$ by conjugation by the above conjugation action. Define $c_{i,j} \equiv s(i)\, s(j)\,s(ij)^{-1}$ which lives in $\ker\pi$. One can check that $(\rho_j, c_{i,j})$ indeed defines a weak $J$-action on $G$. Two different choices of set-theoretic sections of the same group extension are isomorphic. Therefore, we have  
\begin{theorem}
    There is a one-to-one correspondence between isomorphism classes of weak $J$-actions on $G$ and isomorphism classes of group extensions $1\rightarrow G\rightarrow H \rightarrow
 J \rightarrow 1$ for fixed $G$ and $J$.
\end{theorem}
If the group extension splits, then we have a \textbf{\textit{strict $J$-action}} on $G$.

\subsection{Weak \texorpdfstring{$J$}{J}-Actions on \texorpdfstring{$D(G)$}{D(G)} and \texorpdfstring{$\text{Rep}(\calD(G))$}{Rep(D(G))}}\label{subsection - weak J actions}

    Before describing the weak $J$-action on Hopf algebras and their representation categories, it is instructive to have a quick review of the elementary definitions of Hopf algebras.  
    
    Recall that a \textit{\textbf{$\mathbb{K}$-Hopf algebra}} consists of an algebra $A$ over a base field $\mathbb{K}$ with multiplication $\nabla: A\otimes A \rightarrow A$, a comultiplication $\Delta: A\rightarrow A\otimes A$, a unit $\eta:K \rightarrow A$, a counit $\epsilon:A\rightarrow \mathbb{K}$ and a $\mathbb{K}$-linear function $S: A\rightarrow A$ called the \textbf{\textit{antipode}} so that:
    \begin{equation}
        m \circ (\text{id}\otimes S)\circ \Delta = m\circ(S\otimes\text{id})\circ\Delta = \eta\circ \epsilon.
    \end{equation}
    If the algebra contains an invertible element $R$ in $A\otimes A$ so that:
    \begin{itemize}
        \item $R\Delta(x)R^{-1} =(T\circ \Delta)(x)$ for all $x\in A$ and $T$ is a $\mathbb{K}$-linear map $T: A\otimes A \rightarrow A\otimes A$ so that $T(x\otimes y ) = y\otimes x$,
        \item $(\Delta \otimes 1)(R) = R_{12}R_{23}$ and $( 1\otimes \Delta)(R) = R_{13}R_{12}$, where $R_{12} = \phi_{12}(R)$, $R_{13} = \phi_{12}(R)$, $R_{23}= \phi_{23}(R)$, and $\phi_{12}, \phi_{23}, \phi_{13}$ are algebra morphisms $H\otimes H \rightarrow H\otimes H\otimes H$ evaluated by:
        \begin{equation}
            \phi_{12}(x\otimes y) = x\otimes y \otimes 1, \qquad \phi_{12}(x\otimes y) = x\otimes 1 \otimes y,\qquad \phi_{12}(x\otimes y) = 1\otimes x \otimes y,
        \end{equation}
    \end{itemize}
    then this algebra is called a \textbf{\textit{quasitriangular Hopf algebra}}, where $R$ is called the $R$-matrix. Furthermore, if the algebra contains an invertible central element $\nu$ called the \textbf{\textit{ribbon element}} satisfying:
    \begin{equation}
        \begin{split}
            &v^2 = u S(u), \quad S(\nu) = \nu, \quad \epsilon(\nu)=1,\\
            & \Delta(\nu) = (R_{21}R_{12})^{-1}(\nu\otimes\nu).
        \end{split}
    \end{equation}
    where $u = \nabla (S\otimes\text{id})(R_{21})$, then the algebra becomes \textbf{\textit{a ribbon Hopf algebra}}.

     Given any finite group $G$, we can canonically define a Hopf algebra by the Drinfeld double $\mathcal{D}(G)$ construction. The Drinfeld double takes the group algebra $K(G)$ of a finite group $G$ as an input and produces a ribbon Hopf algebra $\mathcal{D}(G)$. The canonical basis for $\mathcal{D}(G)$ is spanned by $(\delta_g\otimes h)_{g,h\in G}$. The multiplication $\nabla$ is defined by:
        \begin{equation}
        \nabla\left( (\delta_g\otimes h),(\delta_{g'}\otimes h')\right)
            =
            \begin{cases}
            \delta_g\otimes hh' & \text{for } g = h g' h^{-1}\\
            0 & \text{else}
            \end{cases}.
        \end{equation}
    The comultiplication is given by:
    \begin{equation}
\Delta(\delta_g\otimes h)
=
\sum_{g' g'' = g}(\delta_{g'}\otimes h)\otimes(\delta_{g''}\otimes h).
\end{equation}
    The unit is $\sum_{g\in G}\delta_g\otimes 1$ and the counit is
$\epsilon(\delta_1\otimes h)=1$ and $\epsilon(\delta_g\otimes h)=0$ for $g\neq 1$ for all $h\in G$. The antipode map is given by
\begin{equation}
S(\delta_g\otimes h) = \delta_{h^{-1}g^{-1}h}\otimes h^{-1}.
\end{equation}
The $R$-matrix is given by: \begin{equation}
R
:=
\sum_{g\in G}\sum_{h\in G}(\delta_g\otimes 1)\otimes(\delta_h\otimes g)
\in D(G)\otimes D(G).
\end{equation}
and the ribbon element is:
\begin{equation}
    \theta
:=
\sum_{g\in G}(\delta_g\otimes g^{-1})
\in \mathcal{D}(G).
\end{equation}

The $J$-action on $G$ naturally induces a $J$-action on the Hopf algebra $\mathcal{D}(G)$. For pedagogical purposes it is helpful to first state the defintion of a weak $J$-action on an algebra $A$.
A \textbf{\textit{weak $J$-action}} on $A$ consists of algebra automorphisms
$\varphi_j \in \mathrm{Aut}(A)$ labeled by $j \in J$, and
invertible elements $c_{i,j} \in A$ labeled by a pair of elements $i,j \in J$, such that for all $i,j,k \in J$ we have:
\begin{equation}
\begin{aligned}
\varphi_i \circ \varphi_j &= \mathrm{Inn}_{c_{i,j}} \circ \varphi_{ij}, \\
\varphi_i(c_{j,k}) \cdot c_{i,jk} &= c_{i,j} \cdot c_{ij,k}, \\
c_{1,1} &= 1.
\end{aligned}
\end{equation}
Here $\mathrm{Inn}_x$ with $x$ an invertible element of $A$ denotes the algebra
automorphism $a \mapsto x a x^{-1}$. A \textbf{\textit{$J$-Hopf algebra}} is an algebra $A$ with a $J$-grading
$A=\bigoplus_{j\in J}A_j$
and a weak $J$-action such that\textbf{\textit{$J$-Hopf algebra}}:
\begin{itemize}
\item
The algebra structure of $A$ restricts to the structure of an associative algebra on each $A_j$ so that $A$ is the direct sum of the components $A_j$ as an algebra.

\item
$J$ acts by homomorphisms of Hopf algebras.

\item
The action of $J$ is compatible with the grading: $\varphi_i(A_j)\subset A_{iji^{-1}}$.
\item
The comultiplication $\Delta:A\to A\otimes A$ respects the grading $\Delta(A_j)\subset \bigoplus_{\substack{p,q\in J\\ pq=j}} A_p\otimes A_q$.
\item
The elements $(c_{i,j})_{i,j\in J}$ are group-like: $\Delta(c_{i,j})=c_{i,j}\otimes c_{i,j}$. 
\end{itemize}
Furthermore, if the $J$-Hopf algebra is a ribbon Hopf algebra, then we need to introduce new compatibility conditions of the $R$-matrix with the $J$-grading. See \cite{Maier2011EquivariantMC} for further details.

If the Hopf algebra of interest is the Drinfeld double $\mathcal{D}(G)$, then the resulting Hopf $J$-algebra with a weak $J$-action on $G$ naturally has the structure of a $J$-Hopf algebra. Together with the modified ribbon structure, one can define the \textbf{\textit{$J$-Drinfeld double}} $\mathcal{D}^J(G)$. The algebraic structure of $\mathcal{D}^J(G)$ can be deduced by treating it as a Hopf subalgebra of $\mathcal{D}(H)$. Especially, the representation category of $\mathcal{D}^J(G)$ is a $J$-equivariant tensor category and it can be given the structure of a modular tensor category. $\text{Rep}(\mathcal{D}(G))$ being $J$-equivariant means $\text{Rep}(\mathcal{D}(G)) = \oplus_{j\in J}\text{Rep}(\mathcal{D}(G))_j$, which corresponds to the defectification step in physics literature. We will provide a definition of a $J$-equivariant category in the next subsection.

\subsection{Orbifold Algebra and Orbifold Category}\label{subsection - J-gauging}

    In this subsection we describe the gauging step for both the $J$-Drinfeld doubles $\mathcal{D}^J(G)$ and their representation categories. 
    
    First we define a $J$-orbifoldization of an $J$-equivariant algebra. Let $A$ be an algebra with a weak $J$-action $(\varphi_j,c_{ij})$.
We endow the vector space $\widehat{A}^{J} := A \otimes K[J]$ with a unital associative multiplication on elements of the form $(a \otimes j)$ with $a \in A$ and $j \in J$:
\begin{equation}
    (a \otimes i)(b \otimes j) \equiv a\varphi_i(b)c_{ij} \otimes ij.
\end{equation}
This algebra is called the \textit{\textbf{orbifold algebra}} $\widehat{A}^{J}$ of the $J$-equivariant algebra $A$
with respect to the weak $J$-action. Especially, if $A$ is a $J$-Hopf algebra, then the orbifold algebra $\widehat{A}^J$ is also a Hopf algebra. 

Let us move on to the representation category side of the story. First we define the \textbf{\textit{$J$- action}} on a category $\mathcal{C}$ , which contains the following data:
\begin{itemize}
\item
A collection of functors $\phi_j\colon C\to C$ labeled by $j\in J$. 
\item A functorial isomorphism $\alpha_{i,j}\colon \phi_i\circ \phi_j \xrightarrow{\sim} \phi_{ij}$
called \textit{\textbf{compositors}} for every pair of $i,j\in J$ satisfying the coherence conditions:
    \begin{equation}
        \alpha_{ij,k}\circ \alpha_{i,j} = \alpha_{i,jk}\circ \phi_i(\alpha_{j,k})\qquad \text{and} \qquad \phi_1=\mathrm{id}
    \end{equation}
\end{itemize}
A \textbf{\textit{$J$-equivariant category}} $C$ is a category with a grading $C = \bigoplus_{j\in J}C_j$ and a categorical action of $J$ subject to the compatibility condition $\phi_i(C_j)\subset C_{iji^{-1}}$. One can also define $J$-equivariant tensor categories with braiding structures, which precisely correspond to the $J$-crossed braided tensor categories $C_J^{\times}$ in \cite{Barkeshli:2014cna}. See \cite{ceccherini2022representation} for further details.

Let $C$ be a $J$-equivariant category.
The \textit{\textbf{orbifold category}} $C^{J}$ of $C$ has the following data:
\begin{itemize}
\item
The objects are pairs $\left(V,(\psi_j)_{j\in J}\right)$ consisting of an object $V\in \text{Obj(C)}$ and a family of isomorphisms
$\psi_j\colon {}^jV\to V$ labeled by $j\in J$ such that $\psi_i\circ {}^i\psi_j=\psi_{ij}\circ \alpha_{i,j}$. 
\item
The morphisms are $f\colon (V,\psi^V_j)\to (W,\psi^W_j)$ morphisms $f\in \text{Hom}_{C}(V,W)$ in $C$ such that $\psi_j\circ {}^j(f)=f\circ \psi_j$ for all $j\in J$ 
\end{itemize}

Now we are ready to state the final result. First, we have the theorem for Hopf algebras:
\begin{theorem}
    The $\mathbb{K}$-linear map:
\begin{equation} \label{theorem - gauging a finite symmetry with weak action in DW theory}
\Psi \colon \widehat{\calD^{J}(G)}{}^J \to \calD(H),
\qquad
(\delta_h \otimes g \otimes j) \mapsto (\delta_h \otimes g\,s(j)),
\end{equation}
is an isomorphism of ribbon algebras. Especially, we have an equivalence of ribbon categories:
    \begin{equation}
       \left( \text{Rep}(\widehat{\mathcal{D}^J(G)})\right){}^J \simeq \text{Rep}(\mathcal{D}(H)),
    \end{equation}
    which implies the equivalence of their modular data. 
\end{theorem}
 We end this appendix with a few comments. Here we switch back to the standard physics terminology:
\begin{itemize}
         \item In an SET context, the weak $J^{(0)}$ action on $\text{Rep}(\mathcal{D}(G))$ makes sense only if both the symmetry fractionalization obstruction $[O_3]\in H^3_{[\rho]}(J, \mathcal{A})$ and obstruction to defectification $[O_4]\in H^4(J,U(1))$ vanish \cite{Barkeshli:2014cna,Cui2015OnGS}. Here $\mathcal{A}$ is the group whose elements are the abelian topological charges of $\mathcal{C}$ with group multiplication defined by fusion.
         \item $[O_3]=0$ implies that $O_3 = \delta_{\rho}\mathfrak{w}_2$, where $\mathfrak{w}_2$ is an $\mathcal{A}$-valued cochain. Different classes of solutions to $O_3 = \delta_{\rho}\mathfrak{w}_2$ differ by an $H_{\rho}^2(J,\mathcal{A})$ torsor. Namely, picking a base point $\mathfrak{w}_2(g,h)$ in the collection of all equivalence classes of solutions to $O_3 = \delta_{\rho}\mathfrak{w}_2$, we can generate all solutions by multiplication with the generator of $H_{\rho}^2(J,\mathcal{A})$ torsors:
        \begin{equation}
            \{\mathfrak{w}_2(g,h), t(g,h)\times \mathfrak{w}_2(g,h), t(g,h)^2\times \mathfrak{w}_2(g,h)\dots\},
        \end{equation}
        where $g,h\in J$. Therefore, symmetry fractionalizations are classified by $H_{\rho}^2(J,\mathcal{A})$ torsors \cite{Barkeshli:2014cna}. For bookkeeping purposes, when assigning $\text{Rep}(\widehat{\mathcal{D}^J(G)})$ a $J$-cross braided structure, we can always choose the one corresponding to $[\mathfrak{w}_2(g,h)]$ with no torsion components stacked.
        \item For a fixed weak $J$-action and a specific choice of symmetry fractionalization class, we still have a $H^3(J,U(1))$ ambiguity, which physically corresponds to the stacking of $J$-SPT phases \cite{Barkeshli:2014cna}. Note that Theorem \ref{theorem - gauging a finite symmetry with weak action in DW theory} did not explicitly go through the standard defectification and equivariantization procedure in $(2+1)$D. The $(2+1)$D theory constructed from directly taking the representation categories of the orbifold algebra $\widehat{D^J(G)}$ should be understood as an SET phase with no additional $J$-SPT stacked.
     \end{itemize}

\section{Clifford's Theory and Character Table}
\label{appendix - Mackey's Theory}

Before introducing the computational theorem in Clifford's theory, it is helpful to introduce a a broader mathematical context which will be later restricted to Clifford's theory. If one is interested in studying relation between the representations of a group $G$ and the representations of its subgroups, one needs to invoke \textbf{\textit{Mackey's theory}}. Mackey's theory in its full general form can be applied to any locally compact separable topological groups and it has a beautiful relation with von Neumann algebra. We refer the interested readers to \cite{mackey1976theory} for further details.

Here we restrict our discussion to finite groups. When we are only interested in the relation between the representations of $G$ and the representations of a normal subgroup $N \triangleleft G$, Mackey's theory is effectively reduced to the \textbf{\textit{Clifford's theory}}. We can naturally display the groups of interest in a short exact sequence:
\begin{equation}\label{eq - group extension}
    1 \rightarrow N \rightarrow G \rightarrow H \rightarrow 1
\end{equation}
where $H \simeq G/\iota(N)$ and $\iota: N \hookrightarrow G$. When this short exact sequence splits and $N$ is abelian, Clifford's theory is reduced to the little group method, which constructs $G$ representations in terms of the data of $N$ and $H$. 

In Sec. \ref{subsection - Clifford theory theorems}, we will review some theorems that allow us to reconstruct the character table of $G$ in Eq. \eqref{eq - group extension} when the short exact sequence splits and both $N$ and $H$ are abelian. As an example, we will construct the character table for the Heisenberg group $H_3(\mathbb{Z}_p)$ in Sec. \ref{subsection - Heisenberg group character table}. The rest of this introduction is dedicated to some elementary definition. The main reference of this appendix is \cite{ceccherini2022representation}.

For any finite group $G$, let $\widehat{G}$ denote the collection of all its irreducible representations. 
For any $\sigma \in \widehat{N}$, denote the collection of $G$ irreducible representations whose restriction to $N$ contains $\sigma$ as:
\begin{equation}
    \widehat{G}(\sigma) \equiv\{ \theta \in \widehat{G}\,|\, \sigma \text{ is contained in } \text{Res}^G_N (\theta)\} \equiv \{\theta \in \widehat{G}: \theta \text{ is contained in } \text{Ind}_N^G(\sigma) \}.
\end{equation}
Since $N \triangleleft G$, $N$ is invariant under $G$ conjugation. This induces a $G$ action on $\sigma \in \widehat{N}$ for all $n\in N$:
\begin{equation}
    {}^g\!\sigma(n) \equiv \sigma(g^{-1} n g)
\end{equation}
For any $\sigma \in \widehat{N}$, the collection of $g\in G$ whose action on $\sigma$ leaves $\sigma$ invariant  up to isomorphism is called the \textit{\textbf{inertia group}} $I_G(\sigma)$ of $\sigma \in \widehat{N}$:
\begin{equation}
    I_G(\sigma) \equiv \{ g\in G \,|\, {}^g\!\sigma \sim \sigma  \},
\end{equation}
which is a subgroup of $G$. In fact, the inertia group is just the stabilizer of $\sigma$ in $G$. Define the set:
\begin{equation}
    \widehat{I}_G(\sigma) \equiv \{\psi\in \widehat{I_G(\sigma)}\,|\, \psi \text{ is contained in } \text{Ind}_N^{I_G(\sigma)}\sigma \}.
\end{equation}
We also define the quotient:
\begin{equation}
    H_G(\sigma)\equiv I_G(\sigma)/N,
\end{equation}
which is a subgroup of $H \simeq G/N$.

These new structures allow us to define useful partitions. Since $H_G(\sigma)$ is a subgroup of $H$, we can define a partition of $H$ into left $H_G(\sigma)$ cosets. Label the set of representatives of this partition as $\mathcal{R} = \mathcal{R}(\sigma)$. The $G$-action on $\widehat{N}$ defines an equivalence relation. Namely $\sigma_1 \approx \sigma_2$ if there exists $g\in G$ such that ${}^g\!\sigma_1\sim\sigma_2$. Let $\Sigma$ be a set of representatives for the quotient space $\widehat{N}/\approx$, then we have a partition of $\widehat{N}$ into $G$ orbits labeled by $\sigma \in 
\Sigma$:
\begin{equation}
    \widehat{N} = \sqcup_{\sigma\in\Sigma} \{ {}^r\!\sigma \,|\, r\in 
    \mathcal{R}(\sigma)\} 
\end{equation}

Finally, we mention two important operations on constructing $G$-representations from $H$-representations and a generic subgroup $K$ of $G$.
If $\psi$ is an $H$-representation, then its \textit{\textbf{inflation}} in $G$ is a $G$-representation defined by:
\begin{equation}
    \overset{\Delta}{\psi}\equiv \psi(gN) = \psi(\pi(g)),\quad \forall\, g\in G,
\end{equation}
where $\pi: G \rightarrow H$ is the projection map in the group extension. Now let $K$ be any subgroup of $G$ and consider $\sigma\in \widehat{K}$. An \textbf{extension} of $\sigma$ to $G$ is a representation $\tilde{\sigma}\in\widehat{G}$ so that $\text{Res}^G_K\tilde{\sigma} = \sigma$. Note that $\tilde{\sigma}$ has the same dimension as $\sigma$ and its existence in general is not guaranteed.

    \subsection{Theorems from Clifford's Theory}\label{subsection - Clifford theory theorems}

    In this subsection, we summarize some useful theorems from Clifford's theory. Let us start with a useful theorem applicable to the extension in Eq. \eqref{eq - group extension}, where $N$ and $H$ need not be abelian. For any $\sigma\in \widehat{N}$ and $\theta \in 
    \widehat{G}(\sigma)$, define the \textbf{inertia index} of $\theta$ with respect to $N$ as:
    \begin{equation}
        l_{\theta} \equiv \dim \text{Hom}_N(\sigma, \text{Res}^G_N\theta),
    \end{equation}
    which is the multiplicity of $\theta$ in $\text{Res}^G_N(\theta)$. We have the \textbf{\textit{Clifford correspondence}}:
    \begin{theorem}
        Let $\sigma\in \widehat{N}$, the maps:
    \begin{equation}
        \begin{split}
            \widehat{I}_G(\sigma) &\rightarrow \widehat{G}(\sigma)\\
            \eta & \mapsto \text{Ind}^G_{I_G(\sigma)}\eta
        \end{split}
    \end{equation}
    are bijections. Furthermore:
    \begin{equation}
        l_{\eta} = l_{\text{Ind}^G_{I_G(\sigma)}\eta},
    \end{equation}
    which means that the inertia index of $\eta\in \widehat{I}_G(\sigma)$ with respect to $N$ equals the inertia index of $\text{Ind}^G_{I_G(\sigma)}\eta$ with respect to $N$. Also we have the isomorphism of $N$-representations:
    \begin{equation}
        \text{Res}_N^{I_G(\sigma)}\eta \sim l_{\eta}\sigma.
    \end{equation}
    \end{theorem}
    Therefore, we can deduce all the irreducible $G$-representations by computing the $\widehat{G}(\sigma)$'s by the Clifford correspondence for all $\sigma \in \widehat{N}$. The set of theoretic unions of all the results equals the set $\widehat{G}$\footnote{Note that $\widehat{G}(\sigma_1)$ and $\widehat{G}(\sigma_2)$ for two non-isomorphic $\sigma_1$, $\sigma_2$ might have nontrivial intersections. This is why we insist on taking the set theoretic union, which eliminates the duplicates by default.}. We refer the readers to \cite{ceccherini2022representation} for further details of this algorithm.

    Now we consider a special case of the extension Eq. \eqref{eq - group extension}  which corresponds to a generalized version of the little group method:
    \begin{theorem}
        When $I_G(\sigma) = G$, $I_G(\sigma) = N$, $H_G(\sigma)$ is abelian and $l_{\theta}=1$, we have:
        \begin{itemize}
            \item Let $\sigma\in \widehat{N}$ and suppose it has an extension $\tilde{\sigma}$ to $I_G(\sigma)$, then:
            \begin{equation}
                \widehat{G}(\sigma) = \{ \text{Ind}^G_{I_G(\sigma)} (\tilde{\sigma} \otimes \overset{\Delta}{\psi}) \,|\, \psi \in \widehat{H_G(\sigma)} \},
            \end{equation}
            where $\overset{\Delta}{\psi}$ denotes the inflation of $\psi$ to $I_G(\sigma)$. This is a refinement of the Clifford correspondence in this special case.
            \item Suppose further that every $\sigma\in\widehat{N}$ has an extension $\tilde{\sigma}$ to $I_G(\sigma)$. Let $\Sigma\subseteq N$ be a set of representatives of $G$ orbits on $\widehat{N}$, then:
            \begin{equation}
                \widehat{G} = \{ \text{Ind}^G_{I_G(\sigma)}(\tilde{\sigma }\otimes \overset{\Delta}{\psi})\, |\, \sigma\in \Sigma\, ,\psi\in \widehat{H_G(\sigma)} \}.
            \end{equation}
            Namely, there is a one-to-one correspondence between the irreducible $\widehat{G}$-representations and the pair $(\sigma,\psi)$. 
        \end{itemize}
    \end{theorem}

    If $G$ fits in a split extension and the normal subgroup is abelian, the above theorem becomes the typical little group method for constructing the representation theory of the \Poincare group:
    \begin{theorem}
        Let $G = B \rtimes H$ with $B$ abelian. For any $\xi\in \widehat{B}$, we have $I_G(\xi) = B\rtimes H_{\xi}$, where $H_{\xi} = \{ h\in H\, |\, {}^h\xi = \xi \}$. Any $\xi \in \widehat{B}$ can be extended to a one dimensional representation $\tilde{\xi}$ of $B\rtimes H_{\xi}$ by setting $\tilde{\xi}(ah) \equiv \xi(a)$ for all $a\in B$ and $h\in H_{\xi}$. In terms of the previous theorem, we can label all the irreducible $G$-representations as:
        \begin{equation}
                \widehat{G} = \{ \text{Ind}^G_{B\rtimes H_{\xi}}(\tilde{\xi}\otimes \overset{\Delta}{\psi})\, |\, \xi\in \Sigma\, ,\psi\in \widehat{H_G(\xi)} \}.
            \end{equation}
    \end{theorem}

    Finally, consider $G = B \rtimes A$ for both $A$ and $B$ abelian. In this case, not only can the irreducible $G$-representations be constructed from the little group method, we can also write down an explicit character formula for the irreducible $G$-representations in terms of the data of $A$ and $B$: 
    \begin{theorem}\label{theorem - character table}
        Let $\Xi$ be a set of representative of $A$-orbits on $\widehat{B}$. For $\xi\in \Xi$, define the stabilizer group $A_{\xi} = \{a\in A \, | \, {}^a\!\xi = \xi\}$. $A_{\xi}$ induces a partition of $A$:
        \begin{equation}
            A = \bigsqcup_{r \in R_{\xi}}rA_{\xi}.
        \end{equation}
        Then all the irreducible representations of $G$ are labeled by:
        \begin{equation}
            \widehat{G} = \{\theta =  \text{Ind}^G_{B\rtimes A_{\xi}}(\tilde{\xi}\otimes \overset{\Delta}{\psi})\, |\, \xi\in \Xi\, ,\psi\in \widehat{H_G(\xi)} \}.
        \end{equation}
        For $G = B\rtimes A$, using the fact that elements in $G$ as a set can be expressed as $ba$ where $b\in B$ and $a\in A$, the character of $\theta$ is given by:
        \begin{equation}
            \chi^{\theta}(ba) = \begin{cases}
                \psi(a)\left(\sum_{r\in\mathcal{R}_{\xi}} {}^r\xi(b)\right),  &\text{if } a\in A_{\xi}; \\
                 0,  &\text{otherwise.}
            \end{cases}
        \end{equation}
    \end{theorem}

\subsection{\texorpdfstring{$H_3(\mathbb{Z}_p)$}{H3(Zp)} Character Table}\label{subsection - Heisenberg group character table}

    In this subsection, we construct the character table of the Heisenberg group $H_3(\mathbb{Z}_p)$ with Theorem \ref{theorem - character table}.  The conjugacy classes were already constructed in the main text. The evaluations of the characters are trivial, so here we will only outline the identification of irreducible representations of $H_3(\mathbb{Z}_p)$. 
    
    Recall that $H_3(\mathbb{Z}_p)\simeq (\mathbb{Z}_p\times \mathbb{Z}_p)\rtimes\mathbb{Z}_p$, where the twist action on $\mathbb{Z}_p\times\mathbb{Z}_p$ was given in the main text. Let $(\chi_a, \chi_b)$ be a generic irreducible representation of $\mathbb{Z}_p\times\mathbb{Z}_p$, where $a$ and $b$ take values in integer mod $p$. The twist induces an action on $\widehat{\mathbb{Z}}_p\times\widehat{\mathbb{Z}}_p$ generated by:
    \begin{equation}
        (\chi_a, \chi_b) \mapsto 
         (\chi_{a-b}, \chi_b )
    \end{equation}
    Since $p$ is a prime number, the $\mathbb{Z}_p$ orbits on $\widehat{\mathbb{Z}}_p\times\widehat{\mathbb{Z}}_p$ is either a singlet or a size-$p$ orbit. The singlets are the trivial representation and $(\chi_a, \chi_0)$ where $a\in \{0,1,\dots, p-1\}$. The size-$p$ orbits are labeled by $(\chi_a, \chi_b)$ for each $b\in\{1,2,\dots, p-1\}$.

    By Theorem \ref{theorem - character table}, given a collection of representatives of the orbits $\Xi$, we need to compute the stabilizer subgroup of $\mathbb{Z}_p$ for each $\xi \in \Xi$. For $H_3(\mathbb{Z}_p)$, the stabilizer is the full $\mathbb{Z}_p$ for size-1 orbits and the trivial group for the size-$p$ orbits. Let $\omega_n$ denote the $\mathbb{Z}_p$ irreducible representations, then all the $H_3(\mathbb{Z}_p)$ irreducible representations are classified by:
    \begin{itemize}
        \item $T^{\omega_n} \equiv \text{Ind}^{H_3(\mathbb{Z}_p)}_{H_3(\mathbb{Z}_p)} (\tilde{T}\otimes \overset{\Delta}{\omega_n})$. There are $p$ irreducible representations of this type.
        \item $(\chi_a,\chi_0)^{\omega_n}\equiv \text{Ind}^{H_3(\mathbb{Z}_p)}_{H_3(\mathbb{Z}_p)}\left( \widetilde{(\chi_a,\chi_0)}\otimes \overset{\Delta}{\omega_n}   \right)$. There are $p(p-1)$ irreducible representations of this type. 
        \item $(\chi_a,\chi_b)\equiv \text{Ind}^{H_3(\mathbb{Z}_p)}_{\mathbb{Z}_p\times\mathbb{Z}_p} \left((\chi_a, \chi_b)\otimes \overset{\Delta}{\omega_0}\right) $. There are $p-1$ irreducible representations of this type labeled by $b\in \{1,2,\dots, p-1\}$ and ``$a$" is a representative of a $\mathbb{Z}_p$ orbit on $\widehat{\mathbb{Z}}_p\times \widehat{\mathbb{Z}}_p$
    \end{itemize}
        The dimension of these irreducible representations can be computed by applying the character formula over the identity element of $H_3(\mathbb{Z}_p)$. The first two types of irreducible representations are 1-dimensional and the remaining irreducible representations are $p$-dimensional. 
    

\end{appendix}

\bibliographystyle{JHEP}
\bibliography{biblio.bib}

@article{Barkeshli:2014cna,
    author = "Barkeshli, Maissam and Bonderson, Parsa and Cheng, Meng and Wang, Zhenghan",
    title = "{Symmetry Fractionalization, Defects, and Gauging of Topological Phases}",
    eprint = "1410.4540",
    archivePrefix = "arXiv",
    primaryClass = "cond-mat.str-el",
    doi = "10.1103/PhysRevB.100.115147",
    journal = "Phys. Rev. B",
    volume = "100",
    number = "11",
    pages = "115147",
    year = "2019"
}

@article{Roumpedakis:2022aik,
    author = "Roumpedakis, Konstantinos and Seifnashri, Sahand and Shao, Shu-Heng",
    title = "{Higher Gauging and Non-invertible Condensation Defects}",
    eprint = "2204.02407",
    archivePrefix = "arXiv",
    primaryClass = "hep-th",
    reportNumber = "YITP-SB-2022-14",
    doi = "10.1007/s00220-023-04706-9",
    journal = "Commun. Math. Phys.",
    volume = "401",
    number = "3",
    pages = "3043--3107",
    year = "2023"
}

@article{Cordova:2024jlk,
    author = "Cordova, Clay and Costa, Davi Bastos and Hsin, Po-Shen",
    title = "{Non-invertible symmetries in finite-group gauge theory}",
    eprint = "2407.07964",
    archivePrefix = "arXiv",
    primaryClass = "cond-mat.str-el",
    doi = "10.21468/SciPostPhys.18.1.019",
    journal = "SciPost Phys.",
    volume = "18",
    number = "1",
    pages = "019",
    year = "2025"
}

@article{Cordova:2024mqg,
    author = "Cordova, Clay and Costa, Davi B. and Hsin, Po-Shen",
    title = "{Non-Invertible Symmetries as Condensation Defects in Finite-Group Gauge Theories}",
    eprint = "2412.16681",
    archivePrefix = "arXiv",
    primaryClass = "cond-mat.str-el",
    month = "12",
    year = "2024"
}

@article{Bergman:2024its,
    author = "Bergman, Oren and Mignosa, Francesco",
    title = "{String theory and the SymTFT of 3d orthosymplectic Chern-Simons theory}",
    eprint = "2412.00184",
    archivePrefix = "arXiv",
    primaryClass = "hep-th",
    doi = "10.1007/JHEP04(2025)047",
    journal = "JHEP",
    volume = "04",
    pages = "047",
    year = "2025"
}

@article{Brennan:2023mmt,
    author = "Brennan, T. Daniel and Hong, Sungwoo",
    title = "{Introduction to Generalized Global Symmetries in QFT and Particle Physics}",
    eprint = "2306.00912",
    archivePrefix = "arXiv",
    primaryClass = "hep-ph",
    month = "6",
    year = "2023"
}

@article{Lan:2014uaa,
    author = "Lan, Tian and Wang, Juven C. and Wen, Xiao-Gang",
    title = "{Gapped Domain Walls, Gapped Boundaries and Topological Degeneracy}",
    eprint = "1408.6514",
    archivePrefix = "arXiv",
    primaryClass = "cond-mat.str-el",
    doi = "10.1103/PhysRevLett.114.076402",
    journal = "Phys. Rev. Lett.",
    volume = "114",
    number = "7",
    pages = "076402",
    year = "2015"
}

@article{He:2016xpi,
    author = "He, Huan and Zheng, Yunqin and von Keyserlingk, Curt",
    title = "{Field theories for gauged symmetry-protected topological phases: Non-Abelian anyons with Abelian gauge group $\mathbb Z_2^{\otimes 3}$}",
    eprint = "1608.05393",
    archivePrefix = "arXiv",
    primaryClass = "cond-mat.str-el",
    doi = "10.1103/PhysRevB.95.035131",
    journal = "Phys. Rev. B",
    volume = "95",
    number = "3",
    pages = "035131",
    year = "2017"
}

@article{Kaidi:2022cpf,
    author = "Kaidi, Justin and Ohmori, Kantaro and Zheng, Yunqin",
    title = "{Symmetry TFTs for Non-invertible Defects}",
    eprint = "2209.11062",
    archivePrefix = "arXiv",
    primaryClass = "hep-th",
    doi = "10.1007/s00220-023-04859-7",
    journal = "Commun. Math. Phys.",
    volume = "404",
    number = "2",
    pages = "1021--1124",
    year = "2023"
}

@article{Dijkgraaf:1989pz,
    author = "Dijkgraaf, Robbert and Witten, Edward",
    title = "{Topological Gauge Theories and Group Cohomology}",
    reportNumber = "THU-89-9, IASSNS-HEP-89-33",
    doi = "10.1007/BF02096988",
    journal = "Commun. Math. Phys.",
    volume = "129",
    pages = "393",
    year = "1990"
}

@article{Coste:2000tq,
    author = "Coste, Antoine and Gannon, Terry and Ruelle, Philippe",
    title = "{Finite group modular data}",
    eprint = "hep-th/0001158",
    archivePrefix = "arXiv",
    doi = "10.1016/S0550-3213(00)00285-6",
    journal = "Nucl. Phys. B",
    volume = "581",
    pages = "679--717",
    year = "2000"
}

@article{Bhardwaj:2017xup,
    author = "Bhardwaj, Lakshya and Tachikawa, Yuji",
    title = "{On finite symmetries and their gauging in two dimensions}",
    eprint = "1704.02330",
    archivePrefix = "arXiv",
    primaryClass = "hep-th",
    reportNumber = "IPMU-17-0049",
    doi = "10.1007/JHEP03(2018)189",
    journal = "JHEP",
    volume = "03",
    pages = "189",
    year = "2018"
}

@article{Gaiotto:2014kfa,
    author = "Gaiotto, Davide and Kapustin, Anton and Seiberg, Nathan and Willett, Brian",
    title = "{Generalized Global Symmetries}",
    eprint = "1412.5148",
    archivePrefix = "arXiv",
    primaryClass = "hep-th",
    doi = "10.1007/JHEP02(2015)172",
    journal = "JHEP",
    volume = "02",
    pages = "172",
    year = "2015"
}

@article{Bah:2025oxi,
    author = "Bah, Ibrahima and Leung, Enoch and Waddleton, Thomas",
    title = "{On the Physics of Higher Condensation Defects}",
    eprint = "2506.04346",
    archivePrefix = "arXiv",
    primaryClass = "hep-th",
    month = "6",
    year = "2025"
}

@article{Bhardwaj:2022yxj,
    author = "Bhardwaj, Lakshya and Bottini, Lea E. and Schafer-Nameki, Sakura and Tiwari, Apoorv",
    title = "{Non-invertible higher-categorical symmetries}",
    eprint = "2204.06564",
    archivePrefix = "arXiv",
    primaryClass = "hep-th",
    doi = "10.21468/SciPostPhys.14.1.007",
    journal = "SciPost Phys.",
    volume = "14",
    number = "1",
    pages = "007",
    year = "2023"
}

@article{Schafer-Nameki:2023jdn,
    author = "Schafer-Nameki, Sakura",
    title = "{ICTP lectures on (non-)invertible generalized symmetries}",
    eprint = "2305.18296",
    archivePrefix = "arXiv",
    primaryClass = "hep-th",
    doi = "10.1016/j.physrep.2024.01.007",
    journal = "Phys. Rept.",
    volume = "1063",
    pages = "1--55",
    year = "2024"
}

@inproceedings{Shao:2023gho,
    author = "Shao, Shu-Heng",
    title = "{What's Done Cannot Be Undone: TASI Lectures on Non-Invertible Symmetries}",
    booktitle = "{Theoretical Advanced Study Institute in Elementary Particle Physics 2023}: {Aspects of Symmetry}",
    eprint = "2308.00747",
    archivePrefix = "arXiv",
    primaryClass = "hep-th",
    reportNumber = "YITP-SB-2023-19",
    month = "8",
    year = "2023"
}

@article{Fuchs:2002cm,
    author = "Fuchs, Jurgen and Runkel, Ingo and Schweigert, Christoph",
    title = "{TFT construction of RCFT correlators 1. Partition functions}",
    eprint = "hep-th/0204148",
    archivePrefix = "arXiv",
    reportNumber = "PAR-LPTHE-02-25",
    doi = "10.1016/S0550-3213(02)00744-7",
    journal = "Nucl. Phys. B",
    volume = "646",
    pages = "353--497",
    year = "2002"
}

@article{Fuchs:2003id,
    author = "Fuchs, Jurgen and Runkel, Ingo and Schweigert, Christoph",
    title = "{TFT construction of RCFT correlators. 2. Unoriented world sheets}",
    eprint = "hep-th/0306164",
    archivePrefix = "arXiv",
    reportNumber = "HU-EP-03-23",
    doi = "10.1016/j.nuclphysb.2003.11.026",
    journal = "Nucl. Phys. B",
    volume = "678",
    pages = "511--637",
    year = "2004"
}

@article{Fuchs:2004dz,
    author = "Fuchs, Jurgen and Runkel, Ingo and Schweigert, Christoph",
    title = "{TFT construction of RCFT correlators. 3. Simple currents}",
    eprint = "hep-th/0403157",
    archivePrefix = "arXiv",
    reportNumber = "HU-EP-04-12, LPTHE-04-05, STR-02-042",
    doi = "10.1016/j.nuclphysb.2004.05.014",
    journal = "Nucl. Phys. B",
    volume = "694",
    pages = "277--353",
    year = "2004"
}

@article{Fuchs:2004xi,
    author = "Fuchs, Jurgen and Runkel, Ingo and Schweigert, Christoph",
    title = "{TFT construction of RCFT correlators IV: Structure constants and correlation functions}",
    eprint = "hep-th/0412290",
    archivePrefix = "arXiv",
    reportNumber = "AEI-2004-124",
    doi = "10.1016/j.nuclphysb.2005.03.018",
    journal = "Nucl. Phys. B",
    volume = "715",
    pages = "539--638",
    year = "2005"
}

@article{Fjelstad:2005ua,
    author = "Fjelstad, Jens and Fuchs, Jurgen and Runkel, Ingo and Schweigert, Christoph",
    title = "{TFT construction of RCFT correlators. V. Proof of modular invariance and factorisation}",
    eprint = "hep-th/0503194",
    archivePrefix = "arXiv",
    reportNumber = "AEI-2005-037, LPTHE-05-07",
    journal = "Theor. Appl. Categor.",
    volume = "16",
    pages = "342--433",
    year = "2006"
}

@inproceedings{Moore:1989vd,
    author = "Moore, Gregory W. and Seiberg, Nathan",
    title = "{Lectures on RCFT}",
    booktitle = "{1989 Banff NATO ASI: Physics, Geometry and Topology}",
    reportNumber = "RU-89-32, YCTP-P13-89",
    pages = "1--129",
    month = "9",
    year = "1989"
}

@article{Choi:2022zal,
    author = "Choi, Yichul and Cordova, Clay and Hsin, Po-Shen and Lam, Ho Tat and Shao, Shu-Heng",
    title = "{Non-invertible Condensation, Duality, and Triality Defects in 3+1 Dimensions}",
    eprint = "2204.09025",
    archivePrefix = "arXiv",
    primaryClass = "hep-th",
    reportNumber = "YITP-SB-2022-16, MIT/CTP-5423, YITP-SB-2022-16, MIT/CTP-5423",
    doi = "10.1007/s00220-023-04727-4",
    journal = "Commun. Math. Phys.",
    volume = "402",
    number = "1",
    pages = "489--542",
    year = "2023"
}

@article{Choi:2021kmx,
    author = "Choi, Yichul and Cordova, Clay and Hsin, Po-Shen and Lam, Ho Tat and Shao, Shu-Heng",
    title = "{Noninvertible duality defects in 3+1 dimensions}",
    eprint = "2111.01139",
    archivePrefix = "arXiv",
    primaryClass = "hep-th",
    reportNumber = "MIT-CTP/5359",
    doi = "10.1103/PhysRevD.105.125016",
    journal = "Phys. Rev. D",
    volume = "105",
    number = "12",
    pages = "125016",
    year = "2022"
}

@article{Choi:2023pdp,
    author = "Choi, Yichul and Forslund, Matthew and Lam, Ho Tat and Shao, Shu-Heng",
    title = "{Quantization of Axion-Gauge Couplings and Noninvertible Higher Symmetries}",
    eprint = "2309.03937",
    archivePrefix = "arXiv",
    primaryClass = "hep-ph",
    reportNumber = "YITP-SB-2023-27, MIT-CTP/5606",
    doi = "10.1103/PhysRevLett.132.121601",
    journal = "Phys. Rev. Lett.",
    volume = "132",
    number = "12",
    pages = "121601",
    year = "2024"
}

@article{Choi:2022fgx,
    author = "Choi, Yichul and Lam, Ho Tat and Shao, Shu-Heng",
    title = "{Non-invertible Gauss law and axions}",
    eprint = "2212.04499",
    archivePrefix = "arXiv",
    primaryClass = "hep-th",
    reportNumber = "MIT-CTP/5504, YITP-SB-2022-39",
    doi = "10.1007/JHEP09(2023)067",
    journal = "JHEP",
    volume = "09",
    pages = "067",
    year = "2023"
}

@article{Choi:2022rfe,
    author = "Choi, Yichul and Lam, Ho Tat and Shao, Shu-Heng",
    title = "{Noninvertible Time-Reversal Symmetry}",
    eprint = "2208.04331",
    archivePrefix = "arXiv",
    primaryClass = "hep-th",
    reportNumber = "YITP-SB-2022-28, MIT-CTP/5457",
    doi = "10.1103/PhysRevLett.130.131602",
    journal = "Phys. Rev. Lett.",
    volume = "130",
    number = "13",
    pages = "131602",
    year = "2023"
}

@article{Choi:2022jqy,
    author = "Choi, Yichul and Lam, Ho Tat and Shao, Shu-Heng",
    title = "{Noninvertible Global Symmetries in the Standard Model}",
    eprint = "2205.05086",
    archivePrefix = "arXiv",
    primaryClass = "hep-th",
    reportNumber = "YITP-SB-2022-21, MIT-CTP/5433",
    doi = "10.1103/PhysRevLett.129.161601",
    journal = "Phys. Rev. Lett.",
    volume = "129",
    number = "16",
    pages = "161601",
    year = "2022"
}

@article{Apruzzi:2021nmk,
    author = "Apruzzi, Fabio and Bonetti, Federico and Garc{\'\i}a Etxebarria, I{\~n}aki and Hosseini, Saghar S. and Schafer-Nameki, Sakura",
    title = "{Symmetry TFTs from String Theory}",
    eprint = "2112.02092",
    archivePrefix = "arXiv",
    primaryClass = "hep-th",
    doi = "10.1007/s00220-023-04737-2",
    journal = "Commun. Math. Phys.",
    volume = "402",
    number = "1",
    pages = "895--949",
    year = "2023"
}

@article{Apruzzi:2022rei,
    author = "Apruzzi, Fabio and Bah, Ibrahima and Bonetti, Federico and Schafer-Nameki, Sakura",
    title = "{Noninvertible Symmetries from Holography and Branes}",
    eprint = "2208.07373",
    archivePrefix = "arXiv",
    primaryClass = "hep-th",
    doi = "10.1103/PhysRevLett.130.121601",
    journal = "Phys. Rev. Lett.",
    volume = "130",
    number = "12",
    pages = "121601",
    year = "2023"
}

@article{Apruzzi:2023uma,
    author = "Apruzzi, Fabio and Bonetti, Federico and Gould, Dewi S. W. and Schafer-Nameki, Sakura",
    title = "{Aspects of categorical symmetries from branes: SymTFTs and generalized charges}",
    eprint = "2306.16405",
    archivePrefix = "arXiv",
    primaryClass = "hep-th",
    doi = "10.21468/SciPostPhys.17.1.025",
    journal = "SciPost Phys.",
    volume = "17",
    number = "1",
    pages = "025",
    year = "2024"
}

@article{Lu:2024ytl,
    author = "Lu, Da-Chuan and Sun, Zhengdi and You, Yi-Zhuang",
    title = "{Realizing triality and $p$-ality by lattice twisted gauging in (1+1)d quantum spin systems}",
    eprint = "2405.14939",
    archivePrefix = "arXiv",
    primaryClass = "cond-mat.str-el",
    doi = "10.21468/SciPostPhys.17.5.136",
    journal = "SciPost Phys.",
    volume = "17",
    number = "5",
    pages = "136",
    year = "2024"
}

@article{Seifnashri:2025fgd,
    author = "Seifnashri, Sahand and Shao, Shu-Heng and Yang, Xinping",
    title = "{Gauging non-invertible symmetries on the lattice}",
    eprint = "2503.02925",
    archivePrefix = "arXiv",
    primaryClass = "cond-mat.str-el",
    reportNumber = "MIT-CTP/5842, YITP-SB-2025-03",
    doi = "10.21468/SciPostPhys.19.2.063",
    journal = "SciPost Phys.",
    volume = "19",
    number = "2",
    pages = "063",
    year = "2025"
}

@article{Gorantla:2024ocs,
    author = "Gorantla, Pranay and Shao, Shu-Heng and Tantivasadakarn, Nathanan",
    title = "{Tensor Networks for Noninvertible Symmetries in 3+1D and Beyond}",
    eprint = "2406.12978",
    archivePrefix = "arXiv",
    primaryClass = "quant-ph",
    reportNumber = "YITP-SB-2024-11",
    doi = "10.1103/p32z-v884",
    journal = "Phys. Rev. X",
    volume = "15",
    number = "4",
    pages = "041006",
    year = "2025"
}

@article{Choi:2024rjm,
    author = "Choi, Yichul and Sanghavi, Yaman and Shao, Shu-Heng and Zheng, Yunqin",
    title = "{Non-invertible and higher-form symmetries in 2+1d lattice gauge theories}",
    eprint = "2405.13105",
    archivePrefix = "arXiv",
    primaryClass = "cond-mat.str-el",
    doi = "10.21468/SciPostPhys.18.1.008",
    journal = "SciPost Phys.",
    volume = "18",
    number = "1",
    pages = "008",
    year = "2025"
}

@article{Seiberg:2024gek,
    author = "Seiberg, Nathan and Seifnashri, Sahand and Shao, Shu-Heng",
    title = "{Non-invertible symmetries and LSM-type constraints on a tensor product Hilbert space}",
    eprint = "2401.12281",
    archivePrefix = "arXiv",
    primaryClass = "cond-mat.str-el",
    reportNumber = "YITP-SB-2024-01",
    doi = "10.21468/SciPostPhys.16.6.154",
    journal = "SciPost Phys.",
    volume = "16",
    pages = "154",
    year = "2024"
}

@article{Bhardwaj:2022maz,
    author = "Bhardwaj, Lakshya and Bottini, Lea E. and Schafer-Nameki, Sakura and Tiwari, Apoorv",
    title = "{Non-invertible symmetry webs}",
    eprint = "2212.06842",
    archivePrefix = "arXiv",
    primaryClass = "hep-th",
    doi = "10.21468/SciPostPhys.15.4.160",
    journal = "SciPost Phys.",
    volume = "15",
    number = "4",
    pages = "160",
    year = "2023"
}

@article{Bhardwaj:2022kot,
    author = "Bhardwaj, Lakshya and Schafer-Nameki, Sakura and Tiwari, Apoorv",
    title = "{Unifying constructions of non-invertible symmetries}",
    eprint = "2212.06159",
    archivePrefix = "arXiv",
    primaryClass = "hep-th",
    doi = "10.21468/SciPostPhys.15.3.122",
    journal = "SciPost Phys.",
    volume = "15",
    number = "3",
    pages = "122",
    year = "2023"
}

@article{Lu:2024lzf,
    author = "Lu, Da-Chuan and Sun, Zhengdi and Zhang, Zipei",
    title = "{Exploring G-ality defects in 2-dim QFTs}",
    eprint = "2406.12151",
    archivePrefix = "arXiv",
    primaryClass = "hep-th",
    doi = "10.1007/JHEP11(2025)081",
    journal = "JHEP",
    volume = "11",
    pages = "081",
    year = "2025"
}

@article{Lu:2022ver,
    author = "Lu, Da-Chuan and Sun, Zhengdi",
    title = "{On triality defects in 2d CFT}",
    eprint = "2208.06077",
    archivePrefix = "arXiv",
    primaryClass = "hep-th",
    doi = "10.1007/JHEP02(2023)173",
    journal = "JHEP",
    volume = "02",
    pages = "173",
    year = "2023"
}

@inproceedings{Costa:2024wks,
    author = "Costa, Davi and others",
    title = "{Simons Lectures on Categorical Symmetries}",
    eprint = "2411.09082",
    archivePrefix = "arXiv",
    primaryClass = "math-ph",
    month = "11",
    year = "2024"
}

@article{Chang:2018iay,
    author = "Chang, Chi-Ming and Lin, Ying-Hsuan and Shao, Shu-Heng and Wang, Yifan and Yin, Xi",
    title = "{Topological Defect Lines and Renormalization Group Flows in Two Dimensions}",
    eprint = "1802.04445",
    archivePrefix = "arXiv",
    primaryClass = "hep-th",
    reportNumber = "CALT-TH-2017-067, CALT-TH 2017-067, PUPT-2546",
    doi = "10.1007/JHEP01(2019)026",
    journal = "JHEP",
    volume = "01",
    pages = "026",
    year = "2019"
}

@article{Brennan:2020ehu,
    author = "Brennan, T. Daniel and Cordova, Clay",
    title = "{Axions, higher-groups, and emergent symmetry}",
    eprint = "2011.09600",
    archivePrefix = "arXiv",
    primaryClass = "hep-th",
    doi = "10.1007/JHEP02(2022)145",
    journal = "JHEP",
    volume = "02",
    pages = "145",
    year = "2022"
}

@article{Cordova:2022qtz,
    author = "Cordova, Clay and Koren, Seth",
    title = "{Higher Flavor Symmetries in the Standard Model}",
    eprint = "2212.13193",
    archivePrefix = "arXiv",
    primaryClass = "hep-ph",
    doi = "10.1002/andp.202300031",
    journal = "Annalen Phys.",
    volume = "535",
    number = "8",
    pages = "2300031",
    year = "2023"
}

@article{Aloni:2024jpb,
    author = "Aloni, Daniel and Garc{\'\i}a-Valdecasas, Eduardo and Reece, Matthew and Suzuki, Motoo",
    title = "{Spontaneously broken (-1)-form U(1) symmetries}",
    eprint = "2402.00117",
    archivePrefix = "arXiv",
    primaryClass = "hep-th",
    doi = "10.21468/SciPostPhys.17.2.031",
    journal = "SciPost Phys.",
    volume = "17",
    number = "2",
    pages = "031",
    year = "2024"
}

@article{Cordova:2019jqi,
    author = "C{\'o}rdova, Clay and Ohmori, Kantaro",
    title = "{Anomaly Constraints on Gapped Phases with Discrete Chiral Symmetry}",
    eprint = "1912.13069",
    archivePrefix = "arXiv",
    primaryClass = "hep-th",
    doi = "10.1103/PhysRevD.102.025011",
    journal = "Phys. Rev. D",
    volume = "102",
    number = "2",
    pages = "025011",
    year = "2020"
}

@article{Gaiotto:2017yup,
    author = "Gaiotto, Davide and Kapustin, Anton and Komargodski, Zohar and Seiberg, Nathan",
    title = "{Theta, Time Reversal, and Temperature}",
    eprint = "1703.00501",
    archivePrefix = "arXiv",
    primaryClass = "hep-th",
    doi = "10.1007/JHEP05(2017)091",
    journal = "JHEP",
    volume = "05",
    pages = "091",
    year = "2017"
}

@article{Komargodski:2017keh,
    author = "Komargodski, Zohar and Seiberg, Nathan",
    title = "{A symmetry breaking scenario for QCD$_{3}$}",
    eprint = "1706.08755",
    archivePrefix = "arXiv",
    primaryClass = "hep-th",
    doi = "10.1007/JHEP01(2018)109",
    journal = "JHEP",
    volume = "01",
    pages = "109",
    year = "2018"
}

@article{Gaiotto:2017tne,
    author = "Gaiotto, Davide and Komargodski, Zohar and Seiberg, Nathan",
    title = "{Time-reversal breaking in QCD$_{4}$, walls, and dualities in 2 + 1 dimensions}",
    eprint = "1708.06806",
    archivePrefix = "arXiv",
    primaryClass = "hep-th",
    doi = "10.1007/JHEP01(2018)110",
    journal = "JHEP",
    volume = "01",
    pages = "110",
    year = "2018"
}

@article{Gomis:2017ixy,
    author = "Gomis, Jaume and Komargodski, Zohar and Seiberg, Nathan",
    title = "{Phases Of Adjoint QCD$_3$ And Dualities}",
    eprint = "1710.03258",
    archivePrefix = "arXiv",
    primaryClass = "hep-th",
    doi = "10.21468/SciPostPhys.5.1.007",
    journal = "SciPost Phys.",
    volume = "5",
    number = "1",
    pages = "007",
    year = "2018"
}

@article{Komargodski:2020mxz,
    author = "Komargodski, Zohar and Ohmori, Kantaro and Roumpedakis, Konstantinos and Seifnashri, Sahand",
    title = "{Symmetries and strings of adjoint QCD$_{2}$}",
    eprint = "2008.07567",
    archivePrefix = "arXiv",
    primaryClass = "hep-th",
    reportNumber = "YITP-SB-20-28",
    doi = "10.1007/JHEP03(2021)103",
    journal = "JHEP",
    volume = "03",
    pages = "103",
    year = "2021"
}

@article{Brennan:2024iau,
    author = "Brennan, T. Daniel and Grewal, Jaipratap Singh and Yang, Eric Y.",
    title = "{Revisiting Scattering Enhancement from the Aharonov-Bohm Effect}",
    eprint = "2411.10526",
    archivePrefix = "arXiv",
    primaryClass = "hep-th",
    doi = "10.1103/cdns-8bw7",
    journal = "Phys. Rev. Lett.",
    volume = "135",
    number = "2",
    pages = "021601",
    year = "2025"
}

@article{Cordova:2018cvg,
    author = "C{\'o}rdova, Clay and Dumitrescu, Thomas T. and Intriligator, Kenneth",
    title = "{Exploring 2-Group Global Symmetries}",
    eprint = "1802.04790",
    archivePrefix = "arXiv",
    primaryClass = "hep-th",
    doi = "10.1007/JHEP02(2019)184",
    journal = "JHEP",
    volume = "02",
    pages = "184",
    year = "2019"
}

@article{Heidenreich:2021xpr,
    author = "Heidenreich, Ben and McNamara, Jacob and Montero, Miguel and Reece, Matthew and Rudelius, Tom and Valenzuela, Irene",
    title = "{Non-invertible global symmetries and completeness of the spectrum}",
    eprint = "2104.07036",
    archivePrefix = "arXiv",
    primaryClass = "hep-th",
    reportNumber = "ACFI-T21-03",
    doi = "10.1007/JHEP09(2021)203",
    journal = "JHEP",
    volume = "09",
    pages = "203",
    year = "2021"
}

@article{Rudelius:2020orz,
    author = "Rudelius, Tom and Shao, Shu-Heng",
    title = "{Topological Operators and Completeness of Spectrum in Discrete Gauge Theories}",
    eprint = "2006.10052",
    archivePrefix = "arXiv",
    primaryClass = "hep-th",
    doi = "10.1007/JHEP12(2020)172",
    journal = "JHEP",
    volume = "12",
    pages = "172",
    year = "2020"
}

@article{Banks:2010zn,
    author = "Banks, Tom and Seiberg, Nathan",
    title = "{Symmetries and Strings in Field Theory and Gravity}",
    eprint = "1011.5120",
    archivePrefix = "arXiv",
    primaryClass = "hep-th",
    doi = "10.1103/PhysRevD.83.084019",
    journal = "Phys. Rev. D",
    volume = "83",
    pages = "084019",
    year = "2011"
}

@article{Kapustin:2014gua,
    author = "Kapustin, Anton and Seiberg, Nathan",
    title = "{Coupling a QFT to a TQFT and Duality}",
    eprint = "1401.0740",
    archivePrefix = "arXiv",
    primaryClass = "hep-th",
    doi = "10.1007/JHEP04(2014)001",
    journal = "JHEP",
    volume = "04",
    pages = "001",
    year = "2014"
}

@book{Simon:2023hdq,
    author = "Simon, Steven H.",
    title = "{Topological Quantum}",
    isbn = "978-0-19-888672-3",
    publisher = "Oxford University Press",
    month = "9",
    year = "2023"
}

@article{Freed:2022qnc,
    author = "Freed, Daniel S. and Moore, Gregory W. and Teleman, Constantin",
    title = "{Topological symmetry in quantum field theory}",
    eprint = "2209.07471",
    archivePrefix = "arXiv",
    primaryClass = "hep-th",
    month = "9",
    year = "2022"
}

@article{Brennan:2024fgj,
    author = "Brennan, T. Daniel and Sun, Zhengdi",
    title = "{A SymTFT for continuous symmetries}",
    eprint = "2401.06128",
    archivePrefix = "arXiv",
    primaryClass = "hep-th",
    doi = "10.1007/JHEP12(2024)100",
    journal = "JHEP",
    volume = "12",
    pages = "100",
    year = "2024"
}

@article{Apruzzi:2024htg,
    author = "Apruzzi, Fabio and Bedogna, Francesco and Dondi, Nicola",
    title = "{SymTh for non-finite symmetries}",
    eprint = "2402.14813",
    archivePrefix = "arXiv",
    primaryClass = "hep-th",
    month = "2",
    year = "2024"
}

@article{Jia:2025jmn,
    author = "Jia, Qiang and Luo, Ran and Tian, Jiahua and Wang, Yi-Nan and Zhang, Yi",
    title = "{Symmetry Topological Field Theory for Flavor Symmetry}",
    eprint = "2503.04546",
    archivePrefix = "arXiv",
    primaryClass = "hep-th",
    month = "3",
    year = "2025"
}

@article{Antinucci:2024zjp,
    author = "Antinucci, Andrea and Benini, Francesco",
    title = "{Anomalies and gauging of U(1) symmetries}",
    eprint = "2401.10165",
    archivePrefix = "arXiv",
    primaryClass = "hep-th",
    reportNumber = "SISSA 01/2024/FISI",
    doi = "10.1103/PhysRevB.111.024110",
    journal = "Phys. Rev. B",
    volume = "111",
    number = "2",
    pages = "024110",
    year = "2025"
}

@article{Bonetti:2024cjk,
    author = "Bonetti, Federico and Del Zotto, Michele and Minasian, Ruben",
    title = "{SymTFTs for Continuous non-Abelian Symmetries}",
    eprint = "2402.12347",
    archivePrefix = "arXiv",
    primaryClass = "hep-th",
    month = "2",
    year = "2024"
}

@article{Pace:2025hpb,
    author = {Pace, Salvatore D. and Aksoy, {\"O}mer M. and Lam, Ho Tat},
    title = "{Spacetime symmetry-enriched SymTFT: from LSM anomalies to modulated symmetries and beyond}",
    eprint = "2507.02036",
    archivePrefix = "arXiv",
    primaryClass = "cond-mat.str-el",
    reportNumber = "MIT-CTP/5884",
    month = "7",
    year = "2025"
}

@article{Apruzzi:2025hvs,
    author = "Apruzzi, Fabio and Dondi, Nicola and Garc{\'\i}a Etxebarria, I{\~n}aki and Lam, Ho Tat and Schafer-Nameki, Sakura",
    title = "{Symmetry TFTs for Continuous Spacetime Symmetries}",
    eprint = "2509.07965",
    archivePrefix = "arXiv",
    primaryClass = "hep-th",
    reportNumber = "MIT-CTP/5921",
    month = "9",
    year = "2025"
}

@article{Atiyah:1989vu,
    author = "Atiyah, M.",
    title = "{Topological quantum field theories}",
    doi = "10.1007/BF02698547",
    journal = "Inst. Hautes Etudes Sci. Publ. Math.",
    volume = "68",
    pages = "175--186",
    year = "1989"
}

@inproceedings{Lurie2006HigherTT,
  title={Higher Topos Theory},
  author={Jacob Lurie},
  year={2006},
  url={https://api.semanticscholar.org/CorpusID:5338947}
}

@article{Lan2017ClassificationO,
  title={Classification of 
(3+1)D
 Bosonic Topological Orders: The Case When Pointlike Excitations Are All Bosons},
  author={Tian Lan and Liang Kong and Xiao-Gang Wen},
  journal={Physical Review X},
  year={2017},
  url={https://api.semanticscholar.org/CorpusID:113397431}
}

@article{Lan2018ClassificationO,
  title={Classification of 
3+1D
 Bosonic Topological Orders (II): The Case When Some Pointlike Excitations Are Fermions},
  author={Tian Lan and Xiao-Gang Wen},
  journal={Physical Review X},
  year={2018},
  url={https://api.semanticscholar.org/CorpusID:118964051}
}

@article{Johnson-Freyd:2020usu,
    author = "Johnson-Freyd, Theo",
    title = "{On the Classification of Topological Orders}",
    eprint = "2003.06663",
    archivePrefix = "arXiv",
    primaryClass = "math.CT",
    doi = "10.1007/s00220-022-04380-3",
    journal = "Commun. Math. Phys.",
    volume = "393",
    number = "2",
    pages = "989--1033",
    year = "2022"
}

@article{Delcamp:2019fdp,
    author = "Delcamp, Clement and Tiwari, Apoorv",
    title = "{On 2-form gauge models of topological phases}",
    eprint = "1901.02249",
    archivePrefix = "arXiv",
    primaryClass = "hep-th",
    doi = "10.1007/JHEP05(2019)064",
    journal = "JHEP",
    volume = "05",
    pages = "064",
    year = "2019"
}

@article{Kapustin:2013uxa,
    author = "Kapustin, Anton and Thorngren, Ryan",
    title = "{Higher Symmetry and Gapped Phases of Gauge Theories}",
    eprint = "1309.4721",
    archivePrefix = "arXiv",
    primaryClass = "hep-th",
    doi = "10.1007/978-3-319-59939-7_5",
    journal = "Prog. Math.",
    volume = "324",
    pages = "177--202",
    year = "2017"
}

@book{HatcherAT,
  address = {Cambridge},
  author = {Hatcher, Allen},
  isbn = {0-521-79160-X; 0-521-79540-0},
  mrclass = {55-01 (55-00)},
  mrnumber = {1867354 (2002k:55001)},
  mrreviewer = {Donald W. Kahn},
  pages = {xii+544},
  publisher = {Cambridge University Press},
  title = {Algebraic topology},
  username = {mwpb479},
  year = 2002
}

@book{Etingof,
  author = {Etingof, Pavel and Gelaki, Shlomo and Nikshych, Dmitri and Ostrik, Victor},
  publisher = {In Mathematical surveys and monographs},
  title = {Tensor categories},
  year = 2015
}

@inproceedings{Brown1982CohomologyOG,
  title={Cohomology of Groups},
  author={Kenneth S. Brown},
  year={1982},
  url={https://api.semanticscholar.org/CorpusID:122284721}
}

@article{Kapustin:2013qsa,
    author = "Kapustin, Anton and Thorngren, Ryan",
    title = "{Topological Field Theory on a Lattice, Discrete Theta-Angles and Confinement}",
    eprint = "1308.2926",
    archivePrefix = "arXiv",
    primaryClass = "hep-th",
    doi = "10.4310/ATMP.2014.v18.n5.a4",
    journal = "Adv. Theor. Math. Phys.",
    volume = "18",
    number = "5",
    pages = "1233--1247",
    year = "2014"
}

@article{Maier2011EquivariantMC,
  title={Equivariant Modular Categories via Dijkgraaf-Witten Theory},
  author={Jennifer Maier and Thomas Nickelsen Nikolaus and Christoph Schweigert},
  journal={Advances in Theoretical and Mathematical Physics},
  year={2011},
  volume={16},
  pages={289-358},
  url={https://api.semanticscholar.org/CorpusID:55084371}
}

@book{mackey1976theory,
  title={The Theory of Unitary Group Representations},
  author={Mackey, G.W. and Kaplansky, I.},
  isbn={9780226500522},
  lccn={76017697},
  series={Chicago lectures in mathematics},
  url={https://books.google.com/books?id=oeZzQgAACAAJ},
  year={1976},
  publisher={University of Chicago Press}
}

@book{ceccherini2022representation,
  title={Representation Theory of Finite Group Extensions: Clifford Theory, Mackey Obstruction, and the Orbit Method},
  author={Ceccherini-Silberstein, T. and Scarabotti, F. and Tolli, F. and Springer Nature},
  isbn={9783031138744},
  series={Springer Monographs in Mathematics},
  url={https://books.google.com/books?id=X3inzwEACAAJ},
  year={2022},
  publisher={Springer International Publishing}
}

@article{Cui2015OnGS,
  title={On Gauging Symmetry of Modular Categories},
  author={Shawn X. Cui and C{\'e}sar Galindo and Julia Yael Plavnik and Zhenghan Wang},
  journal={Communications in Mathematical Physics},
  year={2015},
  volume={348},
  pages={1043-1064},
  url={https://api.semanticscholar.org/CorpusID:119126070}
}

@article{Schreier1926berDE,
  title={{\"U}ber die Erweiterung von Gruppen I},
  author={Otto Schreier},
  journal={Monatshefte f{\"u}r Mathematik und Physik},
  year={1926},
  volume={34},
  pages={165-180},
  url={https://api.semanticscholar.org/CorpusID:124731047}
}

@article{Lurie2009OnTC,
  title={On the Classification of Topological Field Theories},
  author={Jacob Lurie},
  journal={arXiv: Category Theory},
  year={2009},
  url={https://api.semanticscholar.org/CorpusID:115162503}
}

@article{Baez:1995xq,
    author = "Baez, J. C. and Dolan, J.",
    title = "{Higher dimensional algebra and topological quantum field theory}",
    eprint = "q-alg/9503002",
    archivePrefix = "arXiv",
    doi = "10.1063/1.531236",
    journal = "J. Math. Phys.",
    volume = "36",
    pages = "6073--6105",
    year = "1995"
}

@article{Kong2013AnyonCA,
  title={Anyon condensation and tensor categories},
  author={Liang Kong},
  journal={Nuclear Physics},
  year={2013},
  volume={886},
  pages={436-482},
  url={https://api.semanticscholar.org/CorpusID:119596357}
}

@article{Thorngren:2015gtw,
    author = "Thorngren, Ryan and von Keyserlingk, Curt",
    title = "{Higher SPT's and a generalization of anomaly in-flow}",
    eprint = "1511.02929",
    archivePrefix = "arXiv",
    primaryClass = "cond-mat.str-el",
    month = "11",
    year = "2015"
}

@article{Mller2018ParallelTO,
  title={Parallel transport of higher flat gerbes as an extended homotopy quantum field theory},
  author={Lukas M{\"u}ller and Lukas Woike},
  journal={Journal of Homotopy and Related Structures},
  year={2018},
  volume={15},
  pages={113 - 142},
  url={https://api.semanticscholar.org/CorpusID:84840634}
}

@article{Muller:2018doa,
    author = {M{\"u}ller, Lukas and Szabo, Richard J.},
    title = "{{\textquoteright}t Hooft Anomalies of Discrete Gauge Theories and Non-abelian Group Cohomology}",
    eprint = "1811.05446",
    archivePrefix = "arXiv",
    primaryClass = "hep-th",
    reportNumber = "EMPG-18-23",
    doi = "10.1007/s00220-019-03546-w",
    journal = "Commun. Math. Phys.",
    volume = "375",
    number = "3",
    pages = "1581--1627",
    year = "2019"
}

@article{Kapustin:2014zva,
    author = "Kapustin, Anton and Thorngren, Ryan",
    title = "{Anomalies of discrete symmetries in various dimensions and group cohomology}",
    eprint = "1404.3230",
    archivePrefix = "arXiv",
    primaryClass = "hep-th",
    month = "4",
    year = "2014"
}

@article{Wang:2021nrp,
    author = "Wang, Qing-Rui and Ning, Shang-Qiang and Cheng, Meng",
    title = "{Domain Wall Decorations, Anomalies and Spectral Sequences in Bosonic Topological Phases}",
    eprint = "2104.13233",
    archivePrefix = "arXiv",
    primaryClass = "cond-mat.str-el",
    month = "4",
    year = "2021"
}

@article{Etingof2009FusionCA,
  title={Fusion categories and homotopy theory},
  author={Pavel Etingof and Dmitri Nikshych and Victor Ostrik and with an appendix by Ehud Meir},
  journal={arXiv: Quantum Algebra},
  year={2009},
  url={https://api.semanticscholar.org/CorpusID:1727270}
}

@article{Barkeshli:2022edm,
    author = "Barkeshli, Maissam and Chen, Yu-An and Hsin, Po-Shen and Kobayashi, Ryohei",
    title = "{Higher-group symmetry in finite gauge theory and stabilizer codes}",
    eprint = "2211.11764",
    archivePrefix = "arXiv",
    primaryClass = "cond-mat.str-el",
    doi = "10.21468/SciPostPhys.16.4.089",
    journal = "SciPost Phys.",
    volume = "16",
    number = "4",
    pages = "089",
    year = "2024"
}

\end{document}